\documentclass[sigconf, screen, nonacm]{acmart}
\usepackage{color,soul}
\usepackage{xspace}
\usepackage{enumitem}
\usepackage{multirow}

\usepackage[capitalise,noabbrev]{cleveref}
\usepackage{subcaption}  
\usepackage{textcomp}
\usepackage{orcidlink}
\usepackage{tikz} 
\usepackage{fancybox}
\usepackage[title]{appendix}
\PassOptionsToPackage{hyphens}{url}
\usepackage{hyperref} 

\definecolor{lightgray}{gray}{0.95}

\hypersetup{
    colorlinks=true,
    linkcolor=red!50!black,
    citecolor=blue,
    urlcolor=blue!80!black
}
\AtBeginDocument{%
  }

\setcopyright{none}

\newcommand{\ignore}[1]{}

\newcommand{\blue}[1]{{\color{black}#1}}

\definecolor{itemhead}{RGB}{34, 87, 122}

\begin{document}
\title{When Mitigations Backfire: Timing Channel Attacks and Defense for PRAC-Based RowHammer Mitigations}


\author{Jeonghyun Woo\,\orcidlink{0000-0001-5819-0693}}
\affiliation{%
  \institution{The University of British Columbia}
  \department{Department of Electrical and Computer Engineering}
  \city{Vancouver}
  \state{BC}
  \country{Canada}
}
\email{jhwoo36@ece.ubc.ca}

\author{Joyce Qu\,\orcidlink{0009-0009-0104-3276}}
\affiliation{%
  \institution{University of Toronto}
  \department{Department of Computer Science}
  \city{Toronto}
  \state{ON}
  \country{Canada}
}
\email{joyce.qu@mail.utoronto.ca}

\author{Gururaj Saileshwar\,\orcidlink{0000-0003-3542-2548}}
\affiliation{%
  \institution{University of Toronto}
  \department{Department of Computer Science}
  \city{Toronto}
  \state{ON}
  \country{Canada}
}
\email{gururaj@cs.toronto.edu}

\author{Prashant J. Nair\,\orcidlink{0000-0002-1732-4314}}
\affiliation{%
  \institution{The University of British Columbia}
  \department{Department of Electrical and Computer Engineering}
  \city{Vancouver}
  \state{BC}
  \country{Canada}
}
\email{prashantnair@ece.ubc.ca}

\renewcommand{\shortauthors}{Jeonghyun Woo, Joyce Qu, Gururaj Saileshwar, Prashant J. Nair}



\newcommand{\red}[1]{{\color{red}#1}}
\newcommand{\JH}[1]{{\color{blue}\textbf{JH:} #1}}
\newcommand{\topic}[1]{\noindent\textbf{#1:}}

\newcommand{\circled}[1]{\textcircled{\raisebox{0.2pt}{\scriptsize #1}}}

            
\newcommand{\Attackname}[1]{\textit{PRACLeak}}
\newcommand{\Defensename}[1]{\textit{TPRAC}}
\newcommand{\DefensenameS}[1]{\textit{TPRAC}}
\newcommand{\DefensenameP}[1]{\textit{TPRAC}}

\newcommand{\AlertBackoff}{Alert Back-Off\xspace}
\newcommand{\NRH}{$\text{N}_{\text{RH}}$\xspace}
\newcommand{\NBO}{$\text{N}_{\text{BO}}$\xspace}
\newcommand{\ABOACT}{$\text{ABO}_{\text{ACT}}$\xspace}
\newcommand{\ABODELAY}{$\text{ABO}_{\text{Delay}}$\xspace}
\newcommand{\NMIT}{$\text{N}_{\text{mit}}$\xspace}
\newcommand{\NONLINE}{$\text{N}_{\text{online}}$\xspace}
\newcommand{\R}[1]{\text{R}\textsubscript{#1}\xspace}
\newcommand{\RFMAB}{$\text{RFM}_{\text{ab}}$\xspace}
\newcommand{\RFMPB}{$\text{RFM}_{\text{pb}}$\xspace}
\newcommand{\RFMSB}{$\text{RFM}_{\text{sb}}$\xspace}
\newcommand{\RFMSBPB}{$\text{RFM}_{\text{sb/pb}}$\xspace}
\newcommand{\BR}{\text{BR}}
\newcommand{\ALERT}{\text{Alert}}
\newcommand{\RFM}{\text{RFM}}
\newcommand{\ACT}{\text{ACT}}
\newcommand{\PRE}{\text{PRE}}
\newcommand{\TRC}{\text{tRC}}
\newcommand{\TREFW}{\text{tREFW}}
\newcommand{\TREFI}{\text{tREFI}}
\newcommand{\TRFC}{\text{tRFC}}
\newcommand{\REF}{\text{REF}}
\newcommand{\TRFM}{\text{tRFM}}
\newcommand{\TBRFM}{\text{TB-RFM}}
\newcommand{\ABOACBRFM}{\text{ABO+ACB-RFM}}
\newcommand{\ABOONLY}{\text{ABO-Only}}
\newcommand{\ABORFM}{\text{ABO-RFM}}
\newcommand{\ACBRFM}{\text{ACB-RFM}}
\newcommand{\TREF}{\text{TREF}}
\newcommand{\TBWINDOW}{\text{TB-Window}}

\begin{abstract}
Per Row Activation Counting (PRAC) has emerged as a robust framework for mitigating RowHammer (RH) vulnerabilities in modern DRAM systems. However, we uncover a critical vulnerability: a timing channel introduced by the Alert Back-Off (ABO) protocol and Refresh Management (RFM) commands. We present \textit{PRACLeak}, a novel attack that exploits these timing differences to leak sensitive information, such as secret keys from vulnerable AES implementations, by monitoring memory access latencies.

To counter this, we propose \textit{Timing-Safe PRAC (TPRAC)}, a defense that eliminates PRAC-induced timing channels without compromising RH mitigation efficacy. \textit{TPRAC} uses \emph{Timing-Based RFMs}, issued periodically and \emph{independent} of memory activity. It requires only a single-entry in-DRAM mitigation queue per DRAM bank and is compatible with existing DRAM standards. Our evaluations demonstrate that \textit{TPRAC} closes timing channels while incurring only 3.4\% performance overhead at the RH threshold of 1024.
\end{abstract}
\maketitle
\section{Introduction}
Technology scaling has enabled high-capacity DRAMs~\cite{archshield}, but also introduced security and reliability challenges such as RowHammer (RH)~\cite{kim2014flipping, kim2014architectural}. To address this, JEDEC recently introduced Per Row Activation Counting (PRAC) in DDR5~\cite{jedec_ddr5_prac}, which precisely tracks row activations and provides robust RH protection. While PRAC successfully mitigates RH, this paper reveals a new critical security threat: information leakage through timing channels. Our goal is to study the extent of these timing channel risks and to propose a low-cost and secure defense that eliminates them.

The PRAC specification provides robust RH mitigation, even at ultra-low RH thresholds~\cite{qureshi2024moat, Chronus, qprac}. PRAC achieves this by precisely tracking row activations using its per-row counters and enabling the DRAM to request mitigation time from the memory controller via the Alert Back-Off (ABO) protocol~\cite{jedec_ddr5_prac}. When a row's activation counter reaches the Back-Off threshold (\NBO{}), DRAM asserts the Alert, prompting the memory controller to issue a Refresh Management (RFM) command, termed \ABORFM{}, which blocks all memory requests (e.g., for 350ns) to allow RH mitigation. Alternatively, the memory controller may proactively send RFMs when any bank experiences a high number of activations~\cite{jedec_ddr5_prac}; we call these Activation-Based RFMs (\ACBRFM{}s). Although effective against RH, we observe that they can inadvertently introduce a new vulnerability--timing channels due to these mitigation actions.

\begin{figure}[b]
\centering
\vspace{-0.25in}
\includegraphics[width=2.6in,keepaspectratio]{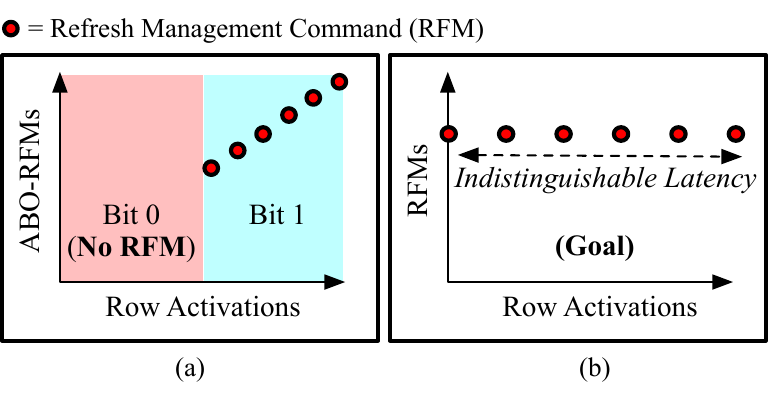}
\vspace{-0.18in}
\caption{Timing Channels in PRAC. Alert Back-Off-triggered RFMs (\ABORFM{}s) introduce activation-dependent latency, enabling timing channels. An adversary can exploit these to transmit `Bit-0' or `Bit-1'. Our goal is to remove this dependency and make RFMs timing independent of activations.}
\Description{}
\label{fig:intro}
\end{figure}

\cref{fig:intro}(a) shows how high row activations can create timing channels in PRAC-enabled systems. As row activation frequencies increase and consequently more rows surpass \NBO{}, ABO protocols are triggered more often. This, in turn, leads to an increasing number of \ABORFM{}s being issued for mitigation. Each triggered \ABORFM{} introduces a 350ns delay at the memory controller, uniformly increasing latency for all concurrent memory accesses to that DRAM channel. These latency variations can be exploited to transmit `Bit-0' or `Bit-1' in a cross-core covert channel.

While memory systems have many sources of latency variations, PRAC-based timing channels are uniquely precise, enabling exact values to be reliably transmitted across processes. For instance, a sender can activate a row exactly the \NBO{}$-X$ times; subsequently, a receiver performing precisely $X$ additional activations on the same row can deterministically trigger an RFM due to ABO protocol (\ABORFM{}). We demonstrate that this capability can be exploited to establish both covert and side channels. Thus, our goal is to prevent such timing channels by ensuring RFMs are issued independently of row activations, as shown in \cref{fig:intro}(b).

This paper observes that all prior RFM-based mitigations--wheth\-er PRAC-based~\cite{Chronus,qureshi2024moat,qprac} or not~\cite{kim2022mithril, jaleel2024pride,MINT}--issue RFMs based on DRAM row activity. Such \emph{activity-dependent} RFMs can be exploited to create timing channels: in PRAC-based mitigations, attackers can observe a victim's row activation patterns via \ABORFM{}s, while in non-PRAC mitigations, they can infer a victim's overall activation levels via Activation-Based RFMs (\ACBRFM{}s). We focus on timing channels arising from PRAC-based mitigations due to their expected widespread adoption in the near future.

This paper makes two key contributions. First, to demonstrate and analyze the threat posed by PRAC-induced timing channels, we introduce a new class of attacks, \Attackname{}, which exploits timing variations caused by ABO protocol in PRAC to leak sensitive information. 
Second, we propose a novel defense, \textit{Timing-Safe PRAC (TPRAC)}, which eliminates these timing channels through a proactive, \emph{activation-independent} mitigation strategy. \Defensename{} is designed to break the link between row activation patterns and memory access latency, effectively neutralizing \Attackname{} attacks while preserving strong RowHammer protection.

\smallskip
\topic{1. The \Attackname{} Attacks} 
Our \Attackname{} attacks are built on two key observations. First, PRAC-based mitigations~\cite{qureshi2024moat,Chronus,qprac} issue RFMs based on DRAM row activation counts. Second, the RFMs triggered by the ABO protocol (\ABORFM{}) introduce latency for all concurrent memory accesses, making both their occurrence and associated delays observable system-wide. When an ABO is initiated by the DRAM, the memory controller issues an RFM All Bank (\RFMAB{}) command. \RFMAB{} halts all memory requests for a predefined period (e.g., 350ns) to allow the DRAM to perform necessary mitigations, increasing the latency of affected memory accesses. This latency impact becomes more pronounced if PRAC is configured to use 2 or 4 RFMs per ABO, extending the delay to 700 or 1400ns. 
These timing differences can be exploited to form a timing channel for leaking information in both cross-process and cross-VM settings. The only requirement is that the attacker and victim share a memory module. \Attackname{} thus enables both covert and side channels, allowing attackers to establish two types of channels:

\smallskip
\begin{itemize}[leftmargin=*,topsep=0pt]
    \item \textbf{Activity-Based Channels}: In this channel, the sender and receiver share the same DRAM channel. The sender transmits a single bit of information over a fixed time window by either activating a row beyond \NBO{} times (to trigger an \ABORFM{}) or not. The receiver monitors the latency of its own memory accesses and decodes the transmitted bit based on the presence (`Bit-1') or absence (`Bit-0') of a latency spike.
    \item \textbf{Activation-Count-Based Channels}: In this channel, the sender and receiver share a single physical DRAM row. This is feasible if (1) the size of DRAM row (e.g., 8KB) exceeds the physical page size (e.g., 4KB), or (2) the memory controller uses address mapping that spreads data from a single page across multiple DRAM banks to enhance bank-level parallelism, which is common in modern systems~\cite{DRAMA2016, wang_2020_dramdig}. This allows two processes to occupy separate pages within the same row. Here, the receiver can leak the exact activation count of the sender. For example, if \NBO{} is 500, and the receiver observes a latency spike after 1 activation, the receiver knows the sender activated that row 499 times. 
    Thus, the sender can transmit any values up to \NBO{} (i.e., $log_{2}(N_{BO})$ bits).
\end{itemize}

Using a PRAC~\cite{Chronus} model in Ramulator2~\cite{ramulator2, kim2015ramulator}, we demonstrate covert channels achieving bitrates of 11–41 Kbps with activity-based channels and 39–124 Kbps with activation-count-based channels in a cross-process setting.  We also demonstrate a side-channel attack on an AES T-Table implementation, leaking secret key bits in under 200 encryptions via the activation-count-based channel.

\smallskip
\topic{2. \Defensename{} Mitigation} \textit{Timing-Safe PRAC (TPRAC)} eliminates PRAC-induced timing channels by replacing \emph{activity-dependent} mitigations with an \emph{activity-independent} strategy. It introduces \textit{Timing-Based RFMs (TB-RFMs)}, issued at fixed time intervals and completely independent of row activation behavior. By decoupling RFMs from memory access patterns, \Defensename{} prevents attackers from inferring activity through latency variations. The time interval between consecutive TB-RFMs (\TBWINDOW{}) is configured to ensure no DRAM row reaches \NBO{}, even under worst-case attack patterns (e.g., Feinting~\cite{ProTRR} or Wave~\cite{wave} attacks). At each \TBRFM{}, \Defensename{} proactively mitigates the most heavily activated row--regardless of whether it has reached \NBO{}--using a \emph{single-entry} mitigation queue per DRAM bank. This preemptive approach eliminates ABO events and the resulting timing channels. \Defensename{} requires no DRAM modifications and is fully compatible with existing interfaces, making it both secure and practical. For example, at the RH threshold of 1024, issuing one \TBRFM{} every 1.6 \TREFI{} suffices to close timing channels with just a 3.4\% performance slowdown.

\smallskip
\topic{Summary of Contributions} 
\begin{itemize}[leftmargin=*,topsep=0pt]
    \item \textbf{Unveiling PRAC Timing Channel Vulnerabilities}: To the best of our knowledge, we are the first to reveal the timing channel vulnerability of any RowHammer mitigations. We demonstrate how timing variations introduced by the ABO protocol of PRAC can be exploited to create a new class of timing channels.
    
    \item \textbf{\Attackname{} Covert-Channel Attacks}: Exploiting ABO-induced timing differences, we demonstrate covert-channel attacks achieving 39–124 Kbps when the sender and receiver share a DRAM row, and 11–41 Kbps without requiring shared rows.
    
    \item \textbf{\Attackname{} Side-Channel Attacks}: Using the same timing channel, we demonstrate a side-channel attack that successfully leaks bits of a secret AES key from a victim running a vulnerable AES T-Table implementation in less than 200 encryptions.

    \item \textbf{\Defensename{} Defense}: We propose \Defensename{}, a low-cost defense that eliminates PRAC-induced timing channels by proactively issuing RFMs at fixed intervals, independent of row activations. \Defensename{} requires only a single-entry mitigation queue per bank.
\end{itemize}

\section{Background and Motivation}\label{sec:background&motivaiton}
\subsection{Threat Model}
We consider a realistic threat model where an adversary and a victim share a DRAM module implementing PRAC-based RowHammer (RH) mitigations~\cite{qureshi2024moat, Chronus, qprac}. The adversary can operate with user-level privileges and run a malicious process to induce Refresh Management (RFM) commands and exploit the resulting timing variations to establish a side or covert channel. By measuring performance metrics such as memory access latency and execution time, the adversary can infer the victim's memory access patterns, enabling them to leak sensitive information (side channel) or transmit secrets (covert channel). We assume that all users can monitor the total number of RFMs within a given time window by issuing memory requests and detecting latency spikes caused by RFMs.

\subsection{The RowHammer Vulnerability}
RowHammer (RH) is a read disturbance issue where rapid activations of DRAM rows (aggressor rows) accelerate charge leakage in neighboring rows (victim rows), leading to bit-flips~\cite{kim2014architectural, kim2014flipping, autorfm_hpca25}. The RH has worsened with continued DRAM scaling, as cells become smaller and closely packed. The RowHammer threshold (\NRH{}), the minimum number of activations required to induce bit flips, has decreased from 69.2K in 2013~\cite{kim2014flipping} to nearly 4.8K in 2020~\cite{kim2020revisitingRH}, and is expected to drop further with future smaller technology nodes.

RH poses a severe security threat and has been exploited in numerous ways~\cite{deepsteal, deepvenom, gruss2016rhjs, cojocar2019eccploit, gruss2018another, vanderveen2016drammer, flipfengshui, prisonbreak}. Example attacks include privilege escalation~\cite{seaborn2015exploiting}, confidential data leakage~\cite{kwong2020rambleed}, degradation of ML model accuracy~\cite{hong2019terminal, deephammer}, and memory performance attacks (Denial-of-Service)~\cite{dapper, sgx-bomb}. As \NRH{} drops, RH can even occur with benign applications, making it a critical reliability concern~\cite{loughlin2022moesi}. 

\subsection{Per Row Activation Counting (PRAC)}\label{subsec:PRAC}
The latest DDR5 specification introduces Per Row Activation Counting (PRAC) to address RH~\cite{jedec_ddr5_prac}. PRAC employs two key mechanisms:

\smallskip
\begin{itemize}[leftmargin=*,topsep=0pt]
\item \textbf{Per-Row Activation Counters}: Each DRAM row includes additional activation counter cells and sense amplifiers. On each activation, the corresponding row’s counter is incremented; this requires a Read-Modify-Write operation during the precharge phase (i.e., when the row is being closed)~\cite{jedec_ddr5_prac}. To support this, Relevant DRAM timings (e.g., tRP and tWR) were adjusted in the PRAC specification~\cite{jedec_ddr5_prac}, which we incorporate in our evaluation.

\item \textbf{Alert Back-Off (ABO) Protocol}: When a row's counter crosses the Back-Off threshold (\NBO{}), the DRAM asserts the \ALERT{} signal to request additional mitigation time from the memory controller. In response, the memory controller permits a limited number of additional activations (\ABOACT{}) before initiating the ABO mitigation period. During the mitigation period, the memory controller issues a predefined number of RFM All Bank (\RFMAB{}) commands, referred to as the PRAC Level (\NMIT{}). Each \RFMAB{} blocks all memory requests for 350ns (t\RFMAB{}), allowing the DRAM to refresh \emph{four} victim rows and \emph{reset} its counter. A subsequent \ALERT{} can only be triggered after a specified number of activations (\ABODELAY{}), which is set equal to \NMIT{}.
\end{itemize}

\smallskip

The PRAC specification also includes proactive Activation-Based RFMs (\ACBRFM{}s), known as Targeted RFMs, which are triggered when the number of activations of a DRAM bank hits a threshold called the Bank Activation threshold (BAT). BAT is typically set below \NBO{} (e.g., 75) to reduce or avoid frequent Alerts~\cite{jedec_ddr5_prac}. Additionally, the specification allows DRAM to perform extra mitigations using the slack time of refresh operations, called Targeted Refresh (\TREF{}), similar to Target Row Refresh (TRR)~\cite{JEDEC-DDR4, hassan2021UTRR}. As we later show in \cref{sec:targeted_ref_design}, \TREF{} can be leveraged to improve the performance of \Defensename{}. \cref{table:prac_params} shows the key PRAC parameters.

\begin{table}[h]
  \centering
  \caption{PRAC Parameters as per JEDEC specification~\cite{jedec_ddr5_prac}}
  \begin{footnotesize}
  \label{table:prac_params}
  \begin{tabular}{lcc}
    \hline
    \textbf{Parameter} & \textbf{Explanation} & \textbf{Value} \\ \hline
    \NMIT{}     & Num \RFM{}s on \ALERT{} (PRAC Level) & 1, 2, or 4 \\ 
    \ABOACT{}  & Max. \ACT{}s to a row from \ALERT{} to \RFM{} & 3 (up to 180ns)  \\ 
    \ABODELAY{}       & Min. \ACT{}s after \RFM{} to \ALERT{} & Same as \NMIT{} (1,2, or 4) \\ 
    t\RFMAB{} &  Duration of RFM All Bank (\RFMAB{}) & 350ns \\\hline  
  \end{tabular}
  \end{footnotesize}
\end{table}

The PRAC specification provides minimal implementation details to allow DRAM vendors flexibility in their implementations. Consequently, PRAC's security and performance heavily depend on implementation decisions, particularly the design of the mitigation queue. For example, recent studies have shown that PRAC implementations using simple FIFO-based mitigation queues are vulnerable to targeted attacks~\cite{qureshi2024moat, qprac}, whereas omitting a mitigation queue entirely can incur severe performance overhead~\cite{qprac}. Therefore, achieving both security and performance in PRAC-enabled systems necessitates careful consideration of the mitigation queue design and its associated management strategy.

\subsection{Performance-Driven Timing Channels}
A timing channel exploits latency differences to transmit information, serving as a \emph{covert channel} for secret communication or a \emph{side channel} for leaking sensitive data. Performance optimizations in modern systems often create such channels by modulating data access latencies~\cite{DRAMA2016, LLCPrimeProbe, Augury, GoFetch}. Mitigation typically involves reducing or disabling these optimizations.

\subsection{PRAC: Security-Driven Timing Channel}\label{subsec:Timing-channel-vulnerability}
PRAC-based timing channels are security-driven~\cite{jedec_ddr5_prac}. Unlike perfo\-rmance-driven timing channels, which can often be mitigated by disabling performance optimizations, disabling PRAC exposes the system to RH attacks. These channels originate from the Alert Back-Off (ABO) protocol: the DRAM asserts the \ALERT{} when a row's activation count reaches \NBO{}. In response, the memory controller issues RFMs for mitigations, and the resulting latency spikes create observable timing variations that can be exploited.

\begin{figure}[b]
\centering
\includegraphics[width=0.9\columnwidth,height=\paperheight,keepaspectratio]{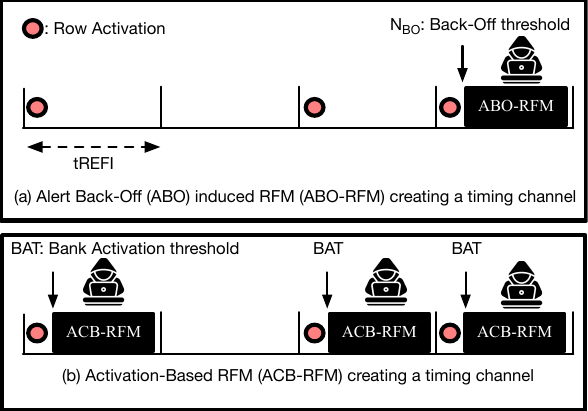}
\caption{The activity-dependent nature of PRAC's \RFM{}s allows attackers to exploit them. (a) Repeated activations of a DRAM row up to the Back-Off threshold (\NBO{}) trigger an RFM due to ABO (\ABORFM{}). (b) Activation-Based RFMs (\ACBRFM{}s), which mitigate proactively at lower thresholds, can avoid \ABORFM{}s but remain exploitable as attackers can still induce \ACBRFM{}s by activating rows within a bank.}
\Description{}
\label{fig:timingattack-motivation}
\end{figure}

\subsection{Pitfalls: ABO-RFM and \ACBRFM{}}\label{subsec:pitfalls}
\cref{fig:timingattack-motivation} illustrates how PRAC-based RH mitigations~\cite{qprac, qureshi2024moat, Chronus} rely on \emph{activity-dependent} RFMs, such as \ABORFM{} and \ACBRFM{}. In the case of \ABORFM{} (\cref{fig:timingattack-motivation}(a)), an attacker can repeatedly activate a DRAM row until its activation count reaches \NBO{}, triggering the ABO protocol. This results in a noticeable latency spike, which can be exploited as a timing channel. In contrast, \ACBRFM{} is triggered when the total number of activations in a bank crosses BAT, potentially avoiding \ABORFM{}s. However, since \ACBRFM{}s are still tied to activation levels, an adversary can exploit them instead of \ABORFM{}s to form timing channels (\cref{fig:timingattack-motivation}(b)).
\section{Demonstrating PRAC's Timing Channels}\label{sec:attack}
In this section, we demonstrate timing channels inherent in PRAC. While our analysis is based on Understanding PRAC (UPRAC)~\cite{UPRAC}, the presented attacks exploit fundamental features of the PRAC—the Alert Back-Off (ABO) protocol and per-row activation counters—ma\-king them applicable to other PRAC-based defenses, including QPRAC~\cite{qprac} and MOAT~\cite{qureshi2024moat}.

We first characterize the timing variations caused by PRAC's ABO protocol (\cref{sec:attack_char}). We then present the proposed \Attackname{} attacks: first, a covert channel between a spy (receiver) and a trojan (sender) (\cref{sec:attack_covert}), followed by a side-channel attack where a spy leaks information from a victim application (\cref{sec:attack_side}).

\subsection{Timing Variation due to \AlertBackoff{}}\label{sec:attack_char}
In PRAC, when any row's activation counter hits the Back-Off threshold (\NBO{}), the DRAM asserts the \ALERT{} signal to request time for RowHammer (RH) mitigations from the memory controller.  In response, the memory controller issues one or more RFM All Bank (\RFMAB{}) commands, each stalling all memory requests to the affected channel for \TRFM{} (typically 350ns). As a result, an ABO event introduces a system-wide latency spike, which can be observed by threads running on other cores.

Because ABO occurrences depend on the activation counts of specific physical rows--which can be influenced by an adversarial thread sharing access to these rows--they can leak information about which row was accessed and how frequently it was activated. 

\begin{figure}[t]
\centering
\includegraphics[width=0.9\columnwidth,height=\paperheight,keepaspectratio]{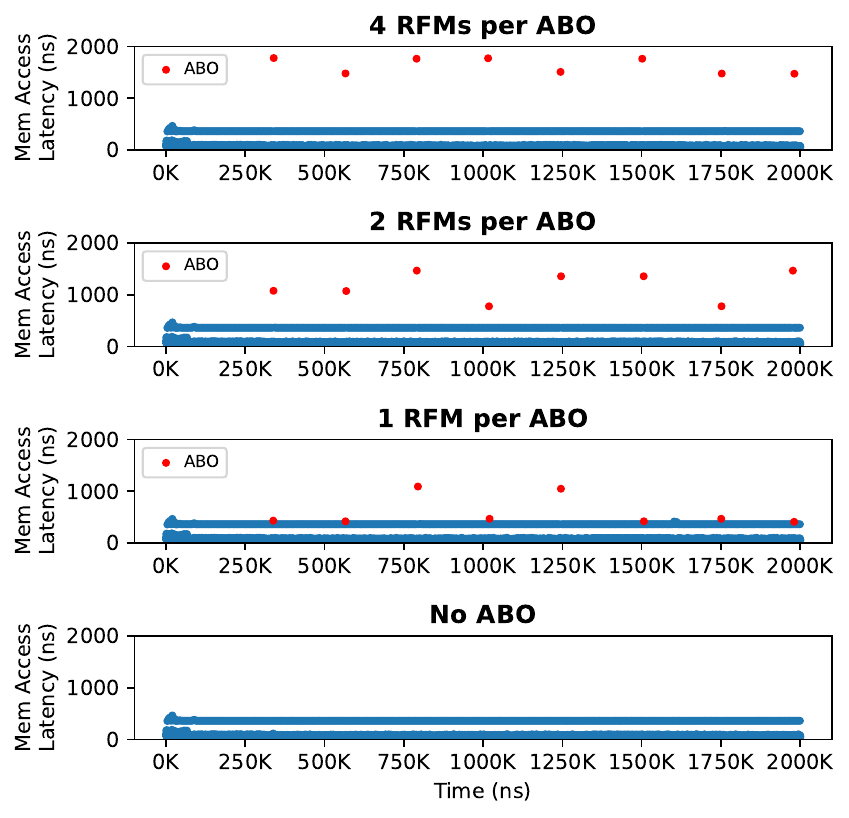}
\caption{Timing variation for memory accesses for an attacker in the presence and absence of a concurrent Alert Back-Off (ABO) due to victim's activity.}
\label{fig:timing_variation}
\end{figure}

\smallskip
\noindent \textbf{Characterization.} \cref{fig:timing_variation} shows how an adversary can detect an ABO by observing increased memory access latency during its occurrence. We assume the attacker (e.g., a process or VM) monitors its memory access latencies on a different core in the same system as a victim, whose repeated activations trigger an ABO. The attacker accesses a different bank in the same channel and performs repeated accesses--either to the same row in an open-page setting or to different rows in a closed-page setting--to avoid increasing activation counts on its own rows and prevent self-induced ABOs.

Without a concurrent ABO from the victim, the attacker's memory access latency remains relatively stable. However, when the victim's activations trigger an ABO, the attacker experiences a notable increase in its memory accesses latency. The PRAC specification allows for a configuration number of \RFM{}s (1, 2, or 4) per ABO. With 1 RFM per ABO, the attacker observes a latency of 545ns on average. This increases to 976ns and 1669ns with 2 and 4 RFMs, respectively, making the ABO induced by victim activity increasingly detectable.

\subsection{PRACLeak Covert-Channel Attacks}\label{sec:attack_covert}
Covert-channel attacks enable a trojan and a spy to communicate covertly by exploiting shared hardware resources. Our attack leverages the timing impact of the ABO on memory access latency to construct two types of covert channels: (1) an activity-based channel, where a single bit is transmitted based on the presence or absence of an ABO within a fixed time window, and (2) an activation-counter-based channel, where multiple bits are transmitted by encoding information into the timing of an ABO occurrence. We refer to these as the \Attackname{} Covert-Channel Attacks.

\smallskip
\noindent \textbf{(1) Activity-Based Covert Channel.} 
In this channel, following the strategy described in \cref{sec:attack_char}, the receiver (spy) repeatedly accesses a single memory location in a loop. Under an open-page policy, the receiver flushes the address from the cache to ensure the row-buffer hits. Under a closed-page policy, it accesses addresses mapped to different rows within a bank to prevent activation counters from reaching \NBO{}, thereby avoiding self-induced ABOs. To transmit a `Bit-1', the sender (trojan) activates a row in a different bank \NBO{} times, typically by alternating accesses to a pair of different rows to cause row-buffer conflicts and guarantee activations. To transmit a `Bit-0', the sender remains idle, avoiding an ABO. When a `Bit-1' is transmitted, the ABO triggered by the sender causes a detectable latency spike in the receiver's memory access, while the absence of such a spike indicates a `Bit-0'. Under a closed-page policy, the transmission period is approximately $\text{N}_\text{BO} \cdot \text{tRC} + \text{tRFM}$, reflecting the time required for \NBO{} activations and ABO servicing.

\smallskip
\noindent \textbf{(2) Activation-Count-Based Covert-Channel.} 
A more sophisticated covert channel can be established if the sender and receiver share the same DRAM row. This is possible when the physical page is smaller than the DRAM row, or when successive cache lines are mapped to different banks for bank-level parallelism~\cite{DRAMA2016, wang_2020_dramdig}, allowing both parties to access the same DRAM row from different physical pages. In this setup, the sender transmits multiple bits by encoding them into the row's activation counter. For example, within a fixed time window, the sender activates the shared row $k$ times (where $k<\text{N}_{\text{BO}}$). The receiver then activates the same row up to \NBO{} times and observes when an ABO-induced delay occurs--specifically, after $\text{N}_\text{BO} - k$ activations--allowing it to infer the value $k$. This enables the sender to transmit $\log(\text{N}_\text{BO})$ bits per time window. Since both the sender and receiver perform up to \NBO{} activations sequentially, the transmission period is nearly double that of the previous attack: approximately $2\cdot \text{N}_\text{BO} \cdot \text{tRC} + \text{tRFM}$. 

\smallskip
\noindent \textbf{Results.}
We implement these attacks in Ramulator2~\cite{ramulator2}, a trace-based DRAM simulator, using spy and trojan traces running on different CPU cores. We use the open-sourced UPRAC~\cite{UPRAC} as our PRAC implementation. While we use PRAC configuration of 4 RFMs per ABO, our bitrates are similar across 1, 2, and 4 RFMs per ABO. \cref{table:covertchannel} shows the transmission period (i.e., time for single transmission)  and bitrates for our two covert channels, evaluated with \NBO{} values ranging from 256 to 1024. At \NBO{} of 256, our activity-based covert channel achieves a transmission period of 24.1$\mu$s and a bitrate of 41.4Kbps. In contrast, our activation-count-based covert channel has a longer transmission period of 64.7~$\mu$s. Still, it achieves a higher bitrate of 123.6 Kbps, as it encodes log(\NBO{}) bits per period compared to the 1 bit per period in the activity-based channel. As \NBO{} increases, bitrates decrease due to longer transmission periods. The error rates are overall negligible ($<0.1\%$).

\begin{table}[!htb]
  \centering
  \caption{Covert Channel Transmission Period and bitrate}
  \vspace{-0.1in}
  \label{table:covertchannel}
  \begin{tabular}{clcc}
    \hline
    \multirow{2}{*}{\textbf{Type}}  & \multirow{2}{*}{\textbf{$\text{N}_\text{BO}$}} & \textbf{Transmission} & \textbf{bitrate} \\ 
                                    && \textbf{Period ($\mu$s)}  & \textbf{(Kbps)}   \\ \hline
    
    \multirow{3}{*}{Activity-Based} & 256  &  24.1     & 41.4 \\
                              & 512  &  46.7    & 21.4 \\
                              & 1024 &  91.8    & 10.9 \\ \hline
    \multirow{3}{*}{Activation-Count-Based} & 256    &  64.7     & 123.6 \\ 
                                        & 512    &  128.0    & 70.3 \\
                                        & 1024   &  257.6    & 38.8 \\ \hline
    
  \end{tabular}
\end{table}

\subsection{PRACLeak Side-Channel Attacks}\label{sec:attack_side}
This section extends the activation-counter-based covert channel into a side-channel attack. We show how an attacker can infer a victim's memory access patterns--and ultimately leak sensitive information--by observing ABO-induced timing variations. We refer to this as the \Attackname{} Side-Channel Attack.

\smallskip
\noindent \textbf{Threat Model.} We assume the attacker and the victim are in different processes or VMs running on different cores of the same system. They share physical DRAM rows, meaning victim secrets are co-located with attacker-controlled data. This can occur either because the DRAM row is larger than a physical page or due to physical-to-DRAM address mappings that stripe successive cache lines across banks for bank-level parallelism, causing data from different processes to map to the same row. The victim is repeatedly executed with attacker-controlled inputs, as in prior attacks~\cite{Spectre,Squip}.

\smallskip
\noindent \textbf{Attack Setup.}
Crypto libraries like OpenSSL and GnuPG provide an AES implementation with T-tables, known to be vulnerable to cache-based side-channel attacks that leak the secret key \cite{Bernstein,tmem:usenix2017,Bonneau}. We use this implementation as our target (victim) to demonstrate our ABO-based side-channel attack. Each of the ten rounds in this AES design performs 16 secret-key-dependent memory accesses to 4 T-tables (4 accesses per table). Each table spans 16 cache lines, with each cache line mapped to a different DRAM row. The T-table lookup indices depend on the input plaintext $p$ and the secret key $k$. For the first round, they are calculated as $x_i = p_i \bigoplus k_i$, where $i$ is the byte position in $p$ or $k$ ($i$ goes from 0 to 15).  

We perform a chosen plaintext attack on the first AES round. The attacker provides plaintexts to the victim for encryption, keeping $p_0$ constant while varying other bytes $p_i$ randomly for $i \neq 0$. This ensures that the index $x_0$ always accesses the same T-table cache line during the first round. The other $x_i$ ($i = $1 to 15) are randomly distributed across all the cache lines. As a result, the cache line corresponding to $x_0$ has more accesses than the others. Concurrently, the attacker repeatedly evicts these cache lines from all cache levels during the first round--either by flushing them (e.g., using clflush)~\cite{FlushReload} when the AES library is shared read-only, or by accessing eviction sets for these cache lines in the inclusive LLC~\cite{LLCPrimeProbe} or the inclusive coherence directory~\cite{DirectoryAttack}. For simplicity, we assume the attacker flushes these addresses from the caches in parallel with the victim's execution. Consequently, the DRAM row corresponding to $x_0$ experiences significantly more memory accesses than the other 15 DRAM rows containing the T-table.

\begin{figure}[t]
\centering
\includegraphics[width=0.45\textwidth]{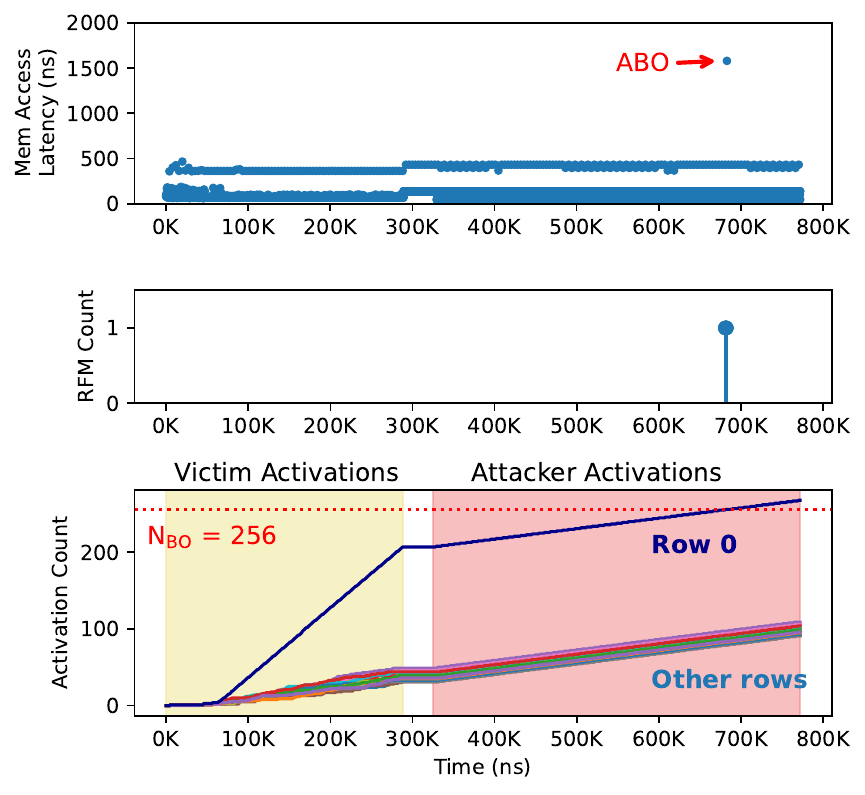}
\vspace{-0.1in}
\caption{\Attackname{} Side-Channel attack on AES T-tables for plaintext byte-0, $p_0 = 0$ with $k_0=0$. During victim activations, the Row-0 activation counts are approximately double those of the other rows. Thus, the attacker observing the high memory access latency of ABO due to activations on Row-0 learns that it is the most activated row.}
\label{fig:sidechannel_raw}
\end{figure}

\smallskip
\noindent \textbf{Attack Mechanism.}
The attacker ensures the victim performs $n$ encryption iterations with different plaintexts and the same secret key. In each iteration, the attacker ensures the victim accesses T-table entries during the first round. Subsequently, the attacker measures the activation counts of different DRAM rows using the ABO-based side channel. It sequentially activates each row ($R_0$, $R_1$,~\dots~, $R_{15}$) once and repeats this in a loop until one of them triggers an ABO. The first row to trigger an ABO is identified as the most frequently activated row, revealing the cache line number within the T-table that had the highest access count. This discloses 4 of the 8 bits of $x_0$ (i.e., the cache line index). Since $p_0$ is known, the attacker recovers 4 of the 8 bits of the key byte $k_0$. Repeating this process while fixing each $p_i$ ($i$ going from 0 to 15) allows the attacker to recover 64 out of 128 bits of the secret AES key.  

\smallskip
\noindent \textbf{Results.}
\cref{fig:sidechannel_raw} shows activation counts for Row-0 and other rows in an attack instance where $p_0 = 0$ and $k_0 = 0$. During the victim execution phase, the victim causes 207 activations to Row-0. In the attacker observation phase, an increase in memory access latency is observed after 49 activations to Row-0 due to an ABO. This is when the combined victim and attacker activations cross \NBO{}$=256$ for Row-0, the most heavily activated row. From this, the attacker learns the top four bits of $k_0$, the first key byte, are 0x0.

\begin{figure}[h]
\centering
\includegraphics[width=0.45\textwidth]{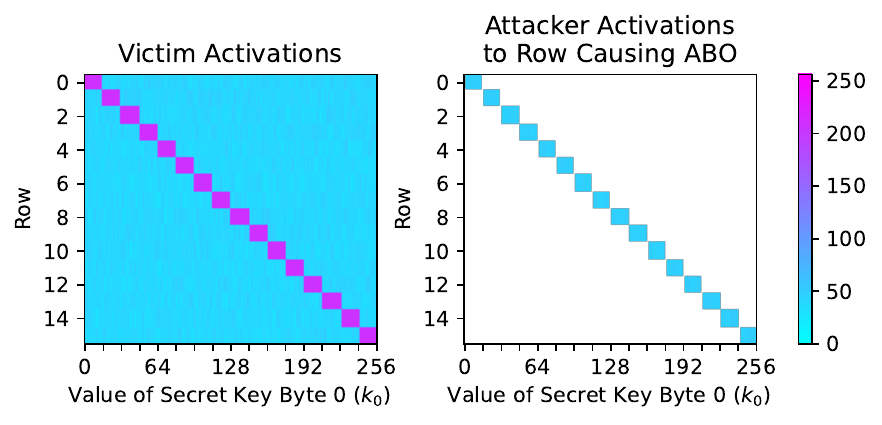}
\vspace{-0.1in}
\caption{Side-Channel attack on AES T-tables for varying key byte-0 $k_0$ values, when plaintext byte-0 $p_0 = 0$. (a)~Number of victim activations for DRAM rows after running 200 encryptions. (b) Number of attacker activations on row causing first ABO -- index of this row leaks out the key bits.}
\label{fig:sidechannel_heatmap}
\end{figure}

Building on this, we generalize our approach to demonstrate that for all secret key bytes $k_i$, information can be leaked from the secret-dependent most-activated row using our \Attackname{} side channel. \cref{fig:sidechannel_heatmap}(a) shows a heatmap of the number of activations across the 16 DRAM rows containing the T-tables after the victim program runs for 200 encryptions. As $k_0$ varies from 0 to 256, the DRAM row with the highest number of activations shifts from Row-0 to Row-15. \cref{fig:sidechannel_heatmap}(b) then shows the first row that triggers an ABO during the attacker's probing phase, along with its activation count. The total activations to this row--by both victim and attacker--sum to exactly 256, and the row index directly reveals the top 4 bits of key byte $k_0$. By varying the fixed plaintext byte $p_i$, the attacker can leak 4 of the 8 bits of each $k_i$, leaking 50\% of the 128-bit AES key.

\section{Timing-Safe PRAC (TPRAC)}\label{sec:mitigation-secruityOpt}
We propose \textit{Timing-Safe PRAC (TPRAC)}, an \emph{activity-independent} Refresh Management (\RFM{}) mechanism that eliminates timing channels in PRAC-based systems. \Defensename{} employs our proposed Timing-Based \RFM{} (\TBRFM{}), where the memory controller issues RFM All Bank (\RFMAB{}) commands at fixed intervals, denoted as \TBWINDOW{}, independent of memory activity. This closes timing channels while maintaining performance balance.

\begin{figure}[h!]
\centering
\includegraphics[width=3in,height=\paperheight,keepaspectratio]{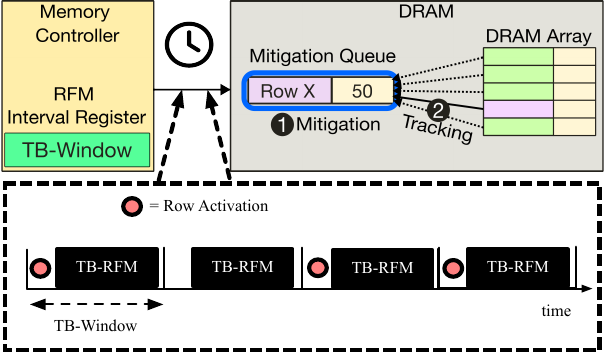}
\vspace{-0.05in}
\caption{Overview of \Defensename{} design. It employs an activity-independent RFM mechanism (Timing-Based \RFM{}) to eliminate timing channels in PRAC-based systems. These RFMs are issued with the fixed time windows called \TBWINDOW{}.}
\Description{}
\vspace{-0.05in}
\label{fig:tgprac-s_design}
\end{figure}

\cref{fig:tgprac-s_design} illustrates how \Defensename{} mitigates PRAC-based timing channels by issuing \TBRFM{}s periodically at each TB-Wind\-ow, configured to eliminate Alert Back-Off-triggered RFMs (ABO-RFMs). \cref{subsec:TGPRAC-S_Security} discusses how to determine an appropriate \TBWINDOW{} based on the worst-case attack pattern. \Defensename{} requires only a single register, the RFM Interval Register, to store \TBWINDOW{}. At the end of each window, it proactively issues an \RFMAB{} (as a \TBRFM{}) regardless of memory activity, preventing attackers from triggering \ABORFM{}s through targeted activations. \Defensename{} also disables RFM postponing~\cite{jedec_ddr5_prac} to prevent attacks that exploit delayed RFMs~\cite{ProTRR}.

\subsection{Underlying PRAC Implementation}\label{subsec:TPRAC_implementation}
Our proposed \TBRFM{} mechanism relies on the underlying PRAC implementation to perform RowHammer (RH) mitigation during each RFM. However, the PRAC specification~\cite{jedec_ddr5_prac} does not define implementation details such as the mitigation queue design. Thus, we outline the potential queue designs suitable for \Defensename{}.

\Defensename{} mitigates the most frequently activated row in each DRAM bank during every RFM, similar to prior secure PRAC designs~\cite{qprac, Chronus, qureshi2024moat}. We propose a frequency-based mitigation queue with a \emph{single-entry} per bank to achieve this practically. Each queue tracks the most highly activated row by storing its address and activation count, replacing the entry when a newly activated row exceeds the current counter (\circled{2} in \cref{fig:tgprac-s_design}). This ensures that the most activated row is always retained. Upon receiving an \RFM{}, \Defensename{} mitigates the row in each bank's queue, regardless of whether it has crossed the Back-Off threshold (\NBO{}) (\circled{1} in \cref{fig:tgprac-s_design}).

\cref{subsec:Mitigation-queue-analysis} shows that \TBRFM{}, combined with this queue design, achieves security equivalent to the idealized PRAC design, UPRAC~\cite{UPRAC}. Similar queue designs are employed in secure PRAC implementations like QPRAC~\cite{qprac} and MOAT~\cite{qureshi2024moat}, making \TBRFM{} readily compatible with these designs.

\subsection{Configuring the \TBWINDOW{} Interval}\label{subsec:TGPRAC-S_Security}
To ensure that \Defensename{} effectively eliminates \ABORFM{}s--and consequently Activation-Based RFMs (\ACBRFM{}s)--and prevents information leakage, we must determine an appropriate \TBWINDOW{} interval for each RowHammer threshold (\NRH{}). Our analysis begins with the idealized PRAC implementation, UPRAC~\cite{UPRAC}, which mitigates the most highly activated row in each DRAM bank during every \RFM{}. We then show that \Defensename{} with our proposed single-entry mitigation queue can achieve equivalent security.

\subsubsection{Worst-Case Attack Pattern and Assumptions}
Following prior studies on RFM~\cite{ProTRR, breakhammer2024}, we adopt the Feinting~\cite{ProTRR} (also known as Wave~\cite{wave}) attack as the basis for our worst-case analysis of \Defensename{}. This attack is mathematically proven to represent the worst-case access pattern for \RFM{}-based mitigations. It proceeds in multiple rounds using a pool of decoy rows and a target row. The decoy rows help bypass \RFM{}s, allowing the attacker to maximize activations on the target row. In each round, all rows in the pool are uniformly activated, and any mitigated rows are removed. This process continues until only the target row remains, enabling the adversary to concentrate all activations on it in the final round. For our security analysis of \Defensename{} under worst-case conditions, we make the following \emph{conservative} assumptions:

\begin{itemize}[leftmargin=*,topsep=1pt]
    \item \textbf{Single \ABORFM{} Leakage}: The adversary can leak information if even a single \ABORFM{} is triggered.
    \item \textbf{Exclusive DRAM Access}: The adversary has exclusive access to the DRAM module used by the victim, eliminating interference or noise from other processes.
    \item \textbf{Full System Knowledge}: The adversary knows all details of \Defensename{} and the PRAC implementation, including the Back-Off threshold (\NBO{}), mitigation queue design, and \TBWINDOW{}.
    \item \textbf{Precise Timing Measurement}: The adversary can accurately measure (1) the total number of \RFM{}s, (2) the exact time at which each \RFM{} is issued, and (3) the duration of each RFM-induced blocking period, by monitoring memory access latencies.
\end{itemize}

\subsubsection{Determining \TBWINDOW{} of \Defensename{}}
The adversary aims to trigger an \ABORFM{} by causing \NBO{} activations to a target row (T$_{\text{ACT}}$). To prevent information leakage, \Defensename{} must ensure this count stays below \NBO{}, as shown in the following equation:
\begin{equation}\label{equ:tgprac_s_safe}
    \text{T}_{\text{ACT}} < \text{N}_{\text{BO}}
\end{equation}

The adversary begins with an initial pool of R$_1$ rows. Since \Defensename{} performs one \TBRFM{} every \TBWINDOW{}, the maximum number of row activations possible within this interval, denoted as ACT$_{\text{TB-Window}}$, is given by:
\begin{equation}
    \text{ACT}_{\text{TB-Window}} = \frac{\text{TB-Window}}{\text{tRC}}
\end{equation}

The number of remaining rows after the first $N-1$ attack rounds (R$_\text{N}$) is computed by subtracting the number of TB-RFMs issued during those rounds from the initial row pool size (R$_1$). Since one \TBRFM{} is issued for every ACT$_{\text{TB-Window}}$ activations, we have:
\begin{equation}\label{equ:tgprac_s_Rn}
    \text{R}_\text{N} = \text{R}_1 - \left\lfloor \frac{\sum_{i = 1}^{{N}-1} \text{R}_{\text{i}}}{\text{ACT}_{\text{TB-Window}}} \right\rfloor, \quad \text{where } N \geq 2
\end{equation}

The total number of attack rounds (AR) can be computed from a given R$_{1}$ using \cref{equ:tgprac_s_Rn}. Since the target row is activated once per round until the final round, and all activations in the final round (ACT$_{\text{TB-Window}}$) go to the target row, the total number of activations to the target row (T$_{\text{ACT}}$) is given by:
\begin{equation}\label{equ:t_act}
    \text{T}_{\text{ACT}} = (\text{AR} - 1) + \text{ACT}_{\text{TB-Window}} 
\end{equation}

Lastly, we evaluate the worst-case number of activations to the target row (T$_\text{MAX}$) under two scenarios:

\smallskip
\begin{itemize}[leftmargin=*, topsep=0pt]
    \item \textbf{With Activation Counter Reset:} Per-row activation counters are reset at every refresh window (\TREFW{}, typically 32ms in DDR5), as proposed in prior work~\cite{qureshi2024moat}. Original Feinting attack shows that the worst-case occurs when R$_1$ equals the number of mitigations (i.e., TB-RFMs) that can be applied during the attack~\cite{ProTRR}. This gives the optimal R$_1$ value (OPT$_{\text{R}_1}$) as:
    \begin{equation}\label{equ:max_r1}
       \text{OPT}_{\text{R}_1} = \frac{\text{MAXACT}_{\text{tREFW}}}{\text{ACT}_{\text{TB-Window}}}     
    \end{equation}
    where MAXACT$_{\text{tREFW}}$ ($\sim$550K) is the maximum number of activations allowed within one \TREFW{}.
    \item \textbf{Without Activation Counter Reset:} Per-row activation counters are reset only when the row is mitigated via an RFM. In this case, we sweep R$_1$ from 1 up to its maximum possible value (128K for the DDR5 32Gb chip) to determine the value of OPT$_{\text{R}_1}$ that maximizes the number of activations to the target row.
\end{itemize}

\begin{figure}[h]
\centering
\includegraphics[width=3in,height=\paperheight,keepaspectratio]{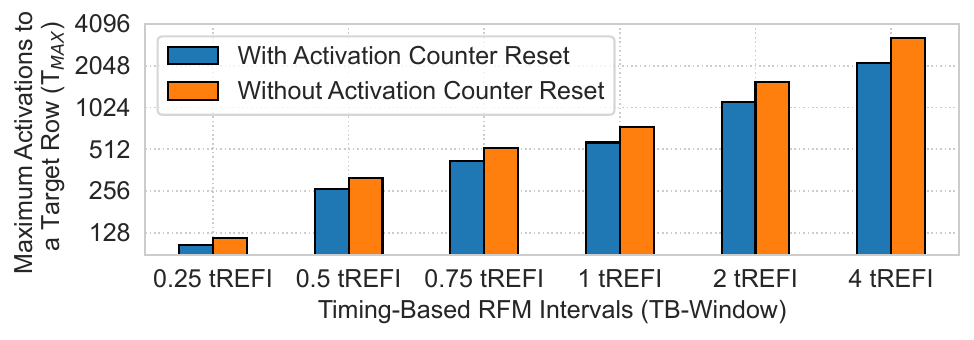}
\vspace{-0.1in}
\caption{Theoretical maximum activations to a target row (T$_{\text{MAX}}$) as the \TBWINDOW{} of \Defensename{} varies, shown for both with and without per-row activation counter reset at each refresh window (\TREFW{}). Results are based on a DDR5 32Gb chip with 128K rows per bank. T$_{\text{MAX}}$ must remain below the Back-Off threshold (\NBO) to avoid Alert Back-Off-triggered \RFM{}s (\ABORFM{}s) and close timing channels.}
\Description{}
\label{fig:tprac_max_act}
\end{figure}

\smallskip
\cref{fig:tprac_max_act} shows the theoretical maximum number of activations to the target row (T$_{\text{MAX}}$) across different \TBWINDOW{} values, comparing scenarios with and without per-row activation counter reset at each \TREFW{}. T$_{\text{MAX}}$ is consistently higher without counter reset due to the larger optimal initial row pool size (OPT$_{\text{R}_1}$). With counter reset, OPT$_{\text{R}_1}$ is constrained by how many \TBRFM{} intervals fit within a single \TREFW{} window. For example, at \TBWINDOW{} of 1~\TREFI{}, only 8192 intervals are possible, limiting OPT$_{\text{R}_1}$ to 8192--almost 16$\times$ smaller than the no-reset case (approximately 128K). As a result, T$_{\text{MAX}}$ reaches 736 without counter reset, compared to 572 with reset. The impact of the reset becomes more pronounced at longer \TBWINDOW{} intervals because fewer \TBRFM{}s fit within \TREFW{}, further reducing OPT$_{\text{R}_1}$. For instance, at \TBWINDOW{} of 4~\TREFI{}, T$_{\text{MAX}}$ reaches 3220 without reset and 2138 with reset. Conversely, at shorter intervals like 0.25~\TREFI{}, the difference narrows, with T${_\text{MAX}}$ values of 118 and 105, respectively.

\begin{figure*}
    \centering
    \includegraphics[width=6.7in,height=\paperheight,keepaspectratio]{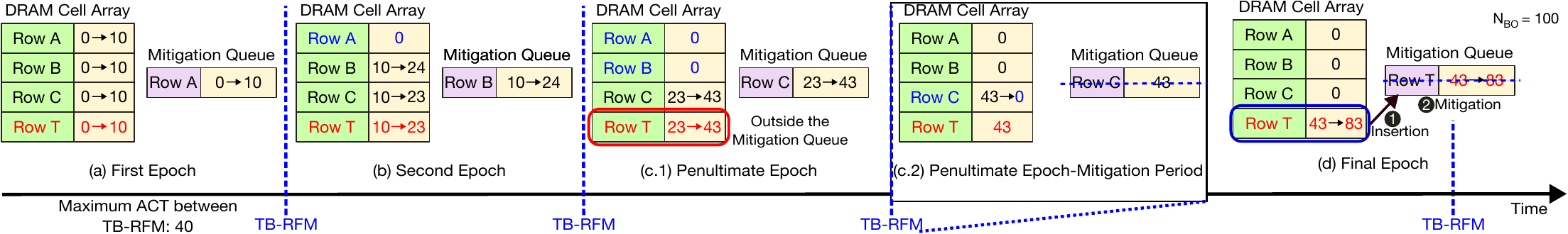}
    \caption{Example of how \Defensename{} prevents Feinting~\cite{ProTRR} or Wave~\cite{wave} attack with a single-entry mitigation queue.} 
    \Description{}
    \label{fig:tgprac-s_defending_feinting}
\end{figure*}

\subsubsection{Impact of the Mitigation Queue Structure of PRAC}\label{subsec:Mitigation-queue-analysis}
We demonstrate that a single-entry, frequency-based mitigation queue is sufficient to match the security guarantees of UPRAC. \cref{fig:tgprac-s_defending_feinting} illustrates through a simple Feinting attack, where the attacker uses three decoy rows alongside a target row (Row T) to bypass \TBRFM{}s and build up activations on Row T. In this example, up to 40 activations are allowed within each \TBWINDOW{}. During the first epoch, all four rows are activated uniformly, as shown by \cref{fig:tgprac-s_defending_feinting}(a). Row A, having the highest activation count, is tracked in the mitigation queue and then mitigated when the \TBRFM{} is issued. In the second epoch, the attacker continues activating Row T and the remaining decoys. Row B accumulates the most activations and is tracked and mitigated at the next \TBRFM{}, as shown by \cref{fig:tgprac-s_defending_feinting}(b). 

\smallskip
\topic{Final Attack Round} At the end of the penultimate epoch (\cref{fig:tgprac-s_defending_feinting}(c.1)), both the target row (Row T) and a decoy row (Row C) have equal activation counts (i.e., 43). Since the mitigation queue holds only a single entry, only one row--Row C--is tracked and subsequently mitigated when the \TBRFM{} is issued, as shown by \cref{fig:tgprac-s_defending_feinting}(c.2). During this mitigation period, memory requests are blocked for 350ns, preventing the attacker from issuing further activations, including to Row T. In the final round (\cref{fig:tgprac-s_defending_feinting}(d)), the attacker focuses activations solely on Row T. As soon as Row T is activated, Row T becomes the most active row and is inserted into the mitigation queue (\circled{1} in \cref{fig:tgprac-s_defending_feinting}). Before it reaches the Back-Off threshold (\NBO{} $= 100$ in this example), the next \TBRFM{} is issued, and Row T is mitigated (\circled{2} in \cref{fig:tgprac-s_defending_feinting}), ensuring no \ABORFM{} is triggered, consistent with our analysis in \cref{subsec:TGPRAC-S_Security}. 

Notably, if the attacker had begun targeting Row T earlier, it would have entered the mitigation queue and been mitigated sooner, further reducing the attack’s effectiveness. This behavior aligns with observations from the original Feinting attack~\cite{ProTRR}.

\smallskip
\topic{Scenario 1 -- Equal Activations}  
In this scenario, the attacker uniformly activates all rows in the initial pool across rounds, including rows that were mitigated in previous epochs. However, this strategy is suboptimal compared to the Feinting attack. Once a row is mitigated, its counter resets to zero (e.g., Row C in \cref{fig:tgprac-s_defending_feinting}(c.2)), so continuing to activate it reduces the total activations available for the target row. For instance, if both Row C and Row T are activated in \cref{fig:tgprac-s_defending_feinting}(d) rather than focusing solely on Row T, the target row receives only 20 activations instead of 40. This observation aligns with the findings from the original Feinting attack~\cite{ProTRR}.

\smallskip
\topic{Scenario 2 -- Delayed Activations} 
In this scenario, the attacker first activates decoy rows (e.g., Row B and Row C) to inflate their activation counts, ensuring they are mitigated when \TBRFM{}s are issued. The attacker then switches to intensively activating the target row (Row T) so it becomes the most activated and enters the mitigation queue. If the same row pool size, number of decoy rows, and activation counts are used, this strategy effectively replicates the Feinting attack. However, it does not improve Feinting's effectiveness and remains secure under \Defensename{}.

\smallskip
\topic{Scenario 3 -- Early and Aggressive Activations} 
In this scenario, the attacker begins by aggressively activating the target row (Row T), aiming to reach the Back-Off threshold (\NBO{}) before any mitigations occur. Row T quickly surpasses other rows in activation count and enters the mitigation queue. However, this strategy is inferior to the Feinting attack. Once Row T is mitigated by a \TBRFM{}, its counter resets, limiting the total number of activations to those between two \TBRFM{} events; this number is significantly lower than what the Feinting attack can achieve. For example, with one \TBRFM{} every \TREFI{}, the attacker can induce at most 60 activations--approximately 12.3$\times$ and 9.5$\times$ fewer than the maximum derived in our analysis without (736) and with (572) counter reset, respectively (\cref{fig:tprac_max_act}).

\smallskip
\topic{Why a Single-Entry Queue is Sufficient} 
A single-entry, freque\-ncy-based mitigation queue is sufficient because it consistently tracks the most frequently activated row at any point in time, ensuring that the highest-risk row is mitigated during each \TBRFM{}. Since \TBRFM{}s are issued at fixed intervals, \emph{independent} of memory access patterns,  the attacker cannot manipulate mitigation timing through activation strategies. As a result, no row can accumulate enough activations to reach \NBO{}, effectively preventing all \ABORFM{}s. Therefore, by combining a single-entry queue with appropriately configured \TBRFM{}s, \Defensename{} achieves security on par with idealized PRAC implementations (UPRAC). This design effectively thwarts a wide range of attack strategies, including balanced, delayed, and aggressive activations, ensuring robust protection even against worst-case patterns like the Feinting attack.

\subsection{Designing TPRAC with Targeted Refreshes}\label{sec:targeted_ref_design}
\Defensename{} can leverage existing Targeted Refreshes (\TREF{}s) to reduce mitigation overhead. Like \TBRFM{}s, each \TREF{} is used to mitigate the most frequently activated row from the mitigation queue. This allows \Defensename{} to skip a scheduled \TBRFM{} if a \TREF{} occurs within the same interval, improving performance without sacrificing security. For example, at \NRH{} of 512, \Defensename{} requires a \TBRFM{} approximately every 0.8~\TREFI{} with counter reset, and every 0.6~\TREFI{} without reset, to eliminate timing channels (\cref{fig:tprac_max_act}). If \TREF{}s are issued once every 2~\TREFI{}, \Defensename{} can skip a \TBRFM{} roughly every 2.5~\TREFI{} (with reset) or 3.3~\TREFI{} (without reset), reducing the total number of RFMs and improving performance.
\section{Evaluation Methodology}\label{sec:eval_methodology}
\topic{Simulation Framework} 
We evaluate our design using ChampSim~\cite{champsim}, a well-known trace-based, out-of-order processor simulator. We integrate ChampSim with the cycle-accurate DRAM simulator Ramulator2~\cite{ramulator2, kim2015ramulator}. \cref{table:system_config} summarizes our simulated system configuration, following prior work~\cite{berti_2022} that models the Intel Sunny Cove microarchitecture. 

The baseline system features a 4-core out-of-order processor with a hashed perceptron branch predictor~\cite{hashed_perceptron}. Each core has private L1 caches: a 32KB instruction cache and a 48KB data cache, with the data cache using the IP-stride prefetcher~\cite{ip-stride}. Each core also has a private 512KB L2 cache, and all cores share an 8MB last-level cache (LLC). The LLC supports 64 MSRHs per core and uses the SPP-PPF prefetcher~\cite{spp-ppf} and SRRIP replacement policy~\cite{srrip}\footnote{We evaluated \Defensename{} with various branch predictor and prefetcher combinations in ChampSim. The result variance is within 1\%, showing that our mitigation is insensitive to different microarchitectural policies.}. 
The memory system comprises a single-channel, quad-rank DDR5 memory totaling 128GB of DRAM. The memory controller uses the Minimalist Open-Page (MOP) address mapping policy~\cite{MOP} and the First Ready First Come First Served (FR-FCFS) scheduler~\cite{zuravleff1997controller, FRFCFS}, following prior work~\cite{Chronus, breakhammer2024, qprac, dapper}. The DDR5 memory is modeled as a 32Gb DDR5-8000B chip, with timing parameters and PRAC-specific adjustments based on the JEDEC specification~\cite{jedec_ddr5_prac}.

\begin{table}[h!]
\begin{center}
\begin{small}
\caption{{System Configuration}}{
\resizebox{\columnwidth}{!}{
\begin{tabular}{c|c}
\toprule
  \multirow{2}{*}{Out-Of-Order Cores} &  4 Cores, 4GHz, 6-issue width, 4-retire width\\ 
  & 352-entry ROB, {hashed perceptron branch predictor~\cite{hashed_perceptron}}          \\
  {L1 Instruction Cache} & {32KB, 8-way, 4 cycles}\\ 
  {L1 Data Cache} & {48KB, 12-way, 5 cycles, IP-stride prefetcher~\cite{ip-stride}} \\    
  {L2 Cache} & {512KB, 8-way, 10 cycles}\\\midrule
  Last Level Cache (Shared)   & 8MB, 16-way, 20 cycles, {SPP-PPF prefetcher~\cite{spp-ppf}, SRRIP~\cite{srrip}}  \\\midrule
  Address Mapping & Minimalist Open-Page (MOP)~\cite{MOP}\\ 
  Scheduling Policy& FR-FCFS~\cite{zuravleff1997controller, FRFCFS} with a cap of 4~\cite{FRFCFS_CAP}\\\midrule
  Memory Type                  & 32Gb DDR5-8000B \\
  DRAM Organization      & 4 Bank x 8 Groups x 4 Ranks x 1 Channel \\
  Rows Per Bank, Size                 & 128K, 8KB \\ 
  tRCD, tCL, tRAS~\cite{lisa} & 16ns, 16ns, 16ns\\
  tRP, tRTP, tWR, \TRC{} & 36ns, 5ns, 10ns, 52ns \\ 
  \TRFC{}, \TREFI{}~\cite{avatar,refpause}     &   410 ns, 3.9$\mu$s \\
  t\ABOACT{}, t\RFMAB{} & 180ns, 350ns \\
\bottomrule

\end{tabular}}
}
\label{table:system_config}
\end{small}
\end{center}
\end{table}

\topic{Evaluated Design} 
We evaluate the performance and energy impact of \Defensename{} by comparing it to a baseline PRAC-enabled DRAM without the Alert Back-Off (ABO) protocol. We extend Ramulator2 to model per-row activation counters, the ABO protocol, our proposed Timing-Based \RFM{} (TB-RFM), and the mitigation queue design and strategy. The \TBRFM{} interval (\TBWINDOW{}) is tuned for each RowHammer (RH) threshold (\NRH{}) to eliminate all ABO-triggered RFMs (ABO-RFMs).
We compare \Defensename{} against the following baselines, which are insecure against timing channel attacks. As the underlying PRAC implementation, we use QPRAC~\cite{qprac}, one of the existing secure PRAC designs~\cite{Chronus, qureshi2024moat, qprac}: 1) \textbf{ABO-Only}, which relies solely on the ABO protocol for RH mitigations, and 2)~\textbf{ABO+ACB-RFM}, which incorporates proactive Activation-Based RFMs (ACB-RFMs) to eliminate \ABORFM{}s, which is the current RFM implementation defined in the JEDEC standard~\cite{jedec_ddr5_prac}. For each \NRH{}, we configure the Bank Activation threshold (BAT) to eliminate \ABORFM{}s under the worst-case Feinting~\cite{ProTRR} (or Wave~\cite{wave}) attack pattern.

\smallskip
\topic{Workloads} We evaluate 50 publicly available workloads from SPEC2006~\cite{SPEC2006}, SPEC2017~\cite{SPEC2017}, and CloudSuite~\cite{CloudSuite} benchmarks. Workloads are categorized into three memory-intensity groups based on their row-buffer misses per kilo instructions (RBMPKI): High (H) for $\text{RBMPKI} \geq 10$, Medium (M) for $1 \leq \text{RBMPKI} < 10$, and Low (L) for $\text{RBMPKI} < 1$, as shown in \cref{table:workload_categorization}. 

\begin{table}[t]
\centering
\caption{{Workload Categorization Based on RBMPKI}}
\label{table:workload_categorization}
\resizebox{\columnwidth}{!}{
\begin{tabular}{c|c}
\toprule
\textbf{RBMPKI} & \textbf{Workloads} \\ \midrule
\multirow{6}{*}{{\shortstack{\textbf{High} \\ \textbf{[10+)}}}} 
& nutch, cassandra, classification, 433.milc, cloud9, \\&  410.bwaves, 470.lbm, 471.omnetpp, 483.xalancbmk, \\&  519.lbm, 520.omnetpp, 649.fotonik3d, 450.soplex, 619.lbm, \\& 429.mcf, 654.roms, 470.lbm, 483.xalancbmk, 471.omnetpp, \\& 605.mcf, 482.sphinx3, 437.leslie3d, 627.cam4, 620.omnetpp, \\& 628.pop2, 607.cactuBSSN, 436.cactusADM, 459.GemsFDTD \\ \hline
\multirow{2}{*}{\shortstack{\textbf{Medium} \\ \textbf{[1, 10)}}}
& 401.bzip2, 657.xz, 602.gcc, 473.astar, \\ & 623.xalancbmk,   464.h264ref, 481.wrf\\\hline
\multirow{5}{*}
{{\shortstack{\textbf{Low} \\ 
\textbf{[0, 1)}}}} & 
631.deepsjeng, 458.sjeng, 456.hmmer, 625.x264, 403.gcc,  \\
 & 444.namd, 603.bwaves, 456.hmmer,  464.h264ref,    \\
 & 638.imagick, 644.nab, 481.wrf, 600.perlbench, 621.wrf,    \\
 & 465.tonto, 447.dealII, 435.gromacs, 641.leela, 454.calculix,  \\
 & 445.gobmk, 453.povray, 416.gamess, 648.exchange2 \\ \bottomrule
\end{tabular}
}
\end{table}

We simulate four-core homogeneous workloads for SPEC2006 and SPEC2017, while each core in CloudSuite runs a distinct thread per workload.
For SPEC workloads, each core executes 250 million instructions, comprising a 50M warm-up followed by 200M instructions used for performance measurement. For CloudSuite, each core executes 25M warm-up followed by 125M measured instructions.

By default, we set \NRH{} to 1024 with one \RFM{} per Alert (PRAC-1) and enable activation counter resets at every \TREFW{}. Additionally, we analyze sensitivity to \NRH{} values from 128 to 4096, PRAC levels with 1, 2, or 4 RFMs per Alert, targeted refresh rates varying from once every 1 to 4 tREFIs, and activation counter reset policies.  Performance is measured using weighted speedup.
\section{Results and Analysis}\label{sec:results}
\begin{figure}[b]
\centering
\includegraphics[width=0.45\textwidth]{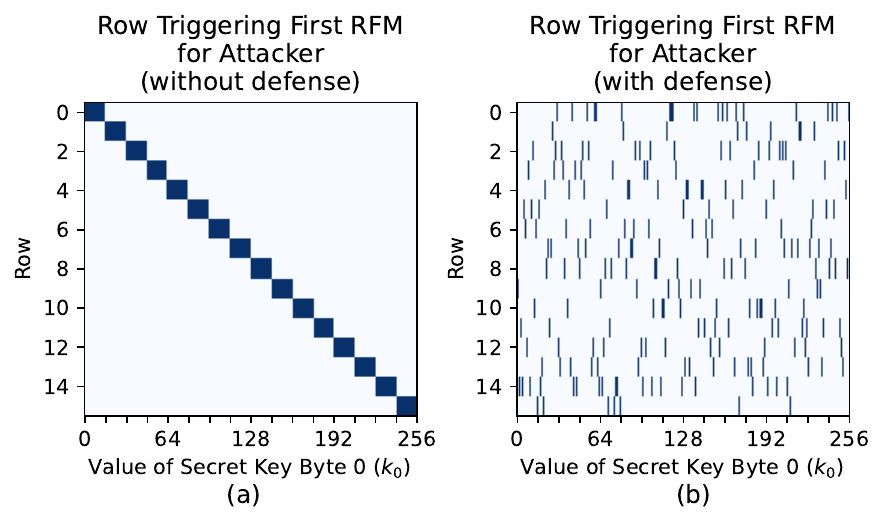}
\caption{Row triggering the first RFM for the attacker. Without our \Defensename{} defense, the row leaks the secret key value, whereas with the defense, the triggered row is random and does not reveal any secret information.}
\label{fig:sidechannel_defense}
\end{figure}

\begin{figure*}[t]
    \centering
    \includegraphics[width=0.95\linewidth, height=\paperheight, keepaspectratio]{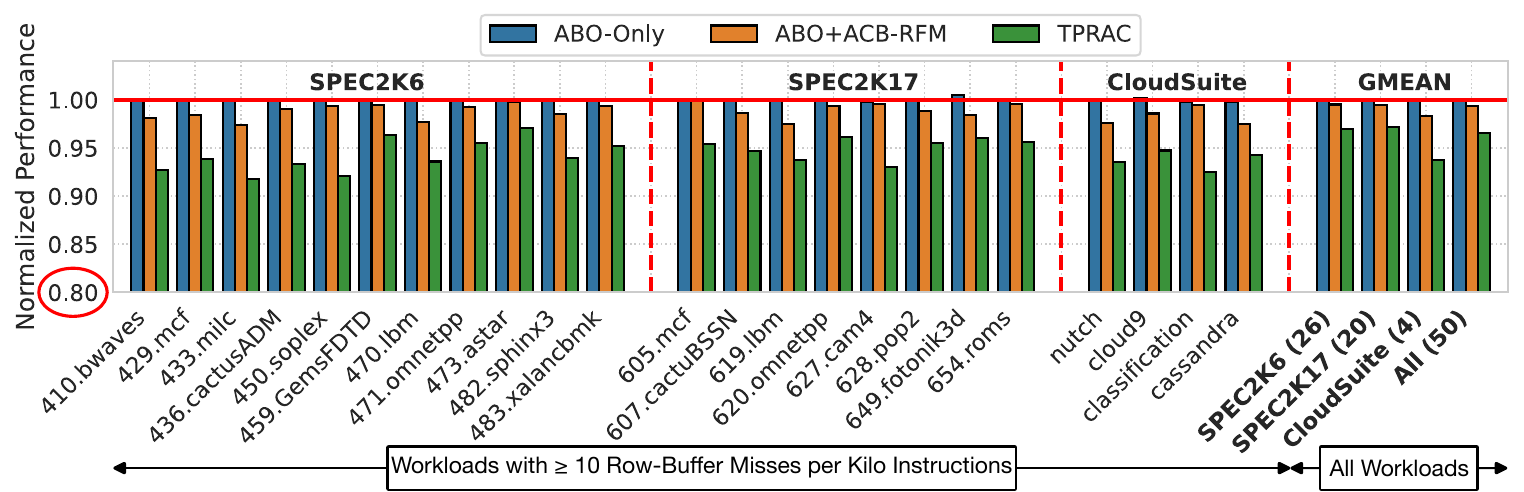}
    \caption{Normalized performance of \Defensename{} compared to insecure baselines, ABO-Only and ABO+ACB-RFM, at a RowHammer threshold of 1024. \Defensename{} incurs a slowdown of only 3.4\% on average due to periodic Timing-Based RFMs (\TBRFM{}s). In contrast, ABO+ACB-RFM and ABO-Only exhibit slowdowns of {0.7\%} and near zero, respectively, due to their lower RFM frequency resulting from their activity-dependent design. However, this activity-dependent RFM method is not secure.}
    \Description{}
    \label{fig:perf_main}
\end{figure*}

\subsection{Empirical Security Validation}\label{sec:TPRACDefenseAES}
To empirically demonstrate the effectiveness of TPRAC's mitigation capability, \cref{fig:sidechannel_defense} shows the results of the \Attackname{} Side-Channel attack on AES T-tables (described in \cref{sec:attack_side}), both with and without our \Defensename{} defense. The figure shows that the DRAM row that triggers the first RFM during the attacker's probing phase reveals the most heavily activated row during the victim execution, which is dependent on the secret key.

In \cref{fig:sidechannel_defense}(a), without the defense, there is a strong correlation between the value of key byte $k_0$ and the row triggering the first RFM, clearly leaking $k_0$. In contrast, \cref{fig:sidechannel_defense}(b) shows that with \Defensename{}, the row triggering the first RFM appears random, leaking no information to the attacker. This is due to two reasons: 
(1) \Defensename{}'s periodic Timing-Based RFMs (\TBRFM{}s) proactively mitigate the most activated rows, fully preventing \ABORFM{}s; and (2) even if the \TBRFM{} interval (\TBWINDOW{}) is misconfigured, the attacker cannot distinguish \TBRFM{}s from \ABORFM{}s, as both incur the same 350ns latency. As a result, no secret information is leaked through memory latency variations.

\subsection{Performance Overhead}
\cref{fig:perf_main} shows the performance of \Defensename{} at a RowHammer threshold (\NRH{}) of 1024, compared to two insecure baselines: ABO-Only and \ABOACBRFM{}. Performance is normalized to a PRAC-enabled DDR5 system without the ABO protocol.\Defensename{} incurs only a 3.4\% average slowdown, as issuing one Timing-Based \RFM{} (\TBRFM{}) every 1.6 \TREFI{} (6.2$\mu\text{s}$) is sufficient to eliminate all ABO-RFMs. The primary source of overhead is reduced DRAM bandwidth--each \TBRFM{} block all banks for 350ns (t\RFMAB{}) every 6.2$\mu\text{s}$, leading to a maximum DRAM bandwidth loss of 5.6\% $\left(\frac{{\text{tRFM}}_{\text{ab}}}{\text{TB-Window}}\right)$. As a result, several memory-intensive workloads experience slowdowns near or slightly above 6\%. 

Another source of slowdown is reduced row-buffer locality. Since each \TBRFM{} requires closing all open rows to issue an \RFMAB{}, row-buffer hit rates may decline. On average, \Defensename{} increases row-buffer misses by only 0.5\%, with negligible impact on most workloads. However, in \emph{433.milc}, a 2.3\% increase in misses, combined with bandwidth loss, results in the highest slowdown of 8.3\%.

\ABOACBRFM{} incurs only a 0.7\% slowdown, as it triggers Activation-Based RFMs (\ACBRFM{}s) only when per-bank activations reach the Bank Activation threshold (BAT), minimizing the bandwidth loss. \ABOONLY{} shows almost no slowdown because \ABORFM{}s are rare at \NRH{} of 1024. These results are consistent with recent works~\cite{Chronus, qureshi2024moat, qprac}. However, both \ABOONLY{} and AB\-O+ACB-RFM remain vulnerable to timing channel attacks due to their \emph{activity-dependent} mitigation strategies.

\begin{figure}[b]
\vspace{-0.2in}
\centering
\includegraphics[width=3in,height=\paperheight,keepaspectratio]{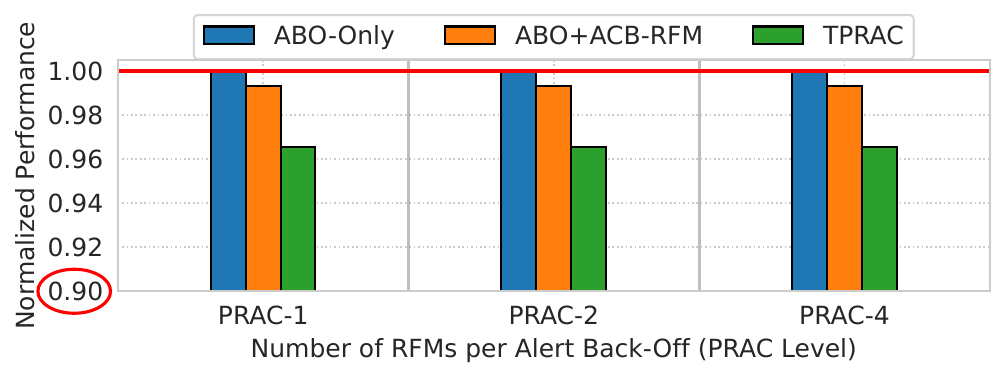}
\vspace{-0.15in}
\caption{
Performance comparison of \Defensename{} and insecure baselines, \ABOONLY{} and \ABOACBRFM{}, across different PRAC levels at a RowHammer threshold of 1024. Performance remains consistent across PRAC levels. \Defensename{} and \ABOACBRFM{} incur slowdowns of 3.4\% and 0.7\%, respectively, due to proactive RFMs reducing DRAM bandwidth. In contrast, \ABOONLY{} shows almost no overhead, as \ABORFM{}s are rare in benign applications at this threshold.}
\Description{}
\label{fig:prac_level_sensitivity}
\end{figure}

\subsection{Sensitivity to PRAC Levels}
\cref{fig:prac_level_sensitivity} compares the performance of \Defensename{} and insecure baselines, \ABOONLY{} and ABO+ACT-RFM, as the PRAC level--the number of RFMs per ABO--varies at \NRH{} of 1024. While higher PRAC level allows DRAM to mitigate more rows per ABO, they also increase DRAM blocking time, which can impact both performance and security when \ABORFM{} occurs.
The results show that PRAC level variation has no impact on the performance of \Defensename{} and ABO+ACB-RFM, since both designs eliminate all \ABORFM{}s. \textit{TPRA\-C} achieves this through periodic Timing-Based RFMs (\TBRFM{}s), while ABO+ACB-RFM uses Activation-Based RFMs (\ACBRFM{}s). Across all PRAC levels, \Defensename{} and \ABOACBRFM{} incur slowdowns of 3.4\% and 0.7\%, respectively. \ABOONLY{} experiences almost no \ABORFM{}s at this threshold, resulting in near-zero slowdown.

\subsection{Sensitivity to Targeted Refreshes}\label{sec:targeted_ref_results}
\begin{figure}[h!]
\centering
\includegraphics[width=3.1in,height=\paperheight,keepaspectratio]{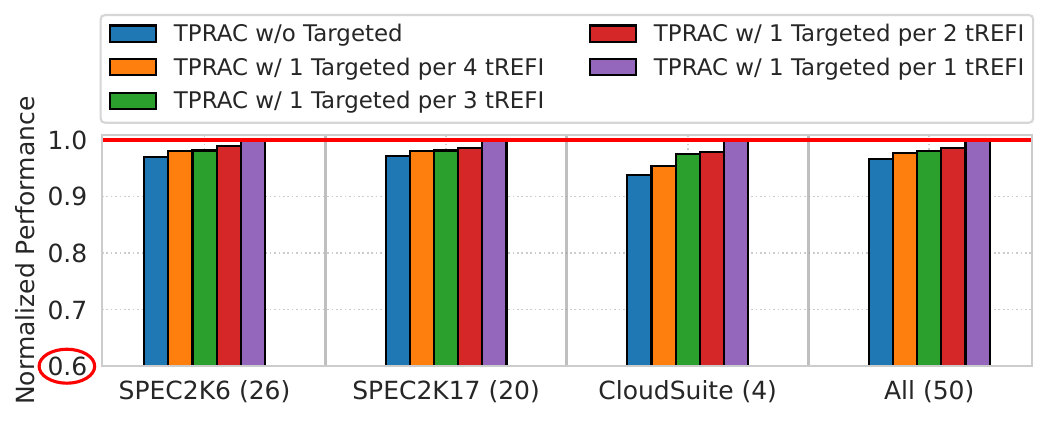}
\caption{
Normalized performance of \Defensename{} as the frequency of Targeted Refreshes (\TREF{}s) varies at \NRH{} of 1024. More frequent \TREF{}s reduce slowdowns by reducing the need for Timing-Based \RFM{}s.  \Defensename{} incurs slowdowns of 3.4\%, 2.4\%, 2.0\%, and 1.4\% with one \TREF{} per 4, 3, and 2 \TREFI{}, respectively, and no overhead with one \TREF{} per \TREFI{}.}
\Description{}
\label{fig:targeted_ref_sensitivity}
\end{figure}

\cref{fig:targeted_ref_sensitivity} shows the performance of \Defensename{} as the frequency of Targeted Refreshes (\TREF{}s) varies from once every four \TREFI{} to once per \TREFI{} at a RowHammer threshold (\NRH{}) of 1024. Increasing \TREF{} frequency improves performance since it allows \Defensename{} to skip more Timing-Based \RFM{}s (\TBRFM{}s). Without \TREF{}, \Defensename{} incurs a 3.4\% slowdown. The slowdown drops to 2.4\%, 2.0\%, and 1.4\% with one \TREF{} every 4, 3, and 2 \TREFI{}s, respectively. At one \TREF{} per \TREFI{}, \Defensename{} incurs no slowdown, as \TREF{}s fully replace \TBRFM{}s, eliminating \ABORFM{}s and preventing timing channels. As discussed in \cref{sec:nbo}, this approach enhances the practicality of \Defensename{} at lower \NRH{} by reducing performance slowdowns.

\begin{figure}[b]
\centering
\includegraphics[width=3.1in,height=\paperheight,keepaspectratio]{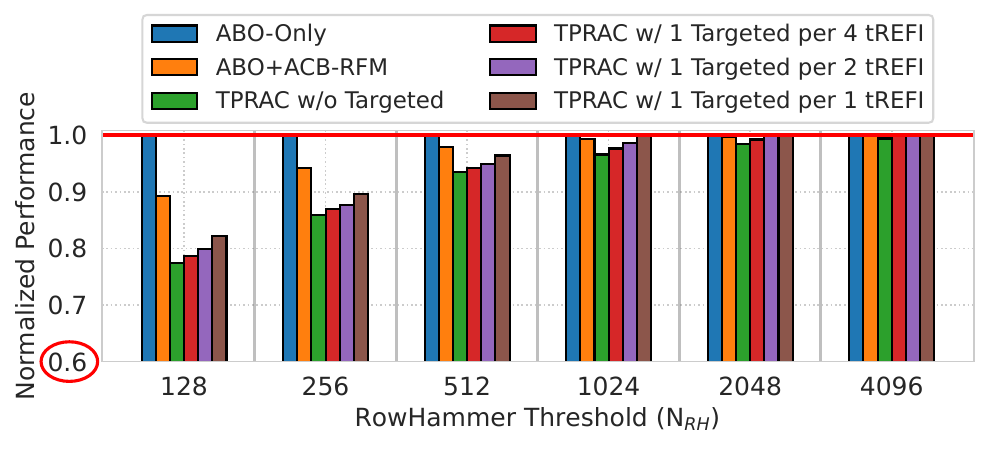}
\caption{Performance comparison of \Defensename{} and insecure baselines, \ABOONLY{} and \ABOACBRFM{}, as \NRH{} varies. \Defensename{} incurs 3.4\%, 1.6\%, and 0.6\% slowdown at \NRH{} of 1024, 2048, and 4096, respectively. However, slowdowns increase to 6.5\%, 14.1\%, and 22.6\% for \NRH{} of 512, 256, and 128, respectively, due to more frequent Timing-Based RFMs (\TBRFM{}s). Co-designing \Defensename{} with Targeted Refreshes improves performance by allowing some \TBRFM{}s to be skipped.
}
\Description{}
\label{fig:nrh_sensitivity}
\end{figure}

\subsection{Sensitivity to RowHammer Threshold}\label{sec:nbo}
\cref{fig:nrh_sensitivity} shows the performance of \Defensename{} as \NRH{} varies from 128 to 4096. At \NRH{} of 4096, 2048, and 1024, \Defensename{} incurs slowdowns of 0.6\%, 1.6\%, and 3.4\%, respectively. Even at an ultra-low \NRH{} of 512, it maintains a moderate overhead of 6.5\%. However, with lower \NRH{} of 256 and 128, the slowdown rises to 14.1\% and 22.6\%, respectively, due to 4.3$\times$ and 6.4$\times$ more frequent \TBRFM{}s needed to eliminate \ABORFM{}s and prevent timing channel attacks.

Co-designing \Defensename{} with Targeted Refreshes (\TREF{}s) improves performance across all evaluated \NRH{} values, as it can skip some \TBRFM{}s by leveraging \TREF{}s. At \NRH{} of 256, this reduce the slowdown from 14.1\% (Without \TREF{}) to 13.0\%, 12.3\%, and 10.5\% with one \TREF{} every 4, 2, and 1 \TREF{}, respectively. 

\ABOACBRFM{} shows similar trends, with slowdowns increasing at lower \NRH{} due to more frequent \ACBRFM{}s issued to eliminate \ABORFM{}s. Its overhead remains lower than \Defensename{}, incurring just 0.7\% at \NRH{} of 1024 and 10.7\% at 128. However, it remains vulnerable to timing channel attacks, and the vulnerability worsens as \NRH{} decreases because more frequent \ACBRFM{}s increase the chance of information leakage. \ABOONLY{} incurs negligible slowdowns across all thresholds, since \ABORFM{}s are rare in benign applications. Nonetheless, like \ABOACBRFM{}, it is insecure against timing channels, with leakage becoming more pronounced at lower \NRH{}, where fewer activations are needed to trigger \ABORFM{}s.

\subsection{Sensitivity to Activation Counter Reset}
\begin{figure}[h!]
\centering
\includegraphics[width=3.1in,height=\paperheight,keepaspectratio]{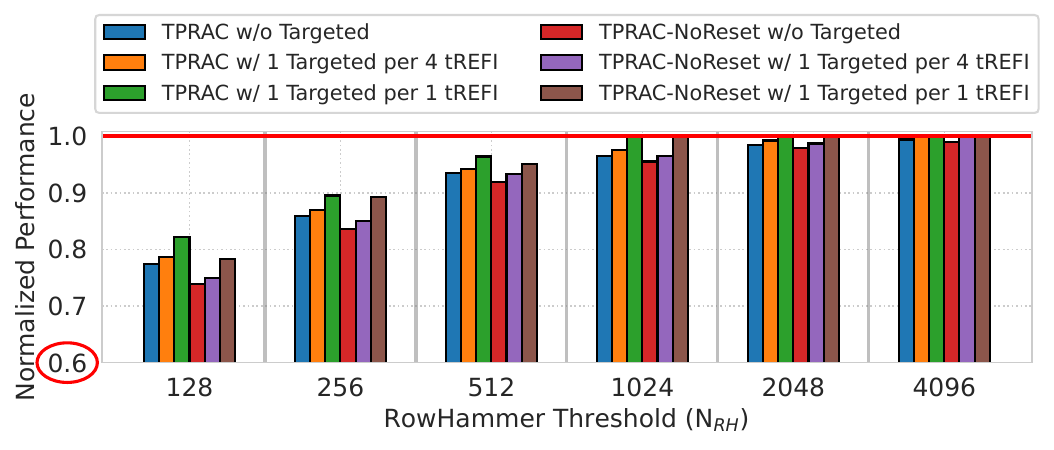}
\caption{Normalized performance of \Defensename{} with and without activation counter reset as \NRH{} varies from 128 to 4096. Resetting counters at each \TREFW{} has negligible impact ($<$ 1\%) at \NRH{} $\geq$ 1024, but offers noticeable performance benefits at lower thresholds due to reduced \TBRFM{} frequency.}
\Description{}
\label{fig:counter_reset_sensitivity}
\end{figure}

\cref{fig:counter_reset_sensitivity} shows the performance of \Defensename{} with and without activation counter reset as \NRH{} varies from 128 to 4096. At \NRH{} of 1024 and above, the impact of counter reset is negligible, with performance differences under 1\%. However, at ultra-low \NRH{} (\NRH{} $\leq$ 512), resetting counters every \TREFW{} improves performance.  For instance, at \NRH{} of 128 without \TREF{}, \Defensename{} without counter reset incurs a 73.9\% slowdown, compared to 77.4\% with reset, resulting in a 3.4\% performance improvement. This improvement arises from a lower \TBRFM{} frequency, as counter resets reduce the attacker’s initial row pool size (see \cref{subsec:TGPRAC-S_Security}). With fewer \TBRFM{}s, DRAM suffers less bandwidth loss, resulting in better overall performance.

\subsection{Energy Overhead}
\cref{table:result-energy_overhead} presents the energy overhead of \Defensename{}. Our \TBRFM{} introduces two main sources of overhead. First, each \TBRFM{} triggers mitigations for the most heavily activated row in the mitigation queue, consisting of \emph{four} victim refreshes and \emph{one} aggressor activation to reset the counter, totaling \emph{five} additional activations per \TBRFM{}. Second, \TBRFM{}s increase overall execution time, contributing to what we refer to as non-mitigation energy overhead.

At \NRH{} of 1024, \Defensename{} incurs an average energy overhead of 7.4\%, primarily due to increased execution time. As \NRH{} decreases, more frequent \TBRFM{}s are required, increasing the overhead. For example, at \NRH{} of 4096, \TBRFM{}s are issued approximately once every seven \TREF{}, leading to only 1\% overhead. In contrast, at ultra-low \NRH{} of 128, \TBRFM{}s are issued every 1$\mu$s--nearly 29$\times$ more often--resulting in significant 44.3\% overhead.

Notably, the JEDEC DDR5 specification defines the minimum Bank Activation threshold (BAT) for \ACBRFM{} as 16~\cite{jedec_ddr5_prac}, which allows issuing one \ACBRFM{} every 0.8$\mu$s. This rate is similar to the 1$\mu$s TB-RFM interval at \NRH{} of 128. As a result, the worst-case energy overhead of \Defensename{} aligns with what is already permissible in the standard. Future work can explore \Defensename{} to reduce energy impact at ultra-low \NRH{}.

\begin{table}[h]
\centering
{\small
\caption{Energy Overhead of \Defensename{}}
\label{table:result-energy_overhead}
\begin{tabular}{c|c|c|c}
\toprule
\multirow{2}{*}{\NRH{}} & \textbf{Mitigation} & \textbf{Non-Mitigation} & \multirow{2}{*}{\textbf{Total}} \\ 
 & \textbf{(RFM)} & \textbf{(Execution Time)} & \\ \midrule
128             &  \blue{18.1\%} &  \blue{26.2\%} &  \blue{44.3\%} \\
256             &  \blue{11.3\%} &  \blue{14.8\%} &  \blue{26.1\%} \\
512             &  \blue{5.9\%} &  \blue{4.5\%} &   \blue{10.4\%}\\
\textbf{1024}   &  \blue{\textbf{2}}\% &  \blue{\textbf{5.4}}\% &   \blue{\textbf{7.4}\%}\\
2048            &   \blue{1\%} &  \blue{1.6\%}&   \blue{2.6\%}\\
4096            &   \blue{0.45}\% &  \blue{0.55\%}&   \blue{1\%}\\
 \bottomrule
\end{tabular}
}
\end{table}

\subsection{Storage Overhead}
\Defensename{} introduces minimal storage overhead by requiring only a single register per memory controller to store the interval for issuing periodic Timing-Based RFMs (\TBRFM{}s). This RFM interval register uses 24 bits (3 bytes), which is sufficient to represent intervals up to approximately half of the \TREFW{} duration.

\section{Alternative Timing Channel Mitigations}
\subsection{Obfuscation-Based Defenses}
An alternative to mitigating PRAC-induced timing channels is to obfuscate the attacker by injecting random noise into memory access latency. The key idea is to introduce timing delays that resemble legitimate \ABORFM{}s (350ns), making it harder for attackers to infer sensitive information. This can be achieved by having either the memory controller issue random \RFMAB{}s or the DRAM probabilistically assert the Alert signals. While this approach does not eliminate \ABORFM{}s like \Defensename{}, it is effective against simple channels that rely solely on memory access latency variations.

An attacker could develop more sophisticated strategies beyond simple latency monitoring. For example, they might profile RFM count distributions over extended periods (e.g., hours or days), collecting data from windows with and without memory activity. The attacker could attempt to infer transmitted bits or secret-dependent behavior by training statistical or machine learning classifiers on this data. If random RFMs are injected with 50\% probability per \TREFI{}, the attacker may typically observe around 4096 RFMs within a \TREFW{} window during idle periods and higher counts during active ones. In extreme cases, observing zero RFMs can definitively indicate a Bit-0 (covert channel) or absence of victim activity (side channel), while counts exceeding 8192 confirm a Bit-1 or victim activity. Nonetheless, the distribution overlap introduced by randomization reduces the reliability of such inferences.

This probabilistic approach offers a flexible trade-off between security and performance. While it does not eliminate all information leakage, it reduces attacker precision, making it suitable in scenarios where occasional leakage is acceptable. It is also appealing for ultra-low \NRH{} values, where \Defensename{} incurs higher performance overhead due to frequent \TBRFM{}s. Future work will further explore how such techniques can complement or enhance \Defensename{}.

\subsection{Bank Partitioning with Per-Bank RFM}
DRAM bank partitioning, where each bank is exclusively assigned to a specific user or process (e.g., per-VM), can provide spatial isolation and prevent inter-bank interference~\cite{FixedService, Siloz}. While this can block some PRAC-based timing channels, such as our activation-count-based channel, it is insufficient to prevent all channels, including our activity-based channel. This is because the PRAC specification only allows memory controllers to issue an RFM All Bank (\RFMAB{}), which blocks the entire DRAM channel for 350ns~\cite{jedec_ddr5_prac}. Consequently, an attacker in a separate bank on the same DRAM module can still observe latency spikes and infer victim activity.

One way to close these timing channels under bank partitioning is to extend the ABO protocol to support a fine-grained Per-Bank \RFM{} (\RFMPB{}) command, which mitigates a specific bank without stalling others. However, this requires changes to the current PRAC specification, such as enabling DRAM to report to the memory controller which bank triggered the Alert signal. In contrast, \Defensename{} eliminates all PRAC-induced timing channels without requiring such modifications or bank partitioning. 

\RFMPB{} could also benefit \Defensename{} by allowing it to issue Per-Bank \TBRFM{}, further reducing DRAM bandwidth loss. We leave the design and evaluation of this extension to future work\footnote{QPRAC~\cite{qprac} also proposes to incorporate \RFMPB{} as part of the ABO protocol to address Denial-of-Service risks. In contrast, our focus is on closing timing channels.}.

\subsection{Hiding RFM-Induced Latency Variations}
Another approach to mitigating PRAC’s timing channels is to hide or eliminate the 
observable latency of RowHammer (RH) mitigation caused by Alert Back-Off-triggered RFMs (\ABORFM{}s). For instance, this can be done by increasing the frequency of periodic retention refreshes and borrowing time from them to perform necessary mitigations, similar to prior Target Row Refresh (TRR). Instead of issuing \Defensename{}'s Timing-Based RFMs (\TBRFM{}s), this method uses additional refreshes to track and mitigate the most frequently activated rows using an in-DRAM mitigation queue and mitigate them, following a mitigation strategy of \Defensename{}.

However, this approach has key limitations: (1) it requires changes to JEDEC specifications, (2) incurs higher performance overhead due to longer blocking time (410ns for Refresh vs. 350ns for \RFM{}), (3) must be conservatively configured for worst-case \NRH{} due to aging and spatial variation~\cite{olgun2024HBM2study, yauglikcci2024spatial}, and (4) complicates DRAM refresh logic by repurposing refreshes for both retention and RH mitigation. In contrast, \Defensename{} uses JEDEC-defined RFM commands, offering a more efficient and standards-compliant solution.
\section{Related Work}\label{sec:related_work}
\subsection{Timing Side Channels}
\smallskip
\noindent \textbf{DRAM Timing Channels.} 
DRAMA~\cite{DRAMA2016} proposes a side channel due to timing variation between row-buffer hits and conflicts. While this leaks the last row or bank accessed, our PRAC-based attacks can leak the history of row accesses, leaking the activation counts for any row accessed in a given time window.  

Other works~\cite{TemporalPartitioning,dagguise,FixedService,3Rank2Bank,Camouflage} discuss side channels due to memory-scheduling decisions in shared memory controllers, and propose spatial~\cite{FixedService, 3Rank2Bank} or temporal partitioning~\cite{TemporalPartitioning}, or memory traffic shaping~\cite{Camouflage,dagguise} to prevent leakage. Such defenses cannot mitigate all PRAC-based timing channels, as the \RFMAB{} command used for mitigation in PRAC stalls the entire DRAM channel, causing delays for all memory accesses despite such defenses. Our solution, TPRAC, closes all PRAC-based timing channels.

\smallskip
\noindent \textbf{Security-Metadata based Timing Channels.} Security metadata can introduce timing side channels. MetaLeak~\cite{MetaLeak} showed that Intel SGX's Merkle trees leak timing information. For instance, victim accesses trigger tree traversals that slow an attacker's traversals. Similarly, split counters~\cite{SplitCounter,saileshwar2018synergy,morphcounter} expose a victim's access counts via shared counter cache lines and overflow behavior. Our attacks also leverage activity-based and counter-based channels, but target PRAC-based row-activation counters. Unlike integrity trees and counter-based encryption~\cite{deuce}, which are now obsolete in new variants of SGX (SGX-v2)~\cite{SGX2}, PRAC is added to the DDR5 DRAM specification and is likely to be widely adopted. Thus, our attacks are more widely applicable.

\smallskip
\noindent \textbf{Timing Channels Due to Performance Optimizations.}
Most timing channels are due to timing variations introduced by performance optimizations.
Among these, cache timing channels~\cite{Bernstein,FlushReload, FlushFlush, LLCPrimeProbe} have been the most well-explored, with the root cause being timing variations due to cache hits and misses. Recently, cache performance optimizations like prefetchers~\cite{GoFetch,Augury,CCS18:PrefetcherChannels} and way-predictors~\cite{takeaway} have also been shown to cause timing variations and side channels. Moreover, processor front-end optimizations like branch prediction~\cite{BranchScope} and speculative execution~\cite{Spectre,Meltdown}, critical for performance, have also been shown to be vulnerable to side channels. Unlike these side channels, which can be mitigated by disabling performance optimizations for sensitive computations (e.g., using fences or uncached memory), our PRAC-based side channels cannot be mitigated by disabling activation counters, as that can lead to vulnerability to RowHammer.

\subsection{Alternative RowHammer Defenses} 
\smallskip
\topic{A. TRR-Based Defenses}
TRR-based defenses mitigate RowHammer (RH) by borrowing time from periodic refreshes to perform mitigation, either probabilistically~\cite{DSAC, HynixPAT} or based on counters tracking frequently activated rows~\cite{hassan2021UTRR}. While these approaches avoid additional latency and are generally immune to timing channels, they have been shown to be ineffective against modern RH attacks~\cite{hassan2021UTRR, PROTEAS, jattke2021blacksmith, jattke2024zenhammer, frigo2020trrespass, HalfDouble}. In contrast, \Defensename{} remains secure against RH attacks while also eliminating timing channels.

\smallskip
\topic{B. RFM-Based In-DRAM Defenses} 
These in-DRAM defenses perform RH mitigations using Activation-Based RFMs (\ACBRFM{}s). Upon receiving an RFM, some approaches apply mitigation probabilistically~\cite{jaleel2024pride, MINT}, while others use tracking structures to determine which rows to mitigate~\cite{kim2022mithril, ProTRR}. Although effective against RH, these techniques are vulnerable to timing channels because they rely on activity-dependent \ACBRFM{}s. In contrast, \Defensename{} employs activity-independent Timing-Based RFMs (\TBRFM{}s) that are both secure against RH and eliminate timing leakage.

\smallskip
\topic{C. Defenses with DRAM Chip and Interface Modifications}
Several prior works propose redesigning the DRAM interface or chip internals to support concurrent mitigations or refreshes during activations~\cite{REGA_SP23, HIRA}. Because they avoid introducing explicit mitigation delays, these designs are resistant to timing channels. However, they require substantial modifications to both DRAM chips and the interface protocol, making deployment difficult. In contrast, \Defensename{} requires no such modifications and achieves timing-channel prevention based on the PRAC specification.

\smallskip
\topic{D. RFM-Based Host-Side Defenses}
Host-side RH defenses mitigate attacks by issuing Directed RFM (DRFM) commands~\cite{jedec_ddr5_prac} to refresh neighboring rows, either probabilistically~\cite{kim2014architectural, kim2014flipping, PROHIT, MRLOC, hammerfilter} or using trackers~\cite{CBT, park2020graphene, qureshi2022hydra, comet, olgun2023abacus, saxena2024start, lee2019twice}.
Tracking-based defenses monitor potential aggressors and issue DRFMs when activity crosses a threshold; this activity-dependent strategy introduces timing leakage. Probabilistic approaches issue DRFMs with a fixed probability on each activation, reducing susceptibility to timing channels, but failing to protect against RH at low \NRH{}~\cite{jaleel2024pride}.

\smallskip
\noindent {\bf E. Alternative Mitigative Actions}: 
Row migration~\cite{saileshwar2022RRS, AQUA, SRS, CROW, ShadowHPCA23, saxena2024rubix} and access throttling techniques~\cite{yauglikcci2021blockhammer, breakhammer2024} mitigate RH by relocating or slowing access to frequently activated rows. However, these methods introduce noticeable latency variations, making them susceptible to timing channels. ECC-based defenses~\cite{faultsim, stuctcoding, xed, citadel1, citadel2, sudoku, ali2022safeguard} provide partial defense for RH but do not address timing leakage. In contrast, \Defensename{} is built on PRAC, preserving its RowHammer protection while eliminating timing channels.
\section{Conclusions}
This paper uncovers a critical security vulnerability in PRAC-based RowHammer mitigations, stemming from timing channels introduced by the Alert Back-Off mechanism and Refresh Management (RFM) commands. We present an attack called \Attackname{}, which exploits these timing channels to create covert and side channels. This enables an adversary sharing a DRAM row with a victim to leak sensitive information -- such as secret keys from vulnerable AES implementations -- by monitoring memory access latencies.

To address this issue, we propose Timing-Safe PRAC (\Defensename{}), an activity-independent RFM mechanism that eliminates timing channels without compromising RowHammer mitigation. \Defensename{} introduces Timing-Based RFMs that are issued periodically and independently of access patterns to eliminate timing leakage. Our evaluation shows that \Defensename{} achieves robust security using only a single-entry per-bank mitigation queue and incurs just 3.4\% performance overhead at a RowHammer threshold of 1024.

\begin{acks}
We thank Stefan Saroiu for sharing insights on PRAC in his keynote at DRAMSec'24 that inspired this work. We also thank Aamer Jaleel for early discussions and feedback, as well as the anonymous reviewers of ISCA 2025 for their valuable comments. We thank UBC's Advanced Research Computing (ARC) team for their support~\cite{sockeye}. This work is supported by the Natural Sciences and Engineering Research Council of Canada (NSERC) under funding numbers RGPIN-2019-05059 and RGPIN-2023-04796, and an NSERC-CSE Research Communities Grant under funding reference number ALLRP-588144-23. The views expressed are those of the authors and do not necessarily reflect those of NSERC, the Communications Security Establishment Canada, or the Government of Canada.
\end{acks}
\begin{appendices}

\section{Artifact Appendix}
\sloppy

\subsection{Abstract}
This artifact supports two key aspects of the paper: (1) the proposed \Attackname{} covert and side channel attacks, which exploit timing variations introduced by PRAC's mitigation actions, and (2) our defense, \Defensename{}, which uses activity-independent Timing-Based RFMs (\TBRFM{}s) to eliminate PRAC-induced timing channels.

For the \Attackname{} evaluations, we provide the Ramulator2~\cite{ramulator2} source code with modifications to support \texttt{clflush} along with required traces for experiments. We also include Bash and Python scripts to generate victim traces for AES T-Table encryption using the Ramulator trace generator, as well as the corresponding sender and receiver traces. Additional scripts are provided to run these traces in Ramulator2 and to reproduce the results shown in \Cref{fig:timing_variation} through \Cref{fig:sidechannel_heatmap} and \Cref{fig:sidechannel_defense}.

For the \Defensename{} evaluations, we provide Python scripts for the security analysis, allowing users to evaluate the maximum number of activations to a row as the \TBRFM{} interval (\TBWINDOW{}) varies and regenerate \cref{fig:tprac_max_act}. For performance evaluations, we include the C++ implementation of the integrated ChampSim and Ramulator2 framework, with the \Defensename{} implementation within Ramulator2. We also provide all evaluated workload traces, along with Bash and Python scripts to run experiments and automate result collation and plot generation (\cref{fig:perf_main} to \cref{fig:counter_reset_sensitivity}).

\subsection{Artifact Check-List (Meta-Information)}
\subsubsection{\Attackname{} Evaluations}
{\small
\begin{itemize}
  \item {\bf Program: } C++ implementations of Ramulator2 with modifications to support \texttt{clflush}. Bash scripts and Python scripts to generate program traces, run experiments, and produce plots.
  \item {\bf Compilation: } Tested with g++ 11.4.0; should also compile with any C++20-compliant compiler.
  \item {\bf Run-time environment: } Tested on Ubuntu 22.04; should run on any Linux distribution with a valid Python3 interpreter.
  \item {\bf Hardware: } Any modern desktop/laptop suffices.
  \item {\bf Output: } ABO timing variation: \Cref{fig:timing_variation};
  \Attackname{} on AES T-Table Encryption: \Cref{fig:sidechannel_raw}, \Cref{fig:sidechannel_heatmap},
  \Cref{fig:sidechannel_defense}
  \item {\bf Experiments: } Instructions for running experiments and parsing results are available in the provided README file.
  \item {\bf How much disk space required (approximately)?: } $\sim 20$ GB.
  \item {\bf How much time is needed to prepare workflow (approximately)?: } Under 5 minutes to install the dependencies.
  \item {\bf How much time is needed to complete experiments (approximately)?: } $\sim 1$ hour.
  \item {\bf Publicly available?:} Yes. GitHub: \url{https://github.com/STAR-Laboratory/PRAC_TC_ISCA25}. Traces: \url{https://doi.org/10.5281/zenodo.15104055}
  \item {\bf Archived (provide DOI)?: } \url{https://doi.org/10.5281/zenodo.15104037}
\end{itemize}
}

\subsubsection{\Defensename{} Evaluations}
{\small
\begin{itemize}
    \item {\bf Program:} C++ implementation of the integrated ChampSim–Ramulator2 framework, including the \Defensename{} implementation within Ramulator2. Accompanied by Bash and Python scripts for running experiments, aggregating results, and generating plots.
    \item {\bf Compilation:}  We highly recommend using \texttt{g++} 12 or newer, as it significantly reduces the compilation time of ChampSim binaries. Tested with \texttt{g++} 11 and 12; should also compile with any C++20-compliant compiler.
    \item {\bf Run-time environment:} We recommend using a modern Linux distribution with support for C++20. For example, Ubuntu 22.04 or later is recommended if you prefer Ubuntu. This artifact has been tested on Ubuntu 22.04 and Rocky Linux 9.4.
    \item {\bf Metrics:} Weighted Speedup.
    \item {\bf Output:} \Defensename{} security analysis: \cref{fig:tprac_max_act}; \Defensename{} performance results: \cref{fig:perf_main} to \cref{fig:counter_reset_sensitivity}.
    \item {\bf Experiments:} Instructions for running experiments and parsing results are available in the provided README file.
  \item {\bf How much disk space required (approximately)?: } $\approx$ 40GB.
  \item {\bf How much time is needed to prepare workflow (approximately)?:} Under an hour to install the dependencies, download the traces, and build all ChampSim binaries.
  \item {\bf How much time is needed to complete experiments (approximately)?: }    
  \begin{itemize}
        \item $\approx$ 1 day (on a cluster with 1000 cores) and $\approx$ 2 days (on an Intel Xeon CPU with 40 cores and 128GB memory) for the security analysis (\cref{fig:tprac_max_act}) and main performance result (\cref{fig:perf_main}).
        \item $\approx$ 2 days (on a cluster with 1000 cores) and $\approx$ 1 week (on an Intel Xeon CPU with 40 cores and 128GB memory) for all performance experiments (\cref{fig:perf_main} to~\cref{fig:counter_reset_sensitivity}).
    \end{itemize}
  \item {\bf Publicly available?:} Yes. GitHub: \url{https://github.com/STAR-Laboratory/PRAC_TC_ISCA25}. Traces: \url{https://doi.org/10.5281/zenodo.15104055}
  \item {\bf Archived (provide DOI)?: } \url{https://doi.org/10.5281/zenodo.15104037}
\end{itemize}
}

\subsection{Description}
\subsubsection{How to access}
The artifact is available at \url{https://github.com/STAR-Laboratory/PRAC_TC_ISCA25}.
\subsubsection{Hardware Recommendations}
\begin{itemize}
    \item \textbf{\Attackname{} Evaluation:} 
    Any modern desktop/laptop should suffice. A laptop with a 2-core CPU and 16GB of memory can perform \Attackname{} analysis within $\sim 1$ hour.
    \item \textbf{\Defensename{} Evaluation: }We strongly recommend using Slurm with a cluster capable of running bulk experiments to accelerate evaluations. If using a personal server, we recommend a machine with at least 40 hardware threads with 128GB of memory to run all evaluations in a reasonable time.
\end{itemize}

\subsubsection{Software Requirements}
\begin{itemize}
    \item \texttt{g++} with C++20 support (tested with \texttt{g++} 11 and 12). We highly recommend using \texttt{g++} 12 or newer, as it significantly reduces the compilation time of ChampSim binaries. 
    \item Python3 (tested with version 3.9 and 3.10).
\end{itemize}

\subsubsection{Traces}
\begin{itemize}
    \item \textbf{\Attackname{} Evaluation: }Our program generates 256 traces for the AES T-Table Encryption program, as well as other supportive traces for covert and side channel evaluations. Traces are available for download at \url{https://doi.org/10.5281/zenodo.15104055}
    \item \textbf{\Defensename{} Evaluation: }We use 50 traces from SPEC2006, SPEC2017, and CloudSuite. Traces are available for download at \url{https://doi.org/10.5281/zenodo.15104055}.
\end{itemize}

\subsection{Installation and Experiment Workflow}
First, clone the GitHub repository:
\noindent
\setlength{\fboxsep}{4pt}
\setlength{\fboxrule}{0.5pt}
\fcolorbox{black}{lightgray}{%
  \begin{minipage}{0.94\linewidth}
    \small\textbf{git clone \url{https://github.com/STAR-Laboratory/PRAC_TC_ISCA25}}
  \end{minipage}
}

\subsubsection{\Attackname{} Evaluation}

No additional setup is required if dependencies are satisfied. To start the experiment:

\smallskip
\noindent
\setlength{\fboxsep}{4pt}
\setlength{\fboxrule}{0.5pt}
\fcolorbox{black}{lightgray}{%
  \begin{minipage}{0.94\linewidth}
    \small\ttfamily
    \textbf{cd PRAC\_TC\_ISCA25/PRACLeak}\\
    \textbf{./run\_artifact.sh}
  \end{minipage}
}

\smallskip

\noindent Alternatively, to avoid regenerating traces and use the provided sample data:

\smallskip
\noindent
\setlength{\fboxsep}{4pt}
\setlength{\fboxrule}{0.5pt}
\fcolorbox{black}{lightgray}{%
  \begin{minipage}{0.94\linewidth}
    \small\ttfamily
    \textbf{cd PRAC\_TC\_ISCA25/PRACLeak}\\
    \textbf{./run\_artifact.sh ----use\_sample}
  \end{minipage}
}

\smallskip

\subsubsection{\Defensename{} Evaluations}
\noindent 1. Setup required library paths:
\smallskip
\noindent
\setlength{\fboxsep}{4pt}
\setlength{\fboxrule}{0.5pt}
\fcolorbox{black}{lightgray}{%
  \begin{minipage}{0.94\linewidth}
    \small\ttfamily
    \textbf{cd PRAC\_TC\_ISCA25/TPRAC}\\
    \textbf{source setup\_lib\_path.sh}
  \end{minipage}
}

\smallskip
    
\noindent 2. Configure the following parameters in \texttt{PRAC\_TC\_ISCA25/TPRAC/run\_artifact.sh}: 
\begin{itemize}[leftmargin=*,topsep=1pt]
    \item Using Slurm:
        \begin{itemize}
            \item SLRUM\_PART\_NAME: Partition name for Slurm jobs. 
            \item SLRUM\_PART\_DEF\_MEM: Default memory size for jobs (recommended: $\geq$ 6GB). 
            \item MAX\_SLRUM\_JOBS: Maximum number of Slurm jobs to submit.
        \end{itemize}
    \item Using a Personal Server:
        \begin{itemize}
            \item PERSONAL\_RUN\_THREADS: Number of parallel threads to use for simulations.
        \end{itemize}    
\end{itemize}

\noindent 3. Run the following commands to install dependencies, build ChampSim and Ramulator2, and execute simulations. If using a personal server, we recommend first running the security analysis (\cref{fig:tprac_max_act}) and main performance experiment (\cref{fig:perf_main}) and reviewing the results before proceeding with the full set of experiments.

\smallskip
\noindent 3.1. Running the security analysis (\cref{fig:tprac_max_act}) and main performance experiment (\cref{fig:perf_main}):

\noindent{Using Slurm}:

\smallskip
\noindent
\setlength{\fboxsep}{4pt}
\setlength{\fboxrule}{0.5pt}
\fcolorbox{black}{lightgray}{%
  \begin{minipage}{0.94\linewidth}
    \small\ttfamily
    \textbf{cd PRAC\_TC\_ISCA25/TPRAC}\\
    \textbf{./run\_artifact.sh ----method slurm ----artifact main}
  \end{minipage}
}

\smallskip
\noindent{Using a Personal Server:}

\smallskip
\noindent
\setlength{\fboxsep}{4pt}
\setlength{\fboxrule}{0.5pt}
\fcolorbox{black}{lightgray}{%
  \begin{minipage}{0.94\linewidth}
    \small\ttfamily
    \textbf{cd PRAC\_TC\_ISCA25/TPRAC}\\
    \textbf{./run\_artifact.sh ----method personal ----artifact main}
  \end{minipage}
}

\smallskip
\noindent 3.2. Running the security analysis (\cref{fig:tprac_max_act}) and all performance experiments (\cref{fig:perf_main} to~\cref{fig:counter_reset_sensitivity}):

\smallskip
\noindent{Using Slurm}:

\smallskip
\noindent
\setlength{\fboxsep}{4pt}
\setlength{\fboxrule}{0.5pt}
\fcolorbox{black}{lightgray}{%
  \begin{minipage}{0.94\linewidth}
    \small\ttfamily
    \textbf{cd PRAC\_TC\_ISCA25/TPRAC}\\
    \textbf{./run\_artifact.sh ----method slurm ----artifact all}
  \end{minipage}
}
    
\smallskip
\noindent{Using a Personal Server:}

\smallskip
\noindent
\setlength{\fboxsep}{4pt}
\setlength{\fboxrule}{0.5pt}
\fcolorbox{black}{lightgray}{%
  \begin{minipage}{0.94\linewidth}
    \small\ttfamily
    \textbf{cd PRAC\_TC\_ISCA25/TPRAC}\\
    \textbf{./run\_artifact.sh ----method personal ----artifact all}
  \end{minipage}
}
\subsection{Evaluation and Expected Results}
\subsubsection{\Attackname{} Evaluations}
We provide \texttt{run\_artifact.sh} to demonstrate \Attackname{} through 5 experiments. Scripts to run them can be found in \texttt{PRAC\_TC\_ISCA25/PRACLeak/scripts/sim\_scripts}, including \texttt{run\_latency.sh}, \texttt{run\_side\_channel.sh}, \texttt{run\_covert\_channel.sh}, \texttt{run\_aes\_no\_defense.sh}, and \texttt{run\_aes\_with\_defense.sh}. After the experiments, raw data will be stored in \texttt{./PRACLeak/results/raw}. 

Moreover, we provide \texttt{plot\_all\_figures.sh} in \texttt{PRAC\_TC\_ISCA25/PRACLeak} to generate \Cref{fig:timing_variation} through \Cref{fig:sidechannel_heatmap} and \Cref{fig:sidechannel_defense}. Generated plots will be stored as PDF files in \texttt{PRAC\_TC\_ISCA25/PRACLeak/results/plots}. Scripts for plotting individual figures can be found in \texttt{PRAC\_TC\_ISCA25/PRACLeak/scripts/plot\_scripts}.

\subsubsection{\Defensename{} Evaluations}
After completing the security analysis and performance experiments using \texttt{run\_artifact.sh}, the results and plots can be regenerated with the provided scripts. Specifically, the artifact includes the \texttt{plot\_main\_figure.sh} and \texttt{plot\_all\_figures.sh} files in the PRAC\_TC\_ISCA25/TPRAC directory. These scripts collate the results (obtained as CSV files in PRAC\_TC\_ISCA25/TPRAC/results/csvs) and generate the plots (obtained as PDF files in PRAC\_TC\_ISCA25/TPRAC/results/plots). The \texttt{plot\_main\_figure.sh} script regenerates \cref{fig:perf_main}, while the \texttt{plot\_all\_figures.sh} script generates \cref{fig:perf_main} to~\cref{fig:counter_reset_sensitivity}. Additionally, we provide scripts to collate results (\texttt{generate\_csv\_fig\#.py}) and generate plots (\texttt{plot\_fig\#.py} or \texttt{plot.ipynb}) for each experiment in the PRAC\_TC\_ISCA25/TPRAC/plot\_scripts directory. Sample result files and plots are available in the PRAC\_TC\_ISCA25/TPRAC/results/sample\_results directory.

\subsection{Experiment Customization: \Defensename{} Performance Evaluations}
We offer easy configuration options for the following parameters: 
1) the evaluated \Defensename{} mechanisms, 
2) the tested PRAC levels,
3) the evaluated RowHammer Thresholds, and 
4) the simulation duration (minimum number of instructions per core during experiments).
These parameters can be customized in the \texttt{PRAC\_TC\_ISCA25/TPRAC/sim\_scripts/run\_config\_fig\#.py} files:
\begin{itemize}[leftmargin=*,topsep=1pt]
    \item \Defensename{} configurations: \texttt{mitigation\_list}.
    \item PRAC levels: \texttt{PRAC\_levels}.
    \item RowHammer Thresholds: \texttt{NRH\_lists}.
    \item Simulation duration: \texttt{NUM\_EXPECTED\_INSTS}.
\end{itemize}

\subsection{Methodology}

Submission, reviewing and badging methodology:

\begin{itemize}
  \item \url{https://www.acm.org/publications/policies/artifact-review-and-badging-current}
  \item \url{https://cTuning.org/ae}
\end{itemize}
\end{appendices}

\balance
\bibliographystyle{ACM-Reference-Format}
\bibliography{main}


\begin{thebibliography}{116}


\ifx \showCODEN    \undefined \def \showCODEN     #1{\unskip}     \fi
\ifx \showDOI      \undefined \def \showDOI       #1{#1}\fi
\ifx \showISBNx    \undefined \def \showISBNx     #1{\unskip}     \fi
\ifx \showISBNxiii \undefined \def \showISBNxiii  #1{\unskip}     \fi
\ifx \showISSN     \undefined \def \showISSN      #1{\unskip}     \fi
\ifx \showLCCN     \undefined \def \showLCCN      #1{\unskip}     \fi
\ifx \shownote     \undefined \def \shownote      #1{#1}          \fi
\ifx \showarticletitle \undefined \def \showarticletitle #1{#1}   \fi
\ifx \showURL      \undefined \def \showURL       {\relax}        \fi
\providecommand\bibfield[2]{#2}
\providecommand\bibinfo[2]{#2}
\providecommand\natexlab[1]{#1}
\providecommand\showeprint[2][]{arXiv:#2}

\bibitem[soc({[n.\,d.]})]%
        {sockeye}
 \bibinfo{year}{[n.\,d.]}\natexlab{}.
\newblock \showarticletitle{U{B}{C} {A}dvanced {R}esearch {C}omputing, "{U}{B}{C} {A}{R}{C} {S}ockeye." {U}{B}{C} {A}dvanced {R}esearch {C}omputing, 2019, doi: 10.14288/{S}{O}{C}{K}{E}{Y}{E}}.
\newblock  (\bibinfo{year}{[n.\,d.]}).
\newblock


\bibitem[Asgari~Khoshouyeh et~al\mbox{.}(2023)]%
        {stuctcoding}
\bibfield{author}{\bibinfo{person}{Ali Asgari~Khoshouyeh}, \bibinfo{person}{Florian Geissler}, \bibinfo{person}{Syed Qutub}, \bibinfo{person}{Michael Paulitsch}, \bibinfo{person}{Prashant Nair}, {and} \bibinfo{person}{Karthik Pattabiraman}.} \bibinfo{year}{2023}\natexlab{}.
\newblock \showarticletitle{Structural Coding: A Low-Cost Scheme to Protect CNNs from Large-Granularity Memory Faults}. In \bibinfo{booktitle}{\emph{Proceedings of the International Conference for High Performance Computing, Networking, Storage and Analysis}} (Denver, CO, USA) \emph{(\bibinfo{series}{SC '23})}. \bibinfo{publisher}{Association for Computing Machinery}, \bibinfo{address}{New York, NY, USA}, Article \bibinfo{articleno}{85}, \bibinfo{numpages}{17}~pages.
\newblock
\showISBNx{9798400701092}
\urldef\tempurl%
\url{https://doi.org/10.1145/3581784.3607084}
\showDOI{\tempurl}


\bibitem[Bernstein(2005)]%
        {Bernstein}
\bibfield{author}{\bibinfo{person}{Daniel~J. Bernstein}.} \bibinfo{year}{2005}\natexlab{}.
\newblock \bibinfo{booktitle}{\emph{Cache-timing attacks on AES}}.
\newblock \bibinfo{type}{{T}echnical {R}eport}.
\newblock


\bibitem[Bhatia et~al\mbox{.}(2019)]%
        {spp-ppf}
\bibfield{author}{\bibinfo{person}{Eshan Bhatia}, \bibinfo{person}{Gino Chacon}, \bibinfo{person}{Seth Pugsley}, \bibinfo{person}{Elvira Teran}, \bibinfo{person}{Paul~V. Gratz}, {and} \bibinfo{person}{Daniel~A. Jim\'{e}nez}.} \bibinfo{year}{2019}\natexlab{}.
\newblock \showarticletitle{Perceptron-based prefetch filtering}. In \bibinfo{booktitle}{\emph{Proceedings of the 46th International Symposium on Computer Architecture}} (Phoenix, Arizona) \emph{(\bibinfo{series}{ISCA '19})}. \bibinfo{publisher}{Association for Computing Machinery}, \bibinfo{address}{New York, NY, USA}, \bibinfo{pages}{1–13}.
\newblock
\showISBNx{9781450366694}
\urldef\tempurl%
\url{https://doi.org/10.1145/3307650.3322207}
\showDOI{\tempurl}


\bibitem[Bonneau and Mironov(2006)]%
        {Bonneau}
\bibfield{author}{\bibinfo{person}{Joseph Bonneau} {and} \bibinfo{person}{Ilya Mironov}.} \bibinfo{year}{2006}\natexlab{}.
\newblock \showarticletitle{Cache-collision timing attacks against AES}. In \bibinfo{booktitle}{\emph{CHES}}.
\newblock


\bibitem[Bostanci et~al\mbox{.}(2024)]%
        {comet}
\bibfield{author}{\bibinfo{person}{F.~Nisa Bostanci}, \bibinfo{person}{ISmail~Emir Yüksel}, \bibinfo{person}{Ataberk Olgun}, \bibinfo{person}{Konstantinos Kanellopoulos}, \bibinfo{person}{Yahya~Can Tuğrul}, \bibinfo{person}{A.~Giray Yağliçi}, \bibinfo{person}{Mohammad Sadrosadati}, {and} \bibinfo{person}{Onur Mutlu}.} \bibinfo{year}{2024}\natexlab{}.
\newblock \showarticletitle{CoMeT: Count-Min-Sketch-based Row Tracking to Mitigate RowHammer at Low Cost}. In \bibinfo{booktitle}{\emph{2024 IEEE International Symposium on High-Performance Computer Architecture (HPCA)}}. \bibinfo{pages}{593--612}.
\newblock
\urldef\tempurl%
\url{https://doi.org/10.1109/HPCA57654.2024.00050}
\showDOI{\tempurl}


\bibitem[Cai et~al\mbox{.}(2024)]%
        {deepvenom}
\bibfield{author}{\bibinfo{person}{Kunbei Cai}, \bibinfo{person}{Md~Hafizul~Islam Chowdhuryy}, \bibinfo{person}{Zhenkai Zhang}, {and} \bibinfo{person}{Fan Yao}.} \bibinfo{year}{2024}\natexlab{}.
\newblock \showarticletitle{DeepVenom: Persistent DNN Backdoors Exploiting Transient Weight Perturbations in Memories}. In \bibinfo{booktitle}{\emph{2024 IEEE Symposium on Security and Privacy (SP)}}. \bibinfo{pages}{2067--2085}.
\newblock
\urldef\tempurl%
\url{https://doi.org/10.1109/SP54263.2024.00223}
\showDOI{\tempurl}


\bibitem[Canpolat et~al\mbox{.}(2024a)]%
        {UPRAC}
\bibfield{author}{\bibinfo{person}{O{\u{g}}uzhan Canpolat}, \bibinfo{person}{A~Giray Ya{\u{g}}l{\i}k{\c{c}}{\i}}, \bibinfo{person}{Geraldo~F Oliveira}, \bibinfo{person}{Ataberk Olgun}, \bibinfo{person}{O{\u{g}}uz Ergin}, {and} \bibinfo{person}{Onur Mutlu}.} \bibinfo{year}{2024}\natexlab{a}.
\newblock \showarticletitle{Understanding the Security Benefits and Overheads of Emerging Industry Solutions to DRAM Read Disturbance}. In \bibinfo{booktitle}{\emph{Workshop on DRAM Security (DRAMSec)}}.
\newblock


\bibitem[Canpolat et~al\mbox{.}(2024b)]%
        {breakhammer2024}
\bibfield{author}{\bibinfo{person}{O\u{g}uzhan Canpolat}, \bibinfo{person}{A.~Giray Ya\u{g}l\i{}k\c{c}\i{}}, \bibinfo{person}{Ataberk Olgun}, \bibinfo{person}{Ismail~Emir Yuksel}, \bibinfo{person}{Yahya~Can Tu\u{g}rul}, \bibinfo{person}{Konstantinos Kanellopoulos}, \bibinfo{person}{O\u{g}uz Ergin}, {and} \bibinfo{person}{Onur Mutlu}.} \bibinfo{year}{2024}\natexlab{b}.
\newblock \showarticletitle{BreakHammer: Enhancing RowHammer Mitigations by Carefully Throttling Suspect Threads}. In \bibinfo{booktitle}{\emph{2024 57th IEEE/ACM International Symposium on Microarchitecture (MICRO)}}. \bibinfo{pages}{915--934}.
\newblock
\urldef\tempurl%
\url{https://doi.org/10.1109/MICRO61859.2024.00072}
\showDOI{\tempurl}


\bibitem[Canpolat et~al\mbox{.}(2025)]%
        {Chronus}
\bibfield{author}{\bibinfo{person}{Oğuzhan Canpolat}, \bibinfo{person}{A.~Giray Yağlıkçı}, \bibinfo{person}{Geraldo~F. Oliveira}, \bibinfo{person}{Ataberk Olgun}, \bibinfo{person}{Nisa Bostancı}, \bibinfo{person}{Ismail~Emir Yuksel}, \bibinfo{person}{Haocong Luo}, \bibinfo{person}{Oğuz Ergin}, {and} \bibinfo{person}{Onur Mutlu}.} \bibinfo{year}{2025}\natexlab{}.
\newblock \showarticletitle{Chronus: Understanding and Securing the Cutting-Edge Industry Solutions to DRAM Read Disturbance}. In \bibinfo{booktitle}{\emph{2025 IEEE International Symposium on High Performance Computer Architecture (HPCA)}}. \bibinfo{pages}{887--905}.
\newblock
\urldef\tempurl%
\url{https://doi.org/10.1109/HPCA61900.2025.00071}
\showDOI{\tempurl}


\bibitem[Chang et~al\mbox{.}(2016)]%
        {lisa}
\bibfield{author}{\bibinfo{person}{Kevin~K. Chang}, \bibinfo{person}{Prashant~J. Nair}, \bibinfo{person}{Donghyuk Lee}, \bibinfo{person}{Saugata Ghose}, \bibinfo{person}{Moinuddin~K. Qureshi}, {and} \bibinfo{person}{Onur Mutlu}.} \bibinfo{year}{2016}\natexlab{}.
\newblock \showarticletitle{Low-Cost Inter-Linked Subarrays (LISA): Enabling fast inter-subarray data movement in DRAM}. In \bibinfo{booktitle}{\emph{2016 IEEE International Symposium on High Performance Computer Architecture (HPCA)}}. \bibinfo{pages}{568--580}.
\newblock
\urldef\tempurl%
\url{https://doi.org/10.1109/HPCA.2016.7446095}
\showDOI{\tempurl}


\bibitem[Chen et~al\mbox{.}(2024)]%
        {GoFetch}
\bibfield{author}{\bibinfo{person}{Boru Chen}, \bibinfo{person}{Yingchen Wang}, \bibinfo{person}{Pradyumna Shome}, \bibinfo{person}{Christopher Fletcher}, \bibinfo{person}{David Kohlbrenner}, \bibinfo{person}{Riccardo Paccagnella}, {and} \bibinfo{person}{Daniel Genkin}.} \bibinfo{year}{2024}\natexlab{}.
\newblock \showarticletitle{{GoFetch}: Breaking {Constant-Time} Cryptographic Implementations Using Data {Memory-Dependent} Prefetchers}. In \bibinfo{booktitle}{\emph{33rd USENIX Security Symposium (USENIX Security 24)}}. \bibinfo{publisher}{USENIX Association}, \bibinfo{address}{Philadelphia, PA}, \bibinfo{pages}{1117--1134}.
\newblock
\showISBNx{978-1-939133-44-1}
\urldef\tempurl%
\url{https://www.usenix.org/conference/usenixsecurity24/presentation/chen-boru}
\showURL{%
\tempurl}


\bibitem[Chen et~al\mbox{.}(2023)]%
        {ip-stride}
\bibfield{author}{\bibinfo{person}{Yun Chen}, \bibinfo{person}{Lingfeng Pei}, {and} \bibinfo{person}{Trevor~E. Carlson}.} \bibinfo{year}{2023}\natexlab{}.
\newblock \showarticletitle{AfterImage: Leaking Control Flow Data and Tracking Load Operations via the Hardware Prefetcher}. In \bibinfo{booktitle}{\emph{Proceedings of the 28th ACM International Conference on Architectural Support for Programming Languages and Operating Systems, Volume 2}} (Vancouver, BC, Canada) \emph{(\bibinfo{series}{ASPLOS 2023})}. \bibinfo{publisher}{Association for Computing Machinery}, \bibinfo{address}{New York, NY, USA}, \bibinfo{pages}{16–32}.
\newblock
\showISBNx{9781450399166}
\urldef\tempurl%
\url{https://doi.org/10.1145/3575693.3575719}
\showDOI{\tempurl}


\bibitem[Chowdhuryy et~al\mbox{.}(2024)]%
        {MetaLeak}
\bibfield{author}{\bibinfo{person}{Md~Hafizul~Islam Chowdhuryy}, \bibinfo{person}{Hao Zheng}, {and} \bibinfo{person}{Fan Yao}.} \bibinfo{year}{2024}\natexlab{}.
\newblock \showarticletitle{MetaLeak: Uncovering Side Channels in Secure Processor Architectures Exploiting Metadata}. In \bibinfo{booktitle}{\emph{2024 ACM/IEEE 51st Annual International Symposium on Computer Architecture (ISCA)}}. \bibinfo{pages}{693--707}.
\newblock
\urldef\tempurl%
\url{https://doi.org/10.1109/ISCA59077.2024.00056}
\showDOI{\tempurl}


\bibitem[Coalson et~al\mbox{.}(2024)]%
        {prisonbreak}
\bibfield{author}{\bibinfo{person}{Zachary Coalson}, \bibinfo{person}{Jeonghyun Woo}, \bibinfo{person}{Shiyang Chen}, \bibinfo{person}{Yu Sun}, \bibinfo{person}{Lishan Yang}, \bibinfo{person}{Prashant Nair}, \bibinfo{person}{Bo Fang}, {and} \bibinfo{person}{Sanghyun Hong}.} \bibinfo{year}{2024}\natexlab{}.
\newblock \bibinfo{title}{PrisonBreak: Jailbreaking Large Language Models with Fewer Than Twenty-Five Targeted Bit-flips}.
\newblock
\newblock
\showeprint[arxiv]{2412.07192}~[cs.CR]
\urldef\tempurl%
\url{https://arxiv.org/abs/2412.07192}
\showURL{%
\tempurl}


\bibitem[Cojocar et~al\mbox{.}(2019)]%
        {cojocar2019eccploit}
\bibfield{author}{\bibinfo{person}{Lucian Cojocar}, \bibinfo{person}{Kaveh Razavi}, \bibinfo{person}{Cristiano Giuffrida}, {and} \bibinfo{person}{Herbert Bos}.} \bibinfo{year}{2019}\natexlab{}.
\newblock \showarticletitle{Exploiting Correcting Codes: On the Effectiveness of ECC Memory Against Rowhammer Attacks}. In \bibinfo{booktitle}{\emph{2019 IEEE Symposium on Security and Privacy (SP)}}. \bibinfo{pages}{55--71}.
\newblock
\urldef\tempurl%
\url{https://doi.org/10.1109/SP.2019.00089}
\showDOI{\tempurl}


\bibitem[Corporation(2006)]%
        {SPEC2006}
\bibfield{author}{\bibinfo{person}{Standard Performance~Evaluation Corporation}.} \bibinfo{year}{2006}\natexlab{}.
\newblock \bibinfo{title}{SPEC CPU2006 Benchmark Suite}.
\newblock
\newblock
\urldef\tempurl%
\url{http://www.spec.org/cpu2006/}
\showURL{%
\tempurl}


\bibitem[Deutsch et~al\mbox{.}(2022)]%
        {dagguise}
\bibfield{author}{\bibinfo{person}{Peter~W. Deutsch}, \bibinfo{person}{Yuheng Yang}, \bibinfo{person}{Thomas Bourgeat}, \bibinfo{person}{Jules Drean}, \bibinfo{person}{Joel~S. Emer}, {and} \bibinfo{person}{Mengjia Yan}.} \bibinfo{year}{2022}\natexlab{}.
\newblock \showarticletitle{DAGguise: mitigating memory timing side channels}. In \bibinfo{booktitle}{\emph{Proceedings of the 27th ACM International Conference on Architectural Support for Programming Languages and Operating Systems}} (Lausanne, Switzerland) \emph{(\bibinfo{series}{ASPLOS '22})}. \bibinfo{publisher}{Association for Computing Machinery}, \bibinfo{address}{New York, NY, USA}, \bibinfo{pages}{329–343}.
\newblock
\showISBNx{9781450392051}
\urldef\tempurl%
\url{https://doi.org/10.1145/3503222.3507747}
\showDOI{\tempurl}


\bibitem[Evtyushkin et~al\mbox{.}(2018)]%
        {BranchScope}
\bibfield{author}{\bibinfo{person}{Dmitry Evtyushkin}, \bibinfo{person}{Ryan Riley}, \bibinfo{person}{Nael~CSE Abu-Ghazaleh}, \bibinfo{person}{ECE}, {and} \bibinfo{person}{Dmitry Ponomarev}.} \bibinfo{year}{2018}\natexlab{}.
\newblock \showarticletitle{BranchScope: A New Side-Channel Attack on Directional Branch Predictor}. In \bibinfo{booktitle}{\emph{Proceedings of the Twenty-Third International Conference on Architectural Support for Programming Languages and Operating Systems}} (Williamsburg, VA, USA) \emph{(\bibinfo{series}{ASPLOS '18})}. \bibinfo{publisher}{Association for Computing Machinery}, \bibinfo{address}{New York, NY, USA}, \bibinfo{pages}{693–707}.
\newblock
\showISBNx{9781450349116}
\urldef\tempurl%
\url{https://doi.org/10.1145/3173162.3173204}
\showDOI{\tempurl}


\bibitem[Fakhrzadehgan et~al\mbox{.}(2022)]%
        {ali2022safeguard}
\bibfield{author}{\bibinfo{person}{Ali Fakhrzadehgan}, \bibinfo{person}{Yale~N. Patt}, \bibinfo{person}{Prashant~J. Nair}, {and} \bibinfo{person}{Moinuddin~K. Qureshi}.} \bibinfo{year}{2022}\natexlab{}.
\newblock \showarticletitle{SafeGuard: Reducing the Security Risk from Row-Hammer via Low-Cost Integrity Protection}. In \bibinfo{booktitle}{\emph{2022 IEEE International Symposium on High-Performance Computer Architecture (HPCA)}}. \bibinfo{pages}{373--386}.
\newblock
\urldef\tempurl%
\url{https://doi.org/10.1109/HPCA53966.2022.00035}
\showDOI{\tempurl}


\bibitem[Ferdman et~al\mbox{.}(2012)]%
        {CloudSuite}
\bibfield{author}{\bibinfo{person}{Michael Ferdman}, \bibinfo{person}{Almutaz Adileh}, \bibinfo{person}{Onur Kocberber}, \bibinfo{person}{Stavros Volos}, \bibinfo{person}{Mohammad Alisafaee}, \bibinfo{person}{Djordje Jevdjic}, \bibinfo{person}{Cansu Kaynak}, \bibinfo{person}{Adrian~Daniel Popescu}, \bibinfo{person}{Anastasia Ailamaki}, {and} \bibinfo{person}{Babak Falsafi}.} \bibinfo{year}{2012}\natexlab{}.
\newblock \showarticletitle{Clearing the clouds: a study of emerging scale-out workloads on modern hardware}. In \bibinfo{booktitle}{\emph{Proceedings of the Seventeenth International Conference on Architectural Support for Programming Languages and Operating Systems}} (London, England, UK) \emph{(\bibinfo{series}{ASPLOS XVII})}. \bibinfo{publisher}{Association for Computing Machinery}, \bibinfo{address}{New York, NY, USA}, \bibinfo{pages}{37–48}.
\newblock
\showISBNx{9781450307598}
\urldef\tempurl%
\url{https://doi.org/10.1145/2150976.2150982}
\showDOI{\tempurl}


\bibitem[Frigo et~al\mbox{.}(2020)]%
        {frigo2020trrespass}
\bibfield{author}{\bibinfo{person}{Pietro Frigo}, \bibinfo{person}{Emanuele Vannacc}, \bibinfo{person}{Hasan Hassan}, \bibinfo{person}{Victor~van der Veen}, \bibinfo{person}{Onur Mutlu}, \bibinfo{person}{Cristiano Giuffrida}, \bibinfo{person}{Herbert Bos}, {and} \bibinfo{person}{Kaveh Razavi}.} \bibinfo{year}{2020}\natexlab{}.
\newblock \showarticletitle{TRRespass: Exploiting the Many Sides of Target Row Refresh}. In \bibinfo{booktitle}{\emph{2020 IEEE Symposium on Security and Privacy (SP)}}. \bibinfo{pages}{747--762}.
\newblock
\urldef\tempurl%
\url{https://doi.org/10.1109/SP40000.2020.00090}
\showDOI{\tempurl}


\bibitem[Gast et~al\mbox{.}(2023)]%
        {Squip}
\bibfield{author}{\bibinfo{person}{Stefan Gast}, \bibinfo{person}{Jonas Juffinger}, \bibinfo{person}{Martin Schwarzl}, \bibinfo{person}{Gururaj Saileshwar}, \bibinfo{person}{Andreas Kogler}, \bibinfo{person}{Simone Franza}, \bibinfo{person}{Markus Köstl}, {and} \bibinfo{person}{Daniel Gruss}.} \bibinfo{year}{2023}\natexlab{}.
\newblock \showarticletitle{SQUIP: Exploiting the Scheduler Queue Contention Side Channel}. In \bibinfo{booktitle}{\emph{2023 IEEE Symposium on Security and Privacy (SP)}}. \bibinfo{pages}{2256--2272}.
\newblock
\urldef\tempurl%
\url{https://doi.org/10.1109/SP46215.2023.10179368}
\showDOI{\tempurl}


\bibitem[Gober et~al\mbox{.}(2022)]%
        {champsim}
\bibfield{author}{\bibinfo{person}{Nathan Gober}, \bibinfo{person}{Gino Chacon}, \bibinfo{person}{Lei Wang}, \bibinfo{person}{Paul~V. Gratz}, \bibinfo{person}{Daniel~A. Jimenez}, \bibinfo{person}{Elvira Teran}, \bibinfo{person}{Seth Pugsley}, {and} \bibinfo{person}{Jinchun Kim}.} \bibinfo{year}{2022}\natexlab{}.
\newblock \bibinfo{title}{The Championship Simulator: Architectural Simulation for Education and Competition}.
\newblock
\newblock
\showeprint[arxiv]{2210.14324}~[cs.AR]
\urldef\tempurl%
\url{https://arxiv.org/abs/2210.14324}
\showURL{%
\tempurl}


\bibitem[Gruss et~al\mbox{.}(2017)]%
        {tmem:usenix2017}
\bibfield{author}{\bibinfo{person}{Daniel Gruss}, \bibinfo{person}{Julian Lettner}, \bibinfo{person}{Felix Schuster}, \bibinfo{person}{Olya Ohrimenko}, \bibinfo{person}{Istvan Haller}, {and} \bibinfo{person}{Manuel Costa}.} \bibinfo{year}{2017}\natexlab{}.
\newblock \showarticletitle{Strong and Efficient Cache {Side-Channel} Protection using Hardware Transactional Memory}. In \bibinfo{booktitle}{\emph{26th USENIX Security Symposium (USENIX Security 17)}}. \bibinfo{publisher}{USENIX Association}, \bibinfo{address}{Vancouver, BC}, \bibinfo{pages}{217--233}.
\newblock
\showISBNx{978-1-931971-40-9}
\urldef\tempurl%
\url{https://www.usenix.org/conference/usenixsecurity17/technical-sessions/presentation/gruss}
\showURL{%
\tempurl}


\bibitem[Gruss et~al\mbox{.}(2018)]%
        {gruss2018another}
\bibfield{author}{\bibinfo{person}{Daniel Gruss}, \bibinfo{person}{Moritz Lipp}, \bibinfo{person}{Michael Schwarz}, \bibinfo{person}{Daniel Genkin}, \bibinfo{person}{Jonas Juffinger}, \bibinfo{person}{Sioli O'Connell}, \bibinfo{person}{Wolfgang Schoechl}, {and} \bibinfo{person}{Yuval Yarom}.} \bibinfo{year}{2018}\natexlab{}.
\newblock \showarticletitle{Another Flip in the Wall of Rowhammer Defenses}. In \bibinfo{booktitle}{\emph{2018 IEEE Symposium on Security and Privacy (SP)}}. \bibinfo{pages}{245--261}.
\newblock
\urldef\tempurl%
\url{https://doi.org/10.1109/SP.2018.00031}
\showDOI{\tempurl}


\bibitem[Gruss et~al\mbox{.}(2016a)]%
        {gruss2016rhjs}
\bibfield{author}{\bibinfo{person}{Daniel Gruss}, \bibinfo{person}{Cl\'{e}mentine Maurice}, {and} \bibinfo{person}{Stefan Mangard}.} \bibinfo{year}{2016}\natexlab{a}.
\newblock \showarticletitle{Rowhammer.js: A Remote Software-Induced Fault Attack in JavaScript}. In \bibinfo{booktitle}{\emph{Proceedings of the 13th International Conference on Detection of Intrusions and Malware, and Vulnerability Assessment - Volume 9721}} (San Sebasti\'{a}n, Spain) \emph{(\bibinfo{series}{DIMVA 2016})}. \bibinfo{publisher}{Springer-Verlag}, \bibinfo{address}{Berlin, Heidelberg}, \bibinfo{pages}{300–321}.
\newblock
\showISBNx{9783319406664}
\urldef\tempurl%
\url{https://doi.org/10.1007/978-3-319-40667-1_15}
\showDOI{\tempurl}


\bibitem[Gruss et~al\mbox{.}(2016b)]%
        {FlushFlush}
\bibfield{author}{\bibinfo{person}{Daniel Gruss}, \bibinfo{person}{Cl\'{e}mentine Maurice}, \bibinfo{person}{Klaus Wagner}, {and} \bibinfo{person}{Stefan Mangard}.} \bibinfo{year}{2016}\natexlab{b}.
\newblock \showarticletitle{Flush+Flush: A Fast and Stealthy Cache Attack}. In \bibinfo{booktitle}{\emph{Proceedings of the 13th International Conference on Detection of Intrusions and Malware, and Vulnerability Assessment - Volume 9721}} (San Sebasti\'{a}n, Spain) \emph{(\bibinfo{series}{DIMVA 2016})}. \bibinfo{publisher}{Springer-Verlag}, \bibinfo{address}{Berlin, Heidelberg}, \bibinfo{pages}{279–299}.
\newblock
\showISBNx{9783319406664}
\urldef\tempurl%
\url{https://doi.org/10.1007/978-3-319-40667-1_14}
\showDOI{\tempurl}


\bibitem[Hassan et~al\mbox{.}(2019)]%
        {CROW}
\bibfield{author}{\bibinfo{person}{Hasan Hassan}, \bibinfo{person}{Minesh Patel}, \bibinfo{person}{Jeremie~S. Kim}, \bibinfo{person}{A.~Giray Yaglikci}, \bibinfo{person}{Nandita Vijaykumar}, \bibinfo{person}{Nika~Mansouri Ghiasi}, \bibinfo{person}{Saugata Ghose}, {and} \bibinfo{person}{Onur Mutlu}.} \bibinfo{year}{2019}\natexlab{}.
\newblock \showarticletitle{CROW: a low-cost substrate for improving DRAM performance, energy efficiency, and reliability}. In \bibinfo{booktitle}{\emph{Proceedings of the 46th International Symposium on Computer Architecture}} (Phoenix, Arizona) \emph{(\bibinfo{series}{ISCA '19})}. \bibinfo{publisher}{Association for Computing Machinery}, \bibinfo{address}{New York, NY, USA}, \bibinfo{pages}{129–142}.
\newblock
\showISBNx{9781450366694}
\urldef\tempurl%
\url{https://doi.org/10.1145/3307650.3322231}
\showDOI{\tempurl}


\bibitem[Hassan et~al\mbox{.}(2021)]%
        {hassan2021UTRR}
\bibfield{author}{\bibinfo{person}{Hasan Hassan}, \bibinfo{person}{Yahya~Can Tugrul}, \bibinfo{person}{Jeremie~S. Kim}, \bibinfo{person}{Victor van~der Veen}, \bibinfo{person}{Kaveh Razavi}, {and} \bibinfo{person}{Onur Mutlu}.} \bibinfo{year}{2021}\natexlab{}.
\newblock \showarticletitle{Uncovering In-DRAM RowHammer Protection Mechanisms:A New Methodology, Custom RowHammer Patterns, and Implications}. In \bibinfo{booktitle}{\emph{MICRO-54: 54th Annual IEEE/ACM International Symposium on Microarchitecture}} (Virtual Event, Greece) \emph{(\bibinfo{series}{MICRO '21})}. \bibinfo{publisher}{Association for Computing Machinery}, \bibinfo{address}{New York, NY, USA}, \bibinfo{pages}{1198–1213}.
\newblock
\showISBNx{9781450385572}
\urldef\tempurl%
\url{https://doi.org/10.1145/3466752.3480110}
\showDOI{\tempurl}


\bibitem[Hong et~al\mbox{.}(2019)]%
        {hong2019terminal}
\bibfield{author}{\bibinfo{person}{Sanghyun Hong}, \bibinfo{person}{Pietro Frigo}, \bibinfo{person}{Yigitcan Kaya}, \bibinfo{person}{Cristiano Giuffrida}, {and} \bibinfo{person}{Tudor Dumitras}.} \bibinfo{year}{2019}\natexlab{}.
\newblock \showarticletitle{Terminal Brain Damage: Exposing the Graceless Degradation in Deep Neural Networks Under Hardware Fault Attacks}. In \bibinfo{booktitle}{\emph{28th USENIX Security Symposium (USENIX Security 19)}}. \bibinfo{publisher}{USENIX Association}, \bibinfo{address}{Santa Clara, CA}, \bibinfo{pages}{497--514}.
\newblock
\showISBNx{978-1-939133-06-9}
\urldef\tempurl%
\url{https://www.usenix.org/conference/usenixsecurity19/presentation/hong}
\showURL{%
\tempurl}


\bibitem[Hong et~al\mbox{.}(2023)]%
        {DSAC}
\bibfield{author}{\bibinfo{person}{Seungki Hong}, \bibinfo{person}{Dongha Kim}, \bibinfo{person}{Jaehyung Lee}, \bibinfo{person}{Reum Oh}, \bibinfo{person}{Changsik Yoo}, \bibinfo{person}{Sangjoon Hwang}, {and} \bibinfo{person}{Jooyoung Lee}.} \bibinfo{year}{2023}\natexlab{}.
\newblock \showarticletitle{Dsac: Low-cost rowhammer mitigation using in-dram stochastic and approximate counting algorithm}.
\newblock \bibinfo{journal}{\emph{arXiv:2302.03591}} (\bibinfo{year}{2023}).
\newblock


\bibitem[Intel(2022)]%
        {SGX2}
\bibfield{author}{\bibinfo{person}{Intel}.} \bibinfo{year}{2022}\natexlab{}.
\newblock \bibinfo{title}{{What does SGX 2.0 (scalable SGX) sacrifice}}.
\newblock \bibinfo{howpublished}{Github}.
\newblock
\urldef\tempurl%
\url{https://github.com/intel/linux-sgx/issues/899}
\showURL{%
\tempurl}


\bibitem[Jaleel et~al\mbox{.}(2024a)]%
        {PROTEAS}
\bibfield{author}{\bibinfo{person}{Aamer Jaleel}, \bibinfo{person}{Stephen~W. Keckler}, {and} \bibinfo{person}{Gururaj Saileshwar}.} \bibinfo{year}{2024}\natexlab{a}.
\newblock \showarticletitle{Probabilistic Tracker Management Policies for Low-Cost and Scalable Rowhammer Mitigation}.
\newblock \bibinfo{journal}{\emph{arXiv:2404.16256}} (\bibinfo{year}{2024}).
\newblock


\bibitem[Jaleel et~al\mbox{.}(2024b)]%
        {jaleel2024pride}
\bibfield{author}{\bibinfo{person}{Aamer Jaleel}, \bibinfo{person}{Gururaj Saileshwar}, \bibinfo{person}{Stephen~W. Keckler}, {and} \bibinfo{person}{Moinuddin Qureshi}.} \bibinfo{year}{2024}\natexlab{b}.
\newblock \showarticletitle{PrIDE: Achieving Secure Rowhammer Mitigation with Low-Cost In-DRAM Trackers}. In \bibinfo{booktitle}{\emph{2024 ACM/IEEE 51st Annual International Symposium on Computer Architecture (ISCA)}}. \bibinfo{pages}{1157--1172}.
\newblock
\urldef\tempurl%
\url{https://doi.org/10.1109/ISCA59077.2024.00087}
\showDOI{\tempurl}


\bibitem[Jaleel et~al\mbox{.}(2010)]%
        {srrip}
\bibfield{author}{\bibinfo{person}{Aamer Jaleel}, \bibinfo{person}{Kevin~B. Theobald}, \bibinfo{person}{Simon~C. Steely}, {and} \bibinfo{person}{Joel Emer}.} \bibinfo{year}{2010}\natexlab{}.
\newblock \showarticletitle{High performance cache replacement using re-reference interval prediction (RRIP)}. In \bibinfo{booktitle}{\emph{Proceedings of the 37th Annual International Symposium on Computer Architecture}} (Saint-Malo, France) \emph{(\bibinfo{series}{ISCA '10})}. \bibinfo{publisher}{Association for Computing Machinery}, \bibinfo{address}{New York, NY, USA}, \bibinfo{pages}{60–71}.
\newblock
\showISBNx{9781450300537}
\urldef\tempurl%
\url{https://doi.org/10.1145/1815961.1815971}
\showDOI{\tempurl}


\bibitem[Jang et~al\mbox{.}(2017)]%
        {sgx-bomb}
\bibfield{author}{\bibinfo{person}{Yeongjin Jang}, \bibinfo{person}{Jaehyuk Lee}, \bibinfo{person}{Sangho Lee}, {and} \bibinfo{person}{Taesoo Kim}.} \bibinfo{year}{2017}\natexlab{}.
\newblock \showarticletitle{SGX-Bomb: Locking Down the Processor via Rowhammer Attack}. In \bibinfo{booktitle}{\emph{Proceedings of the 2nd Workshop on System Software for Trusted Execution}} (Shanghai, China) \emph{(\bibinfo{series}{SysTEX'17})}. \bibinfo{publisher}{Association for Computing Machinery}, \bibinfo{address}{New York, NY, USA}, Article \bibinfo{articleno}{5}, \bibinfo{numpages}{6}~pages.
\newblock
\showISBNx{9781450350976}
\urldef\tempurl%
\url{https://doi.org/10.1145/3152701.3152709}
\showDOI{\tempurl}


\bibitem[Jattke et~al\mbox{.}(2022)]%
        {jattke2021blacksmith}
\bibfield{author}{\bibinfo{person}{Patrick Jattke}, \bibinfo{person}{Victor Van Der~Veen}, \bibinfo{person}{Pietro Frigo}, \bibinfo{person}{Stijn Gunter}, {and} \bibinfo{person}{Kaveh Razavi}.} \bibinfo{year}{2022}\natexlab{}.
\newblock \showarticletitle{BLACKSMITH: Scalable Rowhammering in the Frequency Domain}. In \bibinfo{booktitle}{\emph{2022 IEEE Symposium on Security and Privacy (SP)}}. \bibinfo{pages}{716--734}.
\newblock
\urldef\tempurl%
\url{https://doi.org/10.1109/SP46214.2022.9833772}
\showDOI{\tempurl}


\bibitem[Jattke et~al\mbox{.}(2024)]%
        {jattke2024zenhammer}
\bibfield{author}{\bibinfo{person}{Patrick Jattke}, \bibinfo{person}{Max Wipfli}, \bibinfo{person}{Flavien Solt}, \bibinfo{person}{Michele Marazzi}, \bibinfo{person}{Matej B{\"o}lcskei}, {and} \bibinfo{person}{Kaveh Razavi}.} \bibinfo{year}{2024}\natexlab{}.
\newblock \showarticletitle{{ZenHammer}: Rowhammer Attacks on {AMD} Zen-based Platforms}. In \bibinfo{booktitle}{\emph{33rd USENIX Security Symposium (USENIX Security 24)}}. \bibinfo{publisher}{USENIX Association}, \bibinfo{address}{Philadelphia, PA}, \bibinfo{pages}{1615--1633}.
\newblock
\showISBNx{978-1-939133-44-1}
\urldef\tempurl%
\url{https://www.usenix.org/conference/usenixsecurity24/presentation/jattke}
\showURL{%
\tempurl}


\bibitem[JEDEC(2017)]%
        {JEDEC-DDR4}
\bibfield{author}{\bibinfo{person}{JEDEC}.} \bibinfo{year}{2017}\natexlab{}.
\newblock \showarticletitle{DDR4 SDRAM standard (JESD79-4B)}.
\newblock  (\bibinfo{year}{2017}).
\newblock


\bibitem[{JEDEC}(2024)]%
        {jedec_ddr5_prac}
\bibfield{author}{\bibinfo{person}{{JEDEC}}.} \bibinfo{year}{2024}\natexlab{}.
\newblock \bibinfo{booktitle}{\emph{{JESD79-5C}}}.
\newblock
\newblock
\shownote{\url{https://www.jedec.org/document_search?search_api_views_fulltext=jesd79-5c}}.


\bibitem[Jimenez and Lin(2001)]%
        {hashed_perceptron}
\bibfield{author}{\bibinfo{person}{D.A. Jimenez} {and} \bibinfo{person}{C. Lin}.} \bibinfo{year}{2001}\natexlab{}.
\newblock \showarticletitle{Dynamic branch prediction with perceptrons}. In \bibinfo{booktitle}{\emph{Proceedings HPCA Seventh International Symposium on High-Performance Computer Architecture}}. \bibinfo{pages}{197--206}.
\newblock
\urldef\tempurl%
\url{https://doi.org/10.1109/HPCA.2001.903263}
\showDOI{\tempurl}


\bibitem[Kaseridis et~al\mbox{.}(2011)]%
        {MOP}
\bibfield{author}{\bibinfo{person}{Dimitris Kaseridis}, \bibinfo{person}{Jeffrey Stuecheli}, {and} \bibinfo{person}{Lizy~Kurian John}.} \bibinfo{year}{2011}\natexlab{}.
\newblock \showarticletitle{Minimalist open-page: a DRAM page-mode scheduling policy for the many-core era}. In \bibinfo{booktitle}{\emph{Proceedings of the 44th Annual IEEE/ACM International Symposium on Microarchitecture}} (Porto Alegre, Brazil) \emph{(\bibinfo{series}{MICRO-44})}. \bibinfo{publisher}{Association for Computing Machinery}, \bibinfo{address}{New York, NY, USA}, \bibinfo{pages}{24–35}.
\newblock
\showISBNx{9781450310536}
\urldef\tempurl%
\url{https://doi.org/10.1145/2155620.2155624}
\showDOI{\tempurl}


\bibitem[Kim et~al\mbox{.}(2015)]%
        {kim2014architectural}
\bibfield{author}{\bibinfo{person}{Dae-Hyun Kim}, \bibinfo{person}{Prashant~J. Nair}, {and} \bibinfo{person}{Moinuddin~K. Qureshi}.} \bibinfo{year}{2015}\natexlab{}.
\newblock \showarticletitle{Architectural Support for Mitigating Row Hammering in DRAM Memories}.
\newblock \bibinfo{journal}{\emph{IEEE Computer Architecture Letters}} \bibinfo{volume}{14}, \bibinfo{number}{1} (\bibinfo{year}{2015}), \bibinfo{pages}{9--12}.
\newblock
\urldef\tempurl%
\url{https://doi.org/10.1109/LCA.2014.2332177}
\showDOI{\tempurl}


\bibitem[Kim et~al\mbox{.}(2020)]%
        {kim2020revisitingRH}
\bibfield{author}{\bibinfo{person}{Jeremie~S. Kim}, \bibinfo{person}{Minesh Patel}, \bibinfo{person}{A.~Giray Ya\u{g}l\i{}k\c{c}\i{}}, \bibinfo{person}{Hasan Hassan}, \bibinfo{person}{Roknoddin Azizi}, \bibinfo{person}{Lois Orosa}, {and} \bibinfo{person}{Onur Mutlu}.} \bibinfo{year}{2020}\natexlab{}.
\newblock \showarticletitle{Revisiting RowHammer: an experimental analysis of modern DRAM devices and mitigation techniques}. In \bibinfo{booktitle}{\emph{Proceedings of the ACM/IEEE 47th Annual International Symposium on Computer Architecture}} (Virtual Event) \emph{(\bibinfo{series}{ISCA '20})}. \bibinfo{publisher}{IEEE Press}, \bibinfo{pages}{638–651}.
\newblock
\showISBNx{9781728146614}
\urldef\tempurl%
\url{https://doi.org/10.1109/ISCA45697.2020.00059}
\showDOI{\tempurl}


\bibitem[Kim et~al\mbox{.}(2021)]%
        {hammerfilter}
\bibfield{author}{\bibinfo{person}{Kwangrae Kim}, \bibinfo{person}{Jeonghyun Woo}, \bibinfo{person}{Junsu Kim}, {and} \bibinfo{person}{Ki-Seok Chung}.} \bibinfo{year}{2021}\natexlab{}.
\newblock \showarticletitle{HammerFilter: Robust Protection and Low Hardware Overhead Method for RowHammer}. In \bibinfo{booktitle}{\emph{2021 IEEE 39th International Conference on Computer Design (ICCD)}}. \bibinfo{pages}{212--219}.
\newblock
\urldef\tempurl%
\url{https://doi.org/10.1109/ICCD53106.2021.00043}
\showDOI{\tempurl}


\bibitem[Kim et~al\mbox{.}(2022)]%
        {kim2022mithril}
\bibfield{author}{\bibinfo{person}{Michael~Jaemin Kim}, \bibinfo{person}{Jaehyun Park}, \bibinfo{person}{Yeonhong Park}, \bibinfo{person}{Wanju Doh}, \bibinfo{person}{Namhoon Kim}, \bibinfo{person}{Tae~Jun Ham}, \bibinfo{person}{Jae~W. Lee}, {and} \bibinfo{person}{Jung~Ho Ahn}.} \bibinfo{year}{2022}\natexlab{}.
\newblock \showarticletitle{Mithril: Cooperative Row Hammer Protection on Commodity DRAM Leveraging Managed Refresh}. In \bibinfo{booktitle}{\emph{2022 IEEE International Symposium on High-Performance Computer Architecture (HPCA)}}. \bibinfo{pages}{1156--1169}.
\newblock
\urldef\tempurl%
\url{https://doi.org/10.1109/HPCA53966.2022.00088}
\showDOI{\tempurl}


\bibitem[Kim et~al\mbox{.}(2023)]%
        {HynixPAT}
\bibfield{author}{\bibinfo{person}{Woongrae Kim}, \bibinfo{person}{Chulmoon Jung}, \bibinfo{person}{Seongnyuh Yoo}, \bibinfo{person}{Duckhwa Hong}, \bibinfo{person}{Jeongjin Hwang}, \bibinfo{person}{Jungmin Yoon}, \bibinfo{person}{Ohyong Jung}, \bibinfo{person}{Joonwoo Choi}, \bibinfo{person}{Sanga Hyun}, \bibinfo{person}{Mankeun Kang}, \bibinfo{person}{Sangho Lee}, \bibinfo{person}{Dohong Kim}, \bibinfo{person}{Sanghyun Ku}, \bibinfo{person}{Donhyun Choi}, \bibinfo{person}{Nogeun Joo}, \bibinfo{person}{Sangwoo Yoon}, \bibinfo{person}{Junseok Noh}, \bibinfo{person}{Byeongyong Go}, \bibinfo{person}{Cheolhoe Kim}, \bibinfo{person}{Sunil Hwang}, \bibinfo{person}{Mihyun Hwang}, \bibinfo{person}{Seol-Min Yi}, \bibinfo{person}{Hyungmin Kim}, \bibinfo{person}{Sanghyuk Heo}, \bibinfo{person}{Yeonsu Jang}, \bibinfo{person}{Kyoungchul Jang}, \bibinfo{person}{Shinho Chu}, \bibinfo{person}{Yoonna Oh}, \bibinfo{person}{Kwidong Kim}, \bibinfo{person}{Junghyun Kim}, \bibinfo{person}{Soohwan Kim}, \bibinfo{person}{Jeongtae
  Hwang}, \bibinfo{person}{Sangil Park}, \bibinfo{person}{Junphyo Lee}, \bibinfo{person}{Inchul Jeong}, \bibinfo{person}{Joohwan Cho}, {and} \bibinfo{person}{Jonghwan Kim}.} \bibinfo{year}{2023}\natexlab{}.
\newblock \showarticletitle{A 1.1V 16Gb DDR5 DRAM with Probabilistic-Aggressor Tracking, Refresh-Management Functionality, Per-Row Hammer Tracking, a Multi-Step Precharge, and Core-Bias Modulation for Security and Reliability Enhancement}. In \bibinfo{booktitle}{\emph{2023 IEEE International Solid-State Circuits Conference (ISSCC)}}. \bibinfo{pages}{1--3}.
\newblock
\urldef\tempurl%
\url{https://doi.org/10.1109/ISSCC42615.2023.10067805}
\showDOI{\tempurl}


\bibitem[Kim et~al\mbox{.}(2014)]%
        {kim2014flipping}
\bibfield{author}{\bibinfo{person}{Yoongu Kim}, \bibinfo{person}{Ross Daly}, \bibinfo{person}{Jeremie Kim}, \bibinfo{person}{Chris Fallin}, \bibinfo{person}{Ji~Hye Lee}, \bibinfo{person}{Donghyuk Lee}, \bibinfo{person}{Chris Wilkerson}, \bibinfo{person}{Konrad Lai}, {and} \bibinfo{person}{Onur Mutlu}.} \bibinfo{year}{2014}\natexlab{}.
\newblock \showarticletitle{Flipping bits in memory without accessing them: an experimental study of DRAM disturbance errors}.
\newblock \bibinfo{journal}{\emph{SIGARCH Comput. Archit. News}} \bibinfo{volume}{42}, \bibinfo{number}{3} (\bibinfo{date}{June} \bibinfo{year}{2014}), \bibinfo{pages}{361–372}.
\newblock
\showISSN{0163-5964}
\urldef\tempurl%
\url{https://doi.org/10.1145/2678373.2665726}
\showDOI{\tempurl}


\bibitem[Kim et~al\mbox{.}(2016)]%
        {kim2015ramulator}
\bibfield{author}{\bibinfo{person}{Yoongu Kim}, \bibinfo{person}{Weikun Yang}, {and} \bibinfo{person}{Onur Mutlu}.} \bibinfo{year}{2016}\natexlab{}.
\newblock \showarticletitle{Ramulator: A Fast and Extensible DRAM Simulator}.
\newblock \bibinfo{journal}{\emph{IEEE Computer Architecture Letters}} \bibinfo{volume}{15}, \bibinfo{number}{1} (\bibinfo{year}{2016}), \bibinfo{pages}{45--49}.
\newblock
\urldef\tempurl%
\url{https://doi.org/10.1109/LCA.2015.2414456}
\showDOI{\tempurl}


\bibitem[Kocher et~al\mbox{.}(2019)]%
        {Spectre}
\bibfield{author}{\bibinfo{person}{Paul Kocher}, \bibinfo{person}{Jann Horn}, \bibinfo{person}{Anders Fogh}, \bibinfo{person}{Daniel Genkin}, \bibinfo{person}{Daniel Gruss}, \bibinfo{person}{Werner Haas}, \bibinfo{person}{Mike Hamburg}, \bibinfo{person}{Moritz Lipp}, \bibinfo{person}{Stefan Mangard}, \bibinfo{person}{Thomas Prescher}, \bibinfo{person}{Michael Schwarz}, {and} \bibinfo{person}{Yuval Yarom}.} \bibinfo{year}{2019}\natexlab{}.
\newblock \showarticletitle{Spectre Attacks: Exploiting Speculative Execution}. In \bibinfo{booktitle}{\emph{2019 IEEE Symposium on Security and Privacy (SP)}}. \bibinfo{pages}{1--19}.
\newblock
\urldef\tempurl%
\url{https://doi.org/10.1109/SP.2019.00002}
\showDOI{\tempurl}


\bibitem[Kogler et~al\mbox{.}(2022)]%
        {HalfDouble}
\bibfield{author}{\bibinfo{person}{Andreas Kogler}, \bibinfo{person}{Jonas Juffinger}, \bibinfo{person}{Salman Qazi}, \bibinfo{person}{Yoongu Kim}, \bibinfo{person}{Moritz Lipp}, \bibinfo{person}{Nicolas Boichat}, \bibinfo{person}{Eric Shiu}, \bibinfo{person}{Mattias Nissler}, {and} \bibinfo{person}{Daniel Gruss}.} \bibinfo{year}{2022}\natexlab{}.
\newblock \showarticletitle{{Half-Double}: Hammering From the Next Row Over}. In \bibinfo{booktitle}{\emph{31st USENIX Security Symposium (USENIX Security 22)}}. \bibinfo{publisher}{USENIX Association}, \bibinfo{address}{Boston, MA}, \bibinfo{pages}{3807--3824}.
\newblock
\showISBNx{978-1-939133-31-1}
\urldef\tempurl%
\url{https://www.usenix.org/conference/usenixsecurity22/presentation/kogler-half-double}
\showURL{%
\tempurl}


\bibitem[Kwong et~al\mbox{.}(2020)]%
        {kwong2020rambleed}
\bibfield{author}{\bibinfo{person}{Andrew Kwong}, \bibinfo{person}{Daniel Genkin}, \bibinfo{person}{Daniel Gruss}, {and} \bibinfo{person}{Yuval Yarom}.} \bibinfo{year}{2020}\natexlab{}.
\newblock \showarticletitle{RAMBleed: Reading Bits in Memory Without Accessing Them}. In \bibinfo{booktitle}{\emph{2020 IEEE Symposium on Security and Privacy (SP)}}. \bibinfo{pages}{695--711}.
\newblock
\urldef\tempurl%
\url{https://doi.org/10.1109/SP40000.2020.00020}
\showDOI{\tempurl}


\bibitem[Lee et~al\mbox{.}(2019)]%
        {lee2019twice}
\bibfield{author}{\bibinfo{person}{Eojin Lee}, \bibinfo{person}{Ingab Kang}, \bibinfo{person}{Sukhan Lee}, \bibinfo{person}{G.~Edward Suh}, {and} \bibinfo{person}{Jung~Ho Ahn}.} \bibinfo{year}{2019}\natexlab{}.
\newblock \showarticletitle{TWiCe: preventing row-hammering by exploiting time window counters}. In \bibinfo{booktitle}{\emph{Proceedings of the 46th International Symposium on Computer Architecture}} (Phoenix, Arizona) \emph{(\bibinfo{series}{ISCA '19})}. \bibinfo{publisher}{Association for Computing Machinery}, \bibinfo{address}{New York, NY, USA}, \bibinfo{pages}{385–396}.
\newblock
\showISBNx{9781450366694}
\urldef\tempurl%
\url{https://doi.org/10.1145/3307650.3322232}
\showDOI{\tempurl}


\bibitem[Lipp et~al\mbox{.}(2020)]%
        {takeaway}
\bibfield{author}{\bibinfo{person}{Moritz Lipp}, \bibinfo{person}{Vedad Had\v{z}i\'{c}}, \bibinfo{person}{Michael Schwarz}, \bibinfo{person}{Arthur Perais}, \bibinfo{person}{Cl\'{e}mentine Maurice}, {and} \bibinfo{person}{Daniel Gruss}.} \bibinfo{year}{2020}\natexlab{}.
\newblock \showarticletitle{Take A Way: Exploring the Security Implications of AMD's Cache Way Predictors}. In \bibinfo{booktitle}{\emph{Proceedings of the 15th ACM Asia Conference on Computer and Communications Security}} (Taipei, Taiwan) \emph{(\bibinfo{series}{ASIA CCS '20})}. \bibinfo{publisher}{Association for Computing Machinery}, \bibinfo{address}{New York, NY, USA}, \bibinfo{pages}{813–825}.
\newblock
\showISBNx{9781450367509}
\urldef\tempurl%
\url{https://doi.org/10.1145/3320269.3384746}
\showDOI{\tempurl}


\bibitem[Lipp et~al\mbox{.}(2018)]%
        {Meltdown}
\bibfield{author}{\bibinfo{person}{Moritz Lipp}, \bibinfo{person}{Michael Schwarz}, \bibinfo{person}{Daniel Gruss}, \bibinfo{person}{Thomas Prescher}, \bibinfo{person}{Werner Haas}, \bibinfo{person}{Anders Fogh}, \bibinfo{person}{Jann Horn}, \bibinfo{person}{Stefan Mangard}, \bibinfo{person}{Paul Kocher}, \bibinfo{person}{Daniel Genkin}, \bibinfo{person}{Yuval Yarom}, {and} \bibinfo{person}{Mike Hamburg}.} \bibinfo{year}{2018}\natexlab{}.
\newblock \showarticletitle{Meltdown: Reading Kernel Memory from User Space}. In \bibinfo{booktitle}{\emph{27th USENIX Security Symposium (USENIX Security 18)}}. \bibinfo{publisher}{USENIX Association}, \bibinfo{address}{Baltimore, MD}, \bibinfo{pages}{973--990}.
\newblock
\showISBNx{978-1-939133-04-5}
\urldef\tempurl%
\url{https://www.usenix.org/conference/usenixsecurity18/presentation/lipp}
\showURL{%
\tempurl}


\bibitem[Liu et~al\mbox{.}(2015)]%
        {LLCPrimeProbe}
\bibfield{author}{\bibinfo{person}{Fangfei Liu}, \bibinfo{person}{Yuval Yarom}, \bibinfo{person}{Qian Ge}, \bibinfo{person}{Gernot Heiser}, {and} \bibinfo{person}{Ruby~B. Lee}.} \bibinfo{year}{2015}\natexlab{}.
\newblock \showarticletitle{Last-Level Cache Side-Channel Attacks are Practical}. In \bibinfo{booktitle}{\emph{2015 IEEE Symposium on Security and Privacy}}. \bibinfo{pages}{605--622}.
\newblock
\urldef\tempurl%
\url{https://doi.org/10.1109/SP.2015.43}
\showDOI{\tempurl}


\bibitem[Loughlin et~al\mbox{.}(2023)]%
        {Siloz}
\bibfield{author}{\bibinfo{person}{Kevin Loughlin}, \bibinfo{person}{Jonah Rosenblum}, \bibinfo{person}{Stefan Saroiu}, \bibinfo{person}{Alec Wolman}, \bibinfo{person}{Dimitrios Skarlatos}, {and} \bibinfo{person}{Baris Kasikci}.} \bibinfo{year}{2023}\natexlab{}.
\newblock \showarticletitle{Siloz: Leveraging DRAM Isolation Domains to Prevent Inter-VM Rowhammer}. In \bibinfo{booktitle}{\emph{Proceedings of the 29th Symposium on Operating Systems Principles}} (Koblenz, Germany) \emph{(\bibinfo{series}{SOSP '23})}. \bibinfo{publisher}{Association for Computing Machinery}, \bibinfo{address}{New York, NY, USA}, \bibinfo{pages}{417–433}.
\newblock
\showISBNx{9798400702297}
\urldef\tempurl%
\url{https://doi.org/10.1145/3600006.3613143}
\showDOI{\tempurl}


\bibitem[Loughlin et~al\mbox{.}(2022)]%
        {loughlin2022moesi}
\bibfield{author}{\bibinfo{person}{Kevin Loughlin}, \bibinfo{person}{Stefan Saroiu}, \bibinfo{person}{Alec Wolman}, \bibinfo{person}{Yatin~A. Manerkar}, {and} \bibinfo{person}{Baris Kasikci}.} \bibinfo{year}{2022}\natexlab{}.
\newblock \showarticletitle{MOESI-prime: preventing coherence-induced hammering in commodity workloads}. In \bibinfo{booktitle}{\emph{Proceedings of the 49th Annual International Symposium on Computer Architecture}} (New York, New York) \emph{(\bibinfo{series}{ISCA '22})}. \bibinfo{publisher}{Association for Computing Machinery}, \bibinfo{address}{New York, NY, USA}, \bibinfo{pages}{670–684}.
\newblock
\showISBNx{9781450386104}
\urldef\tempurl%
\url{https://doi.org/10.1145/3470496.3527427}
\showDOI{\tempurl}


\bibitem[Luo et~al\mbox{.}(2024)]%
        {ramulator2}
\bibfield{author}{\bibinfo{person}{Haocong Luo}, \bibinfo{person}{Yahya~Can Tu\u{g}rul}, \bibinfo{person}{F.~Nisa Bostanc\i{}}, \bibinfo{person}{Ataberk Olgun}, \bibinfo{person}{A.~Giray Ya\u{g}l\i{}k\c{c}\i{}}, {and} \bibinfo{person}{Onur Mutlu}.} \bibinfo{year}{2024}\natexlab{}.
\newblock \showarticletitle{Ramulator 2.0: A Modern, Modular, and Extensible DRAM Simulator}.
\newblock \bibinfo{journal}{\emph{IEEE Comput. Archit. Lett.}} \bibinfo{volume}{23}, \bibinfo{number}{1} (\bibinfo{date}{Jan.} \bibinfo{year}{2024}), \bibinfo{pages}{112–116}.
\newblock
\showISSN{1556-6056}
\urldef\tempurl%
\url{https://doi.org/10.1109/LCA.2023.3333759}
\showDOI{\tempurl}


\bibitem[Marazzi et~al\mbox{.}(2022)]%
        {ProTRR}
\bibfield{author}{\bibinfo{person}{Michele Marazzi}, \bibinfo{person}{Patrick Jattke}, \bibinfo{person}{Flavien Solt}, {and} \bibinfo{person}{Kaveh Razavi}.} \bibinfo{year}{2022}\natexlab{}.
\newblock \showarticletitle{ProTRR: Principled yet Optimal In-DRAM Target Row Refresh}. In \bibinfo{booktitle}{\emph{2022 IEEE Symposium on Security and Privacy (SP)}}. \bibinfo{pages}{735--753}.
\newblock
\urldef\tempurl%
\url{https://doi.org/10.1109/SP46214.2022.9833664}
\showDOI{\tempurl}


\bibitem[Marazzi et~al\mbox{.}(2023)]%
        {REGA_SP23}
\bibfield{author}{\bibinfo{person}{Michele Marazzi}, \bibinfo{person}{Flavien Solt}, \bibinfo{person}{Patrick Jattke}, \bibinfo{person}{Kubo Takashi}, {and} \bibinfo{person}{Kaveh Razavi}.} \bibinfo{year}{2023}\natexlab{}.
\newblock \showarticletitle{REGA: Scalable Rowhammer Mitigation with Refresh-Generating Activations}. In \bibinfo{booktitle}{\emph{2023 IEEE Symposium on Security and Privacy (SP)}}. \bibinfo{pages}{1684--1701}.
\newblock
\urldef\tempurl%
\url{https://doi.org/10.1109/SP46215.2023.10179327}
\showDOI{\tempurl}


\bibitem[Mutlu and Moscibroda(2007)]%
        {FRFCFS_CAP}
\bibfield{author}{\bibinfo{person}{Onur Mutlu} {and} \bibinfo{person}{Thomas Moscibroda}.} \bibinfo{year}{2007}\natexlab{}.
\newblock \showarticletitle{Stall-Time Fair Memory Access Scheduling for Chip Multiprocessors}. In \bibinfo{booktitle}{\emph{40th Annual IEEE/ACM International Symposium on Microarchitecture (MICRO 2007)}}. \bibinfo{pages}{146--160}.
\newblock
\urldef\tempurl%
\url{https://doi.org/10.1109/MICRO.2007.21}
\showDOI{\tempurl}


\bibitem[Nair et~al\mbox{.}(2013a)]%
        {refpause}
\bibfield{author}{\bibinfo{person}{Prashant Nair}, \bibinfo{person}{Chia-Chen Chou}, {and} \bibinfo{person}{Moinuddin~K. Qureshi}.} \bibinfo{year}{2013}\natexlab{a}.
\newblock \showarticletitle{A case for Refresh Pausing in DRAM memory systems}. In \bibinfo{booktitle}{\emph{2013 IEEE 19th International Symposium on High Performance Computer Architecture (HPCA)}}. \bibinfo{pages}{627--638}.
\newblock
\urldef\tempurl%
\url{https://doi.org/10.1109/HPCA.2013.6522355}
\showDOI{\tempurl}


\bibitem[Nair et~al\mbox{.}(2019)]%
        {sudoku}
\bibfield{author}{\bibinfo{person}{Prashant~J. Nair}, \bibinfo{person}{Bahar Asgari}, {and} \bibinfo{person}{Moinuddin~K. Qureshi}.} \bibinfo{year}{2019}\natexlab{}.
\newblock \showarticletitle{SuDoku: Tolerating High-Rate of Transient Failures for Enabling Scalable STTRAM}. In \bibinfo{booktitle}{\emph{2019 49th Annual IEEE/IFIP International Conference on Dependable Systems and Networks (DSN)}}. \bibinfo{pages}{388--400}.
\newblock
\urldef\tempurl%
\url{https://doi.org/10.1109/DSN.2019.00048}
\showDOI{\tempurl}


\bibitem[Nair et~al\mbox{.}(2013b)]%
        {archshield}
\bibfield{author}{\bibinfo{person}{Prashant~J. Nair}, \bibinfo{person}{Dae-Hyun Kim}, {and} \bibinfo{person}{Moinuddin~K. Qureshi}.} \bibinfo{year}{2013}\natexlab{b}.
\newblock \showarticletitle{ArchShield: architectural framework for assisting DRAM scaling by tolerating high error rates}. In \bibinfo{booktitle}{\emph{Proceedings of the 40th Annual International Symposium on Computer Architecture}} (Tel-Aviv, Israel) \emph{(\bibinfo{series}{ISCA '13})}. \bibinfo{publisher}{Association for Computing Machinery}, \bibinfo{address}{New York, NY, USA}, \bibinfo{pages}{72–83}.
\newblock
\showISBNx{9781450320795}
\urldef\tempurl%
\url{https://doi.org/10.1145/2485922.2485929}
\showDOI{\tempurl}


\bibitem[Nair et~al\mbox{.}(2014)]%
        {citadel2}
\bibfield{author}{\bibinfo{person}{Prashant~J. Nair}, \bibinfo{person}{David~A. Roberts}, {and} \bibinfo{person}{Moinuddin~K. Qureshi}.} \bibinfo{year}{2014}\natexlab{}.
\newblock \showarticletitle{Citadel: Efficiently Protecting Stacked Memory from Large Granularity Failures}. In \bibinfo{booktitle}{\emph{2014 47th Annual IEEE/ACM International Symposium on Microarchitecture}}. \bibinfo{pages}{51--62}.
\newblock
\urldef\tempurl%
\url{https://doi.org/10.1109/MICRO.2014.57}
\showDOI{\tempurl}


\bibitem[Nair et~al\mbox{.}(2015)]%
        {faultsim}
\bibfield{author}{\bibinfo{person}{Prashant~J. Nair}, \bibinfo{person}{David~A. Roberts}, {and} \bibinfo{person}{Moinuddin~K. Qureshi}.} \bibinfo{year}{2015}\natexlab{}.
\newblock \showarticletitle{FaultSim: A Fast, Configurable Memory-Reliability Simulator for Conventional and 3D-Stacked Systems}.
\newblock \bibinfo{journal}{\emph{ACM Trans. Archit. Code Optim.}} \bibinfo{volume}{12}, \bibinfo{number}{4}, Article \bibinfo{articleno}{44} (\bibinfo{date}{Dec.} \bibinfo{year}{2015}), \bibinfo{numpages}{24}~pages.
\newblock
\showISSN{1544-3566}
\urldef\tempurl%
\url{https://doi.org/10.1145/2831234}
\showDOI{\tempurl}


\bibitem[Nair et~al\mbox{.}(2016a)]%
        {citadel1}
\bibfield{author}{\bibinfo{person}{Prashant~J. Nair}, \bibinfo{person}{David~A. Roberts}, {and} \bibinfo{person}{Moinuddin~K. Qureshi}.} \bibinfo{year}{2016}\natexlab{a}.
\newblock \showarticletitle{Citadel: Efficiently Protecting Stacked Memory from TSV and Large Granularity Failures}.
\newblock \bibinfo{journal}{\emph{ACM Trans. Archit. Code Optim.}} \bibinfo{volume}{12}, \bibinfo{number}{4}, Article \bibinfo{articleno}{49} (\bibinfo{date}{Jan.} \bibinfo{year}{2016}), \bibinfo{numpages}{24}~pages.
\newblock
\showISSN{1544-3566}
\urldef\tempurl%
\url{https://doi.org/10.1145/2840807}
\showDOI{\tempurl}


\bibitem[Nair et~al\mbox{.}(2016b)]%
        {xed}
\bibfield{author}{\bibinfo{person}{Prashant~J. Nair}, \bibinfo{person}{Vilas Sridharan}, {and} \bibinfo{person}{Moinuddin~K. Qureshi}.} \bibinfo{year}{2016}\natexlab{b}.
\newblock \showarticletitle{XED: exposing on-die error detection information for strong memory reliability}. In \bibinfo{booktitle}{\emph{Proceedings of the 43rd International Symposium on Computer Architecture}} (Seoul, Republic of Korea) \emph{(\bibinfo{series}{ISCA '16})}. \bibinfo{publisher}{IEEE Press}, \bibinfo{pages}{341–353}.
\newblock
\showISBNx{9781467389471}
\urldef\tempurl%
\url{https://doi.org/10.1109/ISCA.2016.38}
\showDOI{\tempurl}


\bibitem[Navarro-Torres et~al\mbox{.}(2022)]%
        {berti_2022}
\bibfield{author}{\bibinfo{person}{Agustín Navarro-Torres}, \bibinfo{person}{Biswabandan Panda}, \bibinfo{person}{Jesús Alastruey-Benedé}, \bibinfo{person}{Pablo Ibáñez}, \bibinfo{person}{Víctor Viñals-Yúfera}, {and} \bibinfo{person}{Alberto Ros}.} \bibinfo{year}{2022}\natexlab{}.
\newblock \showarticletitle{Berti: an Accurate Local-Delta Data Prefetcher}. In \bibinfo{booktitle}{\emph{2022 55th IEEE/ACM International Symposium on Microarchitecture (MICRO)}}. \bibinfo{pages}{975--991}.
\newblock
\urldef\tempurl%
\url{https://doi.org/10.1109/MICRO56248.2022.00072}
\showDOI{\tempurl}


\bibitem[Olgun et~al\mbox{.}(2024a)]%
        {olgun2024HBM2study}
\bibfield{author}{\bibinfo{person}{Ataberk Olgun}, \bibinfo{person}{Majd Osseiran}, \bibinfo{person}{A.~Giray Yağlıkçı}, \bibinfo{person}{Yahya~Can Tuğrul}, \bibinfo{person}{Haocong Luo}, \bibinfo{person}{Steve Rhyner}, \bibinfo{person}{Behzad Salami}, \bibinfo{person}{Juan~Gomez Luna}, {and} \bibinfo{person}{Onur Mutlu}.} \bibinfo{year}{2024}\natexlab{a}.
\newblock \showarticletitle{Read Disturbance in High Bandwidth Memory: A Detailed Experimental Study on HBM2 DRAM Chips}. In \bibinfo{booktitle}{\emph{2024 54th Annual IEEE/IFIP International Conference on Dependable Systems and Networks (DSN)}}. \bibinfo{pages}{75--89}.
\newblock
\urldef\tempurl%
\url{https://doi.org/10.1109/DSN58291.2024.00022}
\showDOI{\tempurl}


\bibitem[Olgun et~al\mbox{.}(2024b)]%
        {olgun2023abacus}
\bibfield{author}{\bibinfo{person}{Ataberk Olgun}, \bibinfo{person}{Yahya~Can Tugrul}, \bibinfo{person}{Nisa Bostanci}, \bibinfo{person}{Ismail~Emir Yuksel}, \bibinfo{person}{Haocong Luo}, \bibinfo{person}{Steve Rhyner}, \bibinfo{person}{Abdullah~Giray Yaglikci}, \bibinfo{person}{Geraldo~F. Oliveira}, {and} \bibinfo{person}{Onur Mutlu}.} \bibinfo{year}{2024}\natexlab{b}.
\newblock \showarticletitle{{ABACuS}: {All-Bank} Activation Counters for Scalable and Low Overhead {RowHammer} Mitigation}. In \bibinfo{booktitle}{\emph{33rd USENIX Security Symposium (USENIX Security 24)}}. \bibinfo{publisher}{USENIX Association}, \bibinfo{address}{Philadelphia, PA}, \bibinfo{pages}{1579--1596}.
\newblock
\showISBNx{978-1-939133-44-1}
\urldef\tempurl%
\url{https://www.usenix.org/conference/usenixsecurity24/presentation/olgun}
\showURL{%
\tempurl}


\bibitem[Park et~al\mbox{.}(2020)]%
        {park2020graphene}
\bibfield{author}{\bibinfo{person}{Yeonhong Park}, \bibinfo{person}{Woosuk Kwon}, \bibinfo{person}{Eojin Lee}, \bibinfo{person}{Tae~Jun Ham}, \bibinfo{person}{Jung Ho~Ahn}, {and} \bibinfo{person}{Jae~W. Lee}.} \bibinfo{year}{2020}\natexlab{}.
\newblock \showarticletitle{Graphene: Strong yet Lightweight Row Hammer Protection}. In \bibinfo{booktitle}{\emph{2020 53rd Annual IEEE/ACM International Symposium on Microarchitecture (MICRO)}}. \bibinfo{pages}{1--13}.
\newblock
\urldef\tempurl%
\url{https://doi.org/10.1109/MICRO50266.2020.00014}
\showDOI{\tempurl}


\bibitem[Pessl et~al\mbox{.}(2016)]%
        {DRAMA2016}
\bibfield{author}{\bibinfo{person}{Peter Pessl}, \bibinfo{person}{Daniel Gruss}, \bibinfo{person}{Cl{\'e}mentine Maurice}, \bibinfo{person}{Michael Schwarz}, {and} \bibinfo{person}{Stefan Mangard}.} \bibinfo{year}{2016}\natexlab{}.
\newblock \showarticletitle{{DRAMA}: Exploiting {DRAM} Addressing for {Cross-CPU} Attacks}. In \bibinfo{booktitle}{\emph{25th USENIX Security Symposium (USENIX Security 16)}}. \bibinfo{publisher}{USENIX Association}, \bibinfo{address}{Austin, TX}, \bibinfo{pages}{565--581}.
\newblock
\showISBNx{978-1-931971-32-4}
\urldef\tempurl%
\url{https://www.usenix.org/conference/usenixsecurity16/technical-sessions/presentation/pessl}
\showURL{%
\tempurl}


\bibitem[Qureshi(2025)]%
        {autorfm_hpca25}
\bibfield{author}{\bibinfo{person}{Moinuddin Qureshi}.} \bibinfo{year}{2025}\natexlab{}.
\newblock \showarticletitle{AutoRFM: Scaling Low-Cost in-DRAM Trackers to Ultra-Low Rowhammer Thresholds}. In \bibinfo{booktitle}{\emph{2025 IEEE International Symposium on High Performance Computer Architecture (HPCA)}}. \bibinfo{pages}{991--1004}.
\newblock
\urldef\tempurl%
\url{https://doi.org/10.1109/HPCA61900.2025.00078}
\showDOI{\tempurl}


\bibitem[Qureshi and Qazi(2025)]%
        {qureshi2024moat}
\bibfield{author}{\bibinfo{person}{Moinuddin Qureshi} {and} \bibinfo{person}{Salman Qazi}.} \bibinfo{year}{2025}\natexlab{}.
\newblock \showarticletitle{MOAT: Securely Mitigating Rowhammer with Per-Row Activation Counters}. In \bibinfo{booktitle}{\emph{Proceedings of the 30th ACM International Conference on Architectural Support for Programming Languages and Operating Systems, Volume 1}} (Rotterdam, Netherlands) \emph{(\bibinfo{series}{ASPLOS '25})}. \bibinfo{publisher}{Association for Computing Machinery}, \bibinfo{address}{New York, NY, USA}, \bibinfo{pages}{698–714}.
\newblock
\showISBNx{9798400706981}
\urldef\tempurl%
\url{https://doi.org/10.1145/3669940.3707278}
\showDOI{\tempurl}


\bibitem[Qureshi et~al\mbox{.}(2024)]%
        {MINT}
\bibfield{author}{\bibinfo{person}{Moinuddin Qureshi}, \bibinfo{person}{Salman Qazi}, {and} \bibinfo{person}{Aamer Jaleel}.} \bibinfo{year}{2024}\natexlab{}.
\newblock \showarticletitle{MINT: Securely Mitigating Rowhammer with a Minimalist in-DRAM Tracker}. In \bibinfo{booktitle}{\emph{2024 57th IEEE/ACM International Symposium on Microarchitecture (MICRO)}}. \bibinfo{pages}{899--914}.
\newblock
\urldef\tempurl%
\url{https://doi.org/10.1109/MICRO61859.2024.00071}
\showDOI{\tempurl}


\bibitem[Qureshi et~al\mbox{.}(2022)]%
        {qureshi2022hydra}
\bibfield{author}{\bibinfo{person}{Moinuddin Qureshi}, \bibinfo{person}{Aditya Rohan}, \bibinfo{person}{Gururaj Saileshwar}, {and} \bibinfo{person}{Prashant~J. Nair}.} \bibinfo{year}{2022}\natexlab{}.
\newblock \showarticletitle{Hydra: enabling low-overhead mitigation of row-hammer at ultra-low thresholds via hybrid tracking}. In \bibinfo{booktitle}{\emph{Proceedings of the 49th Annual International Symposium on Computer Architecture}} (New York, New York) \emph{(\bibinfo{series}{ISCA '22})}. \bibinfo{publisher}{Association for Computing Machinery}, \bibinfo{address}{New York, NY, USA}, \bibinfo{pages}{699–710}.
\newblock
\showISBNx{9781450386104}
\urldef\tempurl%
\url{https://doi.org/10.1145/3470496.3527421}
\showDOI{\tempurl}


\bibitem[Qureshi et~al\mbox{.}(2015)]%
        {avatar}
\bibfield{author}{\bibinfo{person}{Moinuddin~K. Qureshi}, \bibinfo{person}{Dae-Hyun Kim}, \bibinfo{person}{Samira Khan}, \bibinfo{person}{Prashant~J. Nair}, {and} \bibinfo{person}{Onur Mutlu}.} \bibinfo{year}{2015}\natexlab{}.
\newblock \showarticletitle{AVATAR: A Variable-Retention-Time (VRT) Aware Refresh for DRAM Systems}. In \bibinfo{booktitle}{\emph{2015 45th Annual IEEE/IFIP International Conference on Dependable Systems and Networks}}. \bibinfo{pages}{427--437}.
\newblock
\urldef\tempurl%
\url{https://doi.org/10.1109/DSN.2015.58}
\showDOI{\tempurl}


\bibitem[Rakin et~al\mbox{.}(2022)]%
        {deepsteal}
\bibfield{author}{\bibinfo{person}{Adnan~Siraj Rakin}, \bibinfo{person}{Md~Hafizul~Islam Chowdhuryy}, \bibinfo{person}{Fan Yao}, {and} \bibinfo{person}{Deliang Fan}.} \bibinfo{year}{2022}\natexlab{}.
\newblock \showarticletitle{DeepSteal: Advanced Model Extractions Leveraging Efficient Weight Stealing in Memories}. In \bibinfo{booktitle}{\emph{2022 IEEE Symposium on Security and Privacy (SP)}}. \bibinfo{pages}{1157--1174}.
\newblock
\urldef\tempurl%
\url{https://doi.org/10.1109/SP46214.2022.9833743}
\showDOI{\tempurl}


\bibitem[Razavi et~al\mbox{.}(2016)]%
        {flipfengshui}
\bibfield{author}{\bibinfo{person}{Kaveh Razavi}, \bibinfo{person}{Ben Gras}, \bibinfo{person}{Erik Bosman}, \bibinfo{person}{Bart Preneel}, \bibinfo{person}{Cristiano Giuffrida}, {and} \bibinfo{person}{Herbert Bos}.} \bibinfo{year}{2016}\natexlab{}.
\newblock \showarticletitle{Flip Feng Shui: Hammering a Needle in the Software Stack}. In \bibinfo{booktitle}{\emph{25th USENIX Security Symposium (USENIX Security 16)}}. \bibinfo{publisher}{USENIX Association}, \bibinfo{address}{Austin, TX}, \bibinfo{pages}{1--18}.
\newblock
\showISBNx{978-1-931971-32-4}
\urldef\tempurl%
\url{https://www.usenix.org/conference/usenixsecurity16/technical-sessions/presentation/razavi}
\showURL{%
\tempurl}


\bibitem[Rixner et~al\mbox{.}(2000)]%
        {FRFCFS}
\bibfield{author}{\bibinfo{person}{S. Rixner}, \bibinfo{person}{W.J. Dally}, \bibinfo{person}{U.J. Kapasi}, \bibinfo{person}{P. Mattson}, {and} \bibinfo{person}{J.D. Owens}.} \bibinfo{year}{2000}\natexlab{}.
\newblock \showarticletitle{Memory access scheduling}. In \bibinfo{booktitle}{\emph{Proceedings of 27th International Symposium on Computer Architecture (IEEE Cat. No.RS00201)}}. \bibinfo{pages}{128--138}.
\newblock
\urldef\tempurl%
\url{https://doi.org/10.1145/339647.339668}
\showDOI{\tempurl}


\bibitem[Saileshwar et~al\mbox{.}(2018b)]%
        {morphcounter}
\bibfield{author}{\bibinfo{person}{Gururaj Saileshwar}, \bibinfo{person}{Prashant~J. Nair}, \bibinfo{person}{Prakash Ramrakhyani}, \bibinfo{person}{Wendy Elsasser}, \bibinfo{person}{Jose~A. Joao}, {and} \bibinfo{person}{Moinuddin~K. Qureshi}.} \bibinfo{year}{2018}\natexlab{b}.
\newblock \showarticletitle{Morphable Counters: Enabling Compact Integrity Trees For Low-Overhead Secure Memories}. In \bibinfo{booktitle}{\emph{2018 51st Annual IEEE/ACM International Symposium on Microarchitecture (MICRO)}}. \bibinfo{pages}{416--427}.
\newblock
\urldef\tempurl%
\url{https://doi.org/10.1109/MICRO.2018.00041}
\showDOI{\tempurl}


\bibitem[Saileshwar et~al\mbox{.}(2018a)]%
        {saileshwar2018synergy}
\bibfield{author}{\bibinfo{person}{Gururaj Saileshwar}, \bibinfo{person}{Prashant~J. Nair}, \bibinfo{person}{Prakash Ramrakhyani}, \bibinfo{person}{Wendy Elsasser}, {and} \bibinfo{person}{Moinuddin~K. Qureshi}.} \bibinfo{year}{2018}\natexlab{a}.
\newblock \showarticletitle{SYNERGY: Rethinking Secure-Memory Design for Error-Correcting Memories}. In \bibinfo{booktitle}{\emph{2018 IEEE International Symposium on High Performance Computer Architecture (HPCA)}}. \bibinfo{pages}{454--465}.
\newblock
\urldef\tempurl%
\url{https://doi.org/10.1109/HPCA.2018.00046}
\showDOI{\tempurl}


\bibitem[Saileshwar et~al\mbox{.}(2022)]%
        {saileshwar2022RRS}
\bibfield{author}{\bibinfo{person}{Gururaj Saileshwar}, \bibinfo{person}{Bolin Wang}, \bibinfo{person}{Moinuddin Qureshi}, {and} \bibinfo{person}{Prashant~J. Nair}.} \bibinfo{year}{2022}\natexlab{}.
\newblock \showarticletitle{Randomized row-swap: mitigating Row Hammer by breaking spatial correlation between aggressor and victim rows}. In \bibinfo{booktitle}{\emph{Proceedings of the 27th ACM International Conference on Architectural Support for Programming Languages and Operating Systems}} (Lausanne, Switzerland) \emph{(\bibinfo{series}{ASPLOS '22})}. \bibinfo{publisher}{Association for Computing Machinery}, \bibinfo{address}{New York, NY, USA}, \bibinfo{pages}{1056–1069}.
\newblock
\showISBNx{9781450392051}
\urldef\tempurl%
\url{https://doi.org/10.1145/3503222.3507716}
\showDOI{\tempurl}


\bibitem[Saxena et~al\mbox{.}(2024)]%
        {saxena2024rubix}
\bibfield{author}{\bibinfo{person}{Anish Saxena}, \bibinfo{person}{Saurav Mathur}, {and} \bibinfo{person}{Moinuddin Qureshi}.} \bibinfo{year}{2024}\natexlab{}.
\newblock \showarticletitle{Rubix: Reducing the Overhead of Secure Rowhammer Mitigations via Randomized Line-to-Row Mapping}. In \bibinfo{booktitle}{\emph{Proceedings of the 29th ACM International Conference on Architectural Support for Programming Languages and Operating Systems, Volume 2}} (La Jolla, CA, USA) \emph{(\bibinfo{series}{ASPLOS '24})}. \bibinfo{publisher}{Association for Computing Machinery}, \bibinfo{address}{New York, NY, USA}, \bibinfo{pages}{1014–1028}.
\newblock
\showISBNx{9798400703850}
\urldef\tempurl%
\url{https://doi.org/10.1145/3620665.3640404}
\showDOI{\tempurl}


\bibitem[Saxena and Qureshi(2024)]%
        {saxena2024start}
\bibfield{author}{\bibinfo{person}{Anish Saxena} {and} \bibinfo{person}{Moinuddin Qureshi}.} \bibinfo{year}{2024}\natexlab{}.
\newblock \showarticletitle{START: Scalable Tracking for any Rowhammer Threshold}. In \bibinfo{booktitle}{\emph{2024 IEEE International Symposium on High-Performance Computer Architecture (HPCA)}}. \bibinfo{pages}{578--592}.
\newblock
\urldef\tempurl%
\url{https://doi.org/10.1109/HPCA57654.2024.00049}
\showDOI{\tempurl}


\bibitem[Saxena et~al\mbox{.}(2022)]%
        {AQUA}
\bibfield{author}{\bibinfo{person}{Anish Saxena}, \bibinfo{person}{Gururaj Saileshwar}, \bibinfo{person}{Prashant~J. Nair}, {and} \bibinfo{person}{Moinuddin Qureshi}.} \bibinfo{year}{2022}\natexlab{}.
\newblock \showarticletitle{AQUA: Scalable Rowhammer Mitigation by Quarantining Aggressor Rows at Runtime}. In \bibinfo{booktitle}{\emph{2022 55th IEEE/ACM International Symposium on Microarchitecture (MICRO)}}. \bibinfo{pages}{108--123}.
\newblock
\urldef\tempurl%
\url{https://doi.org/10.1109/MICRO56248.2022.00022}
\showDOI{\tempurl}


\bibitem[Seaborn and Dullien(2015)]%
        {seaborn2015exploiting}
\bibfield{author}{\bibinfo{person}{Mark Seaborn} {and} \bibinfo{person}{Thomas Dullien}.} \bibinfo{year}{2015}\natexlab{}.
\newblock \showarticletitle{{Exploiting the DRAM rowhammer bug to gain kernel privileges}}.
\newblock \bibinfo{journal}{\emph{Black Hat}}  \bibinfo{volume}{15} (\bibinfo{year}{2015}), \bibinfo{pages}{71}.
\newblock


\bibitem[Seyedzadeh et~al\mbox{.}(2018)]%
        {CBT}
\bibfield{author}{\bibinfo{person}{Seyed~Mohammad Seyedzadeh}, \bibinfo{person}{Alex~K. Jones}, {and} \bibinfo{person}{Rami Melhem}.} \bibinfo{year}{2018}\natexlab{}.
\newblock \showarticletitle{Mitigating wordline crosstalk using adaptive trees of counters}. In \bibinfo{booktitle}{\emph{Proceedings of the 45th Annual International Symposium on Computer Architecture}} (Los Angeles, California) \emph{(\bibinfo{series}{ISCA '18})}. \bibinfo{publisher}{IEEE Press}, \bibinfo{pages}{612–623}.
\newblock
\showISBNx{9781538659847}
\urldef\tempurl%
\url{https://doi.org/10.1109/ISCA.2018.00057}
\showDOI{\tempurl}


\bibitem[Shafiee et~al\mbox{.}(2015)]%
        {FixedService}
\bibfield{author}{\bibinfo{person}{Ali Shafiee}, \bibinfo{person}{Akhila Gundu}, \bibinfo{person}{Manjunath Shevgoor}, \bibinfo{person}{Rajeev Balasubramonian}, {and} \bibinfo{person}{Mohit Tiwari}.} \bibinfo{year}{2015}\natexlab{}.
\newblock \showarticletitle{Avoiding information leakage in the memory controller with fixed service policies}. In \bibinfo{booktitle}{\emph{Proceedings of the 48th International Symposium on Microarchitecture}} (Waikiki, Hawaii) \emph{(\bibinfo{series}{MICRO-48})}. \bibinfo{publisher}{Association for Computing Machinery}, \bibinfo{address}{New York, NY, USA}, \bibinfo{pages}{89–101}.
\newblock
\showISBNx{9781450340342}
\urldef\tempurl%
\url{https://doi.org/10.1145/2830772.2830795}
\showDOI{\tempurl}


\bibitem[Shin et~al\mbox{.}(2018)]%
        {CCS18:PrefetcherChannels}
\bibfield{author}{\bibinfo{person}{Youngjoo Shin}, \bibinfo{person}{Hyung~Chan Kim}, \bibinfo{person}{Dokeun Kwon}, \bibinfo{person}{Ji~Hoon Jeong}, {and} \bibinfo{person}{Junbeom Hur}.} \bibinfo{year}{2018}\natexlab{}.
\newblock \showarticletitle{Unveiling Hardware-based Data Prefetcher, a Hidden Source of Information Leakage}. In \bibinfo{booktitle}{\emph{Proceedings of the 2018 ACM SIGSAC Conference on Computer and Communications Security}} (Toronto, Canada) \emph{(\bibinfo{series}{CCS '18})}. \bibinfo{publisher}{Association for Computing Machinery}, \bibinfo{address}{New York, NY, USA}, \bibinfo{pages}{131–145}.
\newblock
\showISBNx{9781450356930}
\urldef\tempurl%
\url{https://doi.org/10.1145/3243734.3243736}
\showDOI{\tempurl}


\bibitem[Son et~al\mbox{.}(2017)]%
        {PROHIT}
\bibfield{author}{\bibinfo{person}{Mungyu Son}, \bibinfo{person}{Hyunsun Park}, \bibinfo{person}{Junwhan Ahn}, {and} \bibinfo{person}{Sungjoo Yoo}.} \bibinfo{year}{2017}\natexlab{}.
\newblock \showarticletitle{Making DRAM stronger against row hammering}. In \bibinfo{booktitle}{\emph{2017 54th ACM/EDAC/IEEE Design Automation Conference (DAC)}}. \bibinfo{pages}{1--6}.
\newblock
\urldef\tempurl%
\url{https://doi.org/10.1145/3061639.3062281}
\showDOI{\tempurl}


\bibitem[SPEC2017({[n.\,d.]})]%
        {SPEC2017}
SPEC2017 \bibinfo{year}{[n.\,d.]}\natexlab{}.
\newblock \bibinfo{title}{{SPEC CPU2017 Benchmark Suite}}.
\newblock \bibinfo{howpublished}{Standard Performance Evaluation Corporation}.
\newblock
\urldef\tempurl%
\url{http://www.spec.org/cpu2017/}
\showURL{%
\tempurl}


\bibitem[van~der Veen et~al\mbox{.}(2016)]%
        {vanderveen2016drammer}
\bibfield{author}{\bibinfo{person}{Victor van~der Veen}, \bibinfo{person}{Yanick Fratantonio}, \bibinfo{person}{Martina Lindorfer}, \bibinfo{person}{Daniel Gruss}, \bibinfo{person}{Clementine Maurice}, \bibinfo{person}{Giovanni Vigna}, \bibinfo{person}{Herbert Bos}, \bibinfo{person}{Kaveh Razavi}, {and} \bibinfo{person}{Cristiano Giuffrida}.} \bibinfo{year}{2016}\natexlab{}.
\newblock \showarticletitle{Drammer: Deterministic Rowhammer Attacks on Mobile Platforms}. In \bibinfo{booktitle}{\emph{Proceedings of the 2016 ACM SIGSAC Conference on Computer and Communications Security}} (Vienna, Austria) \emph{(\bibinfo{series}{CCS '16})}. \bibinfo{publisher}{Association for Computing Machinery}, \bibinfo{address}{New York, NY, USA}, \bibinfo{pages}{1675–1689}.
\newblock
\showISBNx{9781450341394}
\urldef\tempurl%
\url{https://doi.org/10.1145/2976749.2978406}
\showDOI{\tempurl}


\bibitem[Vicarte et~al\mbox{.}(2022)]%
        {Augury}
\bibfield{author}{\bibinfo{person}{Jose Rodrigo~Sanchez Vicarte}, \bibinfo{person}{Michael Flanders}, \bibinfo{person}{Riccardo Paccagnella}, \bibinfo{person}{Grant Garrett-Grossman}, \bibinfo{person}{Adam Morrison}, \bibinfo{person}{Christopher~W. Fletcher}, {and} \bibinfo{person}{David Kohlbrenner}.} \bibinfo{year}{2022}\natexlab{}.
\newblock \showarticletitle{Augury: Using Data Memory-Dependent Prefetchers to Leak Data at Rest}. In \bibinfo{booktitle}{\emph{2022 IEEE Symposium on Security and Privacy (SP)}}. \bibinfo{pages}{1491--1505}.
\newblock
\urldef\tempurl%
\url{https://doi.org/10.1109/SP46214.2022.9833570}
\showDOI{\tempurl}


\bibitem[Wang et~al\mbox{.}(2020)]%
        {wang_2020_dramdig}
\bibfield{author}{\bibinfo{person}{Minghua Wang}, \bibinfo{person}{Zhi Zhang}, \bibinfo{person}{Yueqiang Cheng}, {and} \bibinfo{person}{Surya Nepal}.} \bibinfo{year}{2020}\natexlab{}.
\newblock \showarticletitle{DRAMDig: a knowledge-assisted tool to uncover DRAM address mapping}. In \bibinfo{booktitle}{\emph{Proceedings of the 57th ACM/EDAC/IEEE Design Automation Conference}} (Virtual Event, USA) \emph{(\bibinfo{series}{DAC '20})}. \bibinfo{publisher}{IEEE Press}, Article \bibinfo{articleno}{89}, \bibinfo{numpages}{6}~pages.
\newblock
\showISBNx{9781450367257}


\bibitem[Wang et~al\mbox{.}(2014)]%
        {TemporalPartitioning}
\bibfield{author}{\bibinfo{person}{Yao Wang}, \bibinfo{person}{Andrew Ferraiuolo}, {and} \bibinfo{person}{G.~Edward Suh}.} \bibinfo{year}{2014}\natexlab{}.
\newblock \showarticletitle{Timing channel protection for a shared memory controller}. In \bibinfo{booktitle}{\emph{2014 IEEE 20th International Symposium on High Performance Computer Architecture (HPCA)}}. \bibinfo{pages}{225--236}.
\newblock
\urldef\tempurl%
\url{https://doi.org/10.1109/HPCA.2014.6835934}
\showDOI{\tempurl}


\bibitem[Wang et~al\mbox{.}(2017)]%
        {3Rank2Bank}
\bibfield{author}{\bibinfo{person}{Yao Wang}, \bibinfo{person}{Benjamin Wu}, {and} \bibinfo{person}{G.~Edward Suh}.} \bibinfo{year}{2017}\natexlab{}.
\newblock \showarticletitle{Secure Dynamic Memory Scheduling Against Timing Channel Attacks}. In \bibinfo{booktitle}{\emph{2017 IEEE International Symposium on High Performance Computer Architecture (HPCA)}}. \bibinfo{pages}{301--312}.
\newblock
\urldef\tempurl%
\url{https://doi.org/10.1109/HPCA.2017.27}
\showDOI{\tempurl}


\bibitem[Wi et~al\mbox{.}(2023)]%
        {ShadowHPCA23}
\bibfield{author}{\bibinfo{person}{Minbok Wi}, \bibinfo{person}{Jaehyun Park}, \bibinfo{person}{Seoyoung Ko}, \bibinfo{person}{Michael~Jaemin Kim}, \bibinfo{person}{Nam Sung~Kim}, \bibinfo{person}{Eojin Lee}, {and} \bibinfo{person}{Jung~Ho Ahn}.} \bibinfo{year}{2023}\natexlab{}.
\newblock \showarticletitle{SHADOW: Preventing Row Hammer in DRAM with Intra-Subarray Row Shuffling}. In \bibinfo{booktitle}{\emph{2023 IEEE International Symposium on High-Performance Computer Architecture (HPCA)}}. \bibinfo{pages}{333--346}.
\newblock
\urldef\tempurl%
\url{https://doi.org/10.1109/HPCA56546.2023.10070966}
\showDOI{\tempurl}


\bibitem[Woo et~al\mbox{.}(2025)]%
        {qprac}
\bibfield{author}{\bibinfo{person}{Jeonghyun Woo}, \bibinfo{person}{Shaopeng~Chris Lin}, \bibinfo{person}{Prashant~J. Nair}, \bibinfo{person}{Aamer Jaleel}, {and} \bibinfo{person}{Gururaj Saileshwar}.} \bibinfo{year}{2025}\natexlab{}.
\newblock \showarticletitle{QPRAC: Towards Secure and Practical PRAC-based Rowhammer Mitigation using Priority Queues}. In \bibinfo{booktitle}{\emph{2025 IEEE International Symposium on High Performance Computer Architecture (HPCA)}}. \bibinfo{pages}{1021--1037}.
\newblock
\urldef\tempurl%
\url{https://doi.org/10.1109/HPCA61900.2025.00080}
\showDOI{\tempurl}


\bibitem[Woo and Nair(2025)]%
        {dapper}
\bibfield{author}{\bibinfo{person}{Jeonghyun Woo} {and} \bibinfo{person}{Prashant~J. Nair}.} \bibinfo{year}{2025}\natexlab{}.
\newblock \showarticletitle{DAPPER: A Performance-Attack-Resilient Tracker for RowHammer Defense}. In \bibinfo{booktitle}{\emph{2025 IEEE International Symposium on High Performance Computer Architecture (HPCA)}}. \bibinfo{pages}{1005--1020}.
\newblock
\urldef\tempurl%
\url{https://doi.org/10.1109/HPCA61900.2025.00079}
\showDOI{\tempurl}


\bibitem[Woo et~al\mbox{.}(2023)]%
        {SRS}
\bibfield{author}{\bibinfo{person}{Jeonghyun Woo}, \bibinfo{person}{Gururaj Saileshwar}, {and} \bibinfo{person}{Prashant~J. Nair}.} \bibinfo{year}{2023}\natexlab{}.
\newblock \showarticletitle{Scalable and Secure Row-Swap: Efficient and Safe Row Hammer Mitigation in Memory Systems}. In \bibinfo{booktitle}{\emph{2023 IEEE International Symposium on High-Performance Computer Architecture (HPCA)}}. \bibinfo{pages}{374--389}.
\newblock
\urldef\tempurl%
\url{https://doi.org/10.1109/HPCA56546.2023.10070999}
\showDOI{\tempurl}


\bibitem[Ya{\u{g}}lik{\c{c}}i et~al\mbox{.}(2021)]%
        {yauglikcci2021blockhammer}
\bibfield{author}{\bibinfo{person}{A.~Giray Ya{\u{g}}lik{\c{c}}i}, \bibinfo{person}{Minesh Patel}, \bibinfo{person}{Jeremie~S. Kim}, \bibinfo{person}{Roknoddin Azizi}, \bibinfo{person}{Ataberk Olgun}, \bibinfo{person}{Lois Orosa}, \bibinfo{person}{Hasan Hassan}, \bibinfo{person}{Jisung Park}, \bibinfo{person}{Konstantinos Kanellopoulos}, \bibinfo{person}{Taha Shahroodi}, \bibinfo{person}{Saugata Ghose}, {and} \bibinfo{person}{Onur Mutlu}.} \bibinfo{year}{2021}\natexlab{}.
\newblock \showarticletitle{BlockHammer: Preventing RowHammer at Low Cost by Blacklisting Rapidly-Accessed DRAM Rows}. In \bibinfo{booktitle}{\emph{2021 IEEE International Symposium on High-Performance Computer Architecture (HPCA)}}. \bibinfo{pages}{345--358}.
\newblock
\urldef\tempurl%
\url{https://doi.org/10.1109/HPCA51647.2021.00037}
\showDOI{\tempurl}


\bibitem[Yan et~al\mbox{.}(2006)]%
        {SplitCounter}
\bibfield{author}{\bibinfo{person}{Chenyu Yan}, \bibinfo{person}{D. Englender}, \bibinfo{person}{M. Prvulovic}, \bibinfo{person}{B. Rogers}, {and} \bibinfo{person}{Yan Solihin}.} \bibinfo{year}{2006}\natexlab{}.
\newblock \showarticletitle{Improving Cost, Performance, and Security of Memory Encryption and Authentication}. In \bibinfo{booktitle}{\emph{33rd International Symposium on Computer Architecture (ISCA'06)}}. \bibinfo{pages}{179--190}.
\newblock
\urldef\tempurl%
\url{https://doi.org/10.1109/ISCA.2006.22}
\showDOI{\tempurl}


\bibitem[Yan et~al\mbox{.}(2019)]%
        {DirectoryAttack}
\bibfield{author}{\bibinfo{person}{Mengjia Yan}, \bibinfo{person}{Read Sprabery}, \bibinfo{person}{Bhargava Gopireddy}, \bibinfo{person}{Christopher Fletcher}, \bibinfo{person}{Roy Campbell}, {and} \bibinfo{person}{Josep Torrellas}.} \bibinfo{year}{2019}\natexlab{}.
\newblock \showarticletitle{Attack Directories, Not Caches: Side Channel Attacks in a Non-Inclusive World}. In \bibinfo{booktitle}{\emph{2019 IEEE Symposium on Security and Privacy (SP)}}. \bibinfo{pages}{888--904}.
\newblock
\urldef\tempurl%
\url{https://doi.org/10.1109/SP.2019.00004}
\showDOI{\tempurl}


\bibitem[Yao et~al\mbox{.}(2020)]%
        {deephammer}
\bibfield{author}{\bibinfo{person}{Fan Yao}, \bibinfo{person}{Adnan~Siraj Rakin}, {and} \bibinfo{person}{Deliang Fan}.} \bibinfo{year}{2020}\natexlab{}.
\newblock \showarticletitle{{DeepHammer}: Depleting the Intelligence of Deep Neural Networks through Targeted Chain of Bit Flips}. In \bibinfo{booktitle}{\emph{29th USENIX Security Symposium (USENIX Security 20)}}. \bibinfo{publisher}{USENIX Association}, \bibinfo{pages}{1463--1480}.
\newblock
\showISBNx{978-1-939133-17-5}
\urldef\tempurl%
\url{https://www.usenix.org/conference/usenixsecurity20/presentation/yao}
\showURL{%
\tempurl}


\bibitem[Yarom and Falkner(2014)]%
        {FlushReload}
\bibfield{author}{\bibinfo{person}{Yuval Yarom} {and} \bibinfo{person}{Katrina Falkner}.} \bibinfo{year}{2014}\natexlab{}.
\newblock \showarticletitle{{FLUSH+ RELOAD: A high resolution, low noise, l3 cache Side-Channel attack}}. In \bibinfo{booktitle}{\emph{23rd USENIX security symposium (USENIX Security)}}. \bibinfo{pages}{719--732}.
\newblock


\bibitem[Yağlikçi et~al\mbox{.}(2022)]%
        {HIRA}
\bibfield{author}{\bibinfo{person}{A.~Giray Yağlikçi}, \bibinfo{person}{Ataberk Olgun}, \bibinfo{person}{Minesh Patel}, \bibinfo{person}{Haocong Luo}, \bibinfo{person}{Hasan Hassan}, \bibinfo{person}{Lois Orosa}, \bibinfo{person}{Oğuz Ergin}, {and} \bibinfo{person}{Onur Mutlu}.} \bibinfo{year}{2022}\natexlab{}.
\newblock \showarticletitle{HiRA: Hidden Row Activation for Reducing Refresh Latency of Off-the-Shelf DRAM Chips}. In \bibinfo{booktitle}{\emph{2022 55th IEEE/ACM International Symposium on Microarchitecture (MICRO)}}. \bibinfo{pages}{815--834}.
\newblock
\urldef\tempurl%
\url{https://doi.org/10.1109/MICRO56248.2022.00062}
\showDOI{\tempurl}


\bibitem[Yağlıkçı et~al\mbox{.}(2021)]%
        {wave}
\bibfield{author}{\bibinfo{person}{Abdullah~Giray Yağlıkçı}, \bibinfo{person}{Jeremie~S. Kim}, \bibinfo{person}{Fabrice Devaux}, {and} \bibinfo{person}{Onur Mutlu}.} \bibinfo{year}{2021}\natexlab{}.
\newblock \bibinfo{title}{Security Analysis of the Silver Bullet Technique for RowHammer Prevention}.
\newblock
\newblock
\showeprint[arxiv]{2106.07084}~[cs.CR]
\urldef\tempurl%
\url{https://arxiv.org/abs/2106.07084}
\showURL{%
\tempurl}


\bibitem[Yağlıkçı et~al\mbox{.}(2024)]%
        {yauglikcci2024spatial}
\bibfield{author}{\bibinfo{person}{Abdullah~Giray Yağlıkçı}, \bibinfo{person}{Yahya~Can Tuğrul}, \bibinfo{person}{Geraldo~F. Oliveira}, \bibinfo{person}{İsmail~Emir Yüksel}, \bibinfo{person}{Ataberk Olgun}, \bibinfo{person}{Haocong Luo}, {and} \bibinfo{person}{Onur Mutlu}.} \bibinfo{year}{2024}\natexlab{}.
\newblock \showarticletitle{Spatial Variation-Aware Read Disturbance Defenses: Experimental Analysis of Real DRAM Chips and Implications on Future Solutions}. In \bibinfo{booktitle}{\emph{2024 IEEE International Symposium on High-Performance Computer Architecture (HPCA)}}. \bibinfo{pages}{560--577}.
\newblock
\urldef\tempurl%
\url{https://doi.org/10.1109/HPCA57654.2024.00048}
\showDOI{\tempurl}


\bibitem[You and Yang(2019)]%
        {MRLOC}
\bibfield{author}{\bibinfo{person}{Jung~Min You} {and} \bibinfo{person}{Joon-Sung Yang}.} \bibinfo{year}{2019}\natexlab{}.
\newblock \showarticletitle{MRLoc: Mitigating Row-hammering based on memory Locality}. In \bibinfo{booktitle}{\emph{2019 56th ACM/IEEE Design Automation Conference (DAC)}}. \bibinfo{pages}{1--6}.
\newblock


\bibitem[Young et~al\mbox{.}(2015)]%
        {deuce}
\bibfield{author}{\bibinfo{person}{Vinson Young}, \bibinfo{person}{Prashant~J. Nair}, {and} \bibinfo{person}{Moinuddin~K. Qureshi}.} \bibinfo{year}{2015}\natexlab{}.
\newblock \showarticletitle{DEUCE: Write-Efficient Encryption for Non-Volatile Memories}. In \bibinfo{booktitle}{\emph{Proceedings of the Twentieth International Conference on Architectural Support for Programming Languages and Operating Systems}} (Istanbul, Turkey) \emph{(\bibinfo{series}{ASPLOS '15})}. \bibinfo{publisher}{Association for Computing Machinery}, \bibinfo{address}{New York, NY, USA}, \bibinfo{pages}{33–44}.
\newblock
\showISBNx{9781450328357}
\urldef\tempurl%
\url{https://doi.org/10.1145/2694344.2694387}
\showDOI{\tempurl}


\bibitem[Zhou et~al\mbox{.}(2017)]%
        {Camouflage}
\bibfield{author}{\bibinfo{person}{Yanqi Zhou}, \bibinfo{person}{Sameer Wagh}, \bibinfo{person}{Prateek Mittal}, {and} \bibinfo{person}{David Wentzlaff}.} \bibinfo{year}{2017}\natexlab{}.
\newblock \showarticletitle{Camouflage: Memory Traffic Shaping to Mitigate Timing Attacks}. In \bibinfo{booktitle}{\emph{2017 IEEE International Symposium on High Performance Computer Architecture (HPCA)}}. \bibinfo{pages}{337--348}.
\newblock
\urldef\tempurl%
\url{https://doi.org/10.1109/HPCA.2017.36}
\showDOI{\tempurl}


\bibitem[Zuravleff and Robinson(1997)]%
        {zuravleff1997controller}
\bibfield{author}{\bibinfo{person}{William~K Zuravleff} {and} \bibinfo{person}{Timothy Robinson}.} \bibinfo{year}{1997}\natexlab{}.
\newblock \bibinfo{title}{Controller for a synchronous DRAM that maximizes throughput by allowing memory requests and commands to be issued out of order}.
\newblock
\newblock
\newblock
\shownote{US Patent 5,630,096}.


\end{thebibliography}

\end{document}